# A Pyroxenite Mantle on Mercury? Experimental Insights from Enstatite Chondrite Melting at Pressures up to 5 GPa

Asmaa BOUJIBAR[1,2*], Kevin RIGHTER[2,3], Emmanuel FONTAINE[4,5], Max COLLINET[6-7], Sarah LAMBART[8], Larry R. NITTLER[9], Kellye M. PANDO[2,10]

[1]Geology Dept., Dept. of Physics & Astronomy, Advanced Materials Science & Engineering Center, Western Washington University, 516 High St, Bellingham, WA, 98225 USA

[2]NASA Johnson Space Center, 2101 E Nasa Pkwy, Houston, TX, 77058 USA

[3]Dept. of Earth and Environmental Sciences, University of Rochester, 120 Trustee Rd, Rochester, NY, 14620 USA

[4]CNRM, Université de Toulouse, Météo-France, CNRS, 42 Av. Gaspard Coriolis, Toulouse, 31057 France

[5]Laboratoire de Météorologie Physique, Université Clermont Auvergne, 4 Av. Blaise Pascal, Aubière, 63178 France

[6]Dept. of Earth, Atmospheric & Planetary Sciences, Massachusetts Institute of Technology, 77 Massachusetts Avenue, 55-101 Cambridge, MA, 02139 USA

[7]Département de Géologie, Université de Namur, Rue de Bruxelles 61, B-5000 Namur, Belgique

[8]Dept. of Geology & Geophysics, University of Utah, 115 S 1460 E, Salt Lake City, UT, 84112 USA

[9]School of Earth and Space Exploration, Arizona State University, PSF 618 Mail Code 6004, Tempe, AZ, 85287 USA

[10]Jacobs, NASA Johnson Space Center, 2101 E Nasa Pkwy, Houston, TX, 77058 USA

[*]Corresponding author (boujiba@wwu.edu).

## Highlights:

- Enstatite stability expands due to Ca-S and Mg-S complexes in silicate melts.
- Sulfides are richer in Mg and Ca at low temperature and when silicate melts are rich in silica, respectively.
- Ca- and Mg-rich sulfides in equilibrium with silicates generate more variety of silicate melt compositions.
- High-pressure melts (3–5 GPa) are Mg-rich, resembling Mercury's high-magnesium region.
- Low-pressure melts (0.5–1 GPa) are Si-rich, similar to the northern volcanic plains.

## Keywords:

# Abstract

Enstatite chondrites are potential source material for the accretion of Mercury due to their reduced nature and enrichment in volatile elements. Understanding their melting properties is therefore important to better assess a scenario where Mercury formed from these chondrites. Here, we present experimental data on the partial melting of a modified EH4 Indarch enstatite chondrite, which was adjusted to have 18% more metallic Si than $SiO_2$ in mass, yielding an oxygen fugacity of 3.7 ±0.6 below iron-wüstite redox buffer and 12 wt% Si in the metal. Experiments were performed from 0.5 to 5 GPa using piston cylinder and multi-anvil apparatuses. Results indicate that the stability field of enstatite expands relative to olivine. This expansion is likely due to the presence of Ca-S and Mg-S complexes in the silicate melt, which enhance $SiO_2$ activity and promote enstatite crystallization. Silicate melts present a correlation between Ca and S concentrations, like the global patterns seen on Mercury's surface but with higher sulfur abundances. Additionally, sulfides show enrichment in Mg and Ca, up to 22 and 13 wt% respectively, the main remaining cations being Fe, Cr and Mn. These high Mg and Ca contents are observed at low temperatures and high silica content in the silicate melt, respectively. Partial melting of this reduced EH4 chondrite yields a large range of silicate melt compositions, due to the Mg- and Ca-rich sulfides which act as significant residual phases. High-pressure melts (2 to 5 GPa, 160-400 km depth in Mercury) are Mg-rich, similar to those in Mercury's high-magnesium region (HMR), while low-pressure melts (0.5 to 1 GPa, 40-80 km depth) are Si-rich, comparable to the northern volcanic plains (NVP). Results suggest that a large fraction of Mercury's surface aligns compositionally with these melts, implying that Mercury's mantle could predominantly have a pyroxenitic composition. However, regions with differing compositions, such as aluminum-rich areas, like the Caloris basin, suggest local variability in mantle geochemistry. The HMR chemistry indicates melting at pressures up to the base of Mercury's mantle, possibly due to a large impact. Our study also explores whether the surface compositions could result from mixing processes like impact gardening or polybaric melting and magma mixing. The findings suggest that areas such as the intercrater plains and heavily cratered regions could be mixtures of melts from different depths, ranging from 0.5 to 5 GPa, which corresponds to the core-mantle boundary to lower pressures. Overall, our results show that if Mercury formed from materials similar to enstatite chondrites, batch melting of its primitive pyroxenite mantle would yield magmas with compositions resembling those of most rocks observed on the surface. While the exact olivine content of the mantle remains uncertain, the residual mantle is likely enstatite-rich due to the extensive stability of enstatite relative to olivine in sulfur-rich reduced systems.

# 1. Introduction

Mercury is the least oxidized terrestrial planet of our Solar System with the largest iron core, comprising about 70 the planet's mass (Smith et al. 2012, Hauck et al. 2013), and the lowest FeO concentration on its surface, averaging 1.5 wt% (Weider et al. 2014). This suggests that Mercury's building materials were highly reduced during its accretion, with iron primarily in its metallic form. Data from the X-ray spectrometer (XRS) aboard the MESSENGER (MErcury Surface, Space ENvironment, GEochemistry, and Ranging) spacecraft have provided significant insights into Mercury's surface composition (e.g. Nittler et al. 2020), revealing unusually high sulfur concentrations compared to other terrestrial planets, averaging 2.6 wt% S (Nittler et al. 2011, Weider et al. 2015). This further supports Mercury's reduced state, as experimental studies show that sulfur partitions into silicate melts under low oxygen fugacity, and into metal at high oxygen fugacity (e.g. Boujibar et al. 2014, Namur et al., 2016a, McCubbin et al. 2012). In contrast, Earth and Mars are more oxidized, with lower sulfur concentrations in their crusts and mantles (< 0.1 wt%), and likely sulfur-enriched cores (e.g. Boujibar et al. 2014, Righter et al. 2021, Steenstra & van Westrenen 2018).

Data from MESSENGER XRS enabled detailed mapping of Mercury's surface, revealing major elemental compositions that highlight chemical heterogeneities (Weider et al. 2015, Nittler et al. 2020). These compositions include a variety of rock types, ranging from basalts, to basaltic-andesites, to andesites, to dacites and rhyolites in the Total Alkali-Silica (TAS) classification (Le Maitre et al., 2002). We note that based on the high MgO content, this rock suite was also proposed to be reclassified as picrites, komatiites and boninites (Peplowski & Stockstill-Cahill 2019). This highlights significant complexity in the differentiation of Mercury's crust (Charlier et al. 2013). Additionally, imaging of the planet's surface has revealed a range of volcanic features, including pyroclastic deposits and extensive effusive volcanism with deposits that are hundreds of meters to several kilometers thick (Head et al., 2011; Denevi et al., 2013; Marchi et al., 2013; Thomas et al., 2014; Klimczak et al., 2012; Byrne et al., 2013). Crater size–frequency analyses identified two major stages of volcanism: an early stage associated with heavily cratered volcanic terrains (∼4.2 Ga or older), and a second major volcanic episode that produced smooth volcanic terrains covering ~ 27% the planet's surface (4.2 to 3.5 Ga), with the latter having, on average, a lower Mg/Si ratio (e.g. Marchi et al. 2013, Byrne et al. 2016).

Mercury likely formed a primary graphite crust through graphite flotation in its primordial magma ocean stage (Vander Kaaden & McCubbin 2015). The potential existence of this crust is supported by the detection of low-reflectance material on the surface (Peplowski et al. 2016). However, much of this primary graphite has been destroyed by impacts or buried beneath volcanic deposits, which form the secondary silicate crust. The average carbon abundance in the crust is debated and has been suggested to represent from less than 1 wt% to 4-5 wt% (e.g. Peplowski et al. 2015, Xu et al. 2024). Thus, most of Mercury's current surface was formed through the differentiation and melting of a silicate mantle, resulting in a crust with variable thickness (ranging from 19 to 42 km) and density (Buoninfante et al. 2023, Padovan et al. 2015, Sori 2018). Previous studies based on high-pressure, high-temperature experiments combined with thermodynamic modeling suggest that these volcanic deposits originated from melts formed by partial melting of a lherzolitic source region with varying contents of clinopyroxene, and other Na-bearing phases (Namur et al. 2016b). Additionally, Vander Kaaden & McCubbin (2016) proposed that the silicic northern smooth volcanic plains likely formed from partial melting of a plagioclase-bearing mantle source.

The chemical composition of Mercury's mantle and its possible similarity to the silicate fraction of chondrites remain uncertain. Given its low oxygen fugacity, previous studies have suggested that Mercury's building materials might resemble EH enstatite chondrites or CB Bencubbin chondrites (Malavergne et al. 2010, Taylor & Scott 2003). Stable isotopic compositions suggest EH and CB chondrites formed in the inner and outer Solar System, respectively (Rüfenacht et al. 2023), which favor a scenario of formation of Mercury with EH chondrites. There is a lack of comprehensive studies on the melting products of these meteoritic compositions, which limits our ability to test chondritic models for Mercury's silicate composition. In particular, previous estimates of Mercury's mantle composition employed a methodology usually applied to basalt generation on Earth, which assumes that magmas are in equilibrium with a peridotite when they are segregated from the mantle (Namur et al. 2016b, Nittler et al. 2018). Hence, Namur et al (2016) estimation of the pressure of equilibration with the residual mantle is based on the assumption that the melt was in equilibrium with the olivine and enstatite before segregation. If the magma was only saturated in enstatite, this would correspond to higher pressure of magma segregation (see Fig. 3 in Namur et al., 2016). Therefore, here we employ a forward model approach by conducting a systematic study on the melting properties of an enstatite-rich hypothetical mantle, which is necessary to assess whether the compositions seen on Mercury's surface can be generated in conditions that satisfy meteorite data. Additionally, Mercury's S-rich composition implies that sulfide could have saturated in magmas (Namur et al. 2016a), leading to high S contents which significantly influenced the planet's differentiation. For example, if Mg- and Ca-bearing sulfides are sufficiently abundant, they could alter the chemical composition of magmas generated from mantle melting. The stability of sulfides would have significant implications for Mercury's internal heat generation, as sulfides can incorporate substantial amounts of heat-producing elements (U, Th and K) (Boujibar et al. 2019, Boukaré et al. 2019).

To date, only two systematic studies have reported experimental chemical compositions of silicate partial melts derived from EH4 enstatite chondrites: one at 1 bar (McCoy et al. 1999) and another at 1 GPa (Berthet et al. 2009). In the 1 bar experiments, the oxygen fugacity ($f$O$_2$) was controlled using external Cr and V metal buffers (McCoy et al. 1999). For the 1 GPa experiments, $f$O$_2$ was controlled with the addition of Si metal powder to the starting meteorite powder (Berthet et al. 2009). However, in the latter study, the oxidation of Si metal to SiO$_2$ during the experiment caused variations in the SiO$_2$ content of the silicate fraction, complicating the study of melting properties for a consistent composition. In the present work, we examine the melting products of EH4 enstatite chondrites from 0.5 to 5 GPa (pressures up to the core-mantle boundary), while carefully controlling both $f$O$_2$ and silicate composition. Our approach utilizes a synthetic powder similar to EH4 chondrites, with a bulk oxygen content lower than the original chondrites, resulting in a Si/SiO$_2$ of 0.18. This methodology enables us to compare Mercury's surface composition with the melting products of the silicate fraction of enstatite chondrites, providing insights into Mercury's differentiation. In this study, experimental results and thermodynamic modeling are used to constrain the compositions of silicate melts and sulfides. We then discuss the role of sulfur and sulfides in the melting properties of enstatite chondrites. By comparing experimental results with Mercury's surface composition, we discuss the mantle's composition and Mercury's evolution during magma ocean crystallization.

## 2. Methods

### 2.1 Starting composition:

We used a synthetic starting material resembling Indarch EH4 enstatite chondrites, but with an oxygen concentration lower than that of natural EH4 chondrites (Berthet et al. 2009, Wiik 1956). All major and minor elemental ratios but those involving oxygen, were fixed to match those of the Indarch EH4 chondrite (Table 1). Reducing the O content resulted in a metal and sulfide phase mass fraction of 50 wt%, which is larger than that of the Indarch chondrite but smaller than Mercury's core mass fraction ($\sim$ 68 %). This reduction was achieved by lowering the SiO$_2$ content at the expense of metallic Si, which was increased while maintaining a bulk Si concentration similar to that of the initial chondrite. A challenge with using Si metal is its inevitable partial oxidation, which affects achieving a reasonably constant oxygen fugacity and chemical composition of the silicate. Partial oxidation of Si enriches the silicate fraction in SiO$_2$, which can alter the mineralogical assemblage and increase the fraction of orthopyroxene, and potentially quartz if the system becomes oversaturated in SiO$_2$ (absent in our experimental charges but seen in Berthet et al. (2009)). In our experiments, all run products contain a liquid Fe-rich metal phase, with a relatively constant Si content across different samples (11.6 $\pm$0.8 wt%, see Table 4), ensuring that all samples experienced similar levels of Si oxidation. Considering a chondritic composition ensured that elemental ratios of the silicate assemblage remained constant, by monitoring the metal composition. In addition, the enrichment of Fe metal allowed FeO concentrations in the silicate melt to reach values closer to the $\sim$1.5 wt% FeO observed on Mercury's surface than in previous studies (0.5 $\pm$0.3 wt% compared to 0.08-0.09 wt% in Namur et al.

2016b). Mercury's mantle FeO content was reached during metal-silicate chemical exchanges in the magma ocean during core formation. The relatively high abundance of Fe Mercury's surface, compared to equally reduced experiments, suggests the presence of Fe-rich sulfides and possibly Fe metal in Mercury's mantle (Malavergne et al. 2014), which proportions are difficult to estimate but can be approached with our experiments. In fact, in our sulfide-saturated experiments, the calculated Fe/Si ratios of the combined silicate-sulfide assemblages are 0.006-0.07at 0.5 GPa, 0.007 to 0.04 at 1 GPa, 0.003 to 0.15 at 2 GPa, 0.1 to 0.16 at 3 GPa and 0.28 to 0.35 at 5 GPa. These Fe/Si ranges are consistent with the 0.02-0.1 measured by MESSENGER (Weider et al. 2014, Nittler et al., 2020).

## 2.2 Experimental methods:

We used high-purity powders of metals (Fe, Ni, Si, Co), oxides ($SiO_2$, MgO, $Al_2O_3$, CaO, $Cr_2O_3$, $MnO_2$, $TiO_2$), carbonates ($Na_2CO_3$, $K_2CO_3$) and sulfide (FeS). The metal-sulfide powders were mixed dry without any solvent, while the oxides and carbonates were dried at 1000 °C and 350 °C overnight, respectively, then finely mixed under ethanol and slowly decarbonated overnight. The silicate powder was then thoroughly mixed dry with the metal-sulfide mixture and stored in a desiccator. High-pressure and high-temperature experiments were conducted using two non-end loaded piston cylinders and a Walker-type multi-anvil press at NASA Johnson Space Center. The experiments were performed at 0.5, 1, 3 and 5 GPa and temperatures between 1250 and 1880 °C. For all experiments, sample powders were contained in graphite capsules, and temperature was measured with a W/Re thermocouple. For the piston cylinder experiments, we used 13- and 10-mm assemblages with a graphite heater, magnesia and alumina rods and barium carbonate sleeve pressure medium either wrapped in lead foil or covered with $MoS_2$ paste. For the multi-anvil press experiments, heating was achieved using a Re heater, and we used magnesia, alumina and zirconia rods and sleeves, a lanthanum chromite thermal insulator, $Cr_2O_3$-doped MgO octahedral pressure medium of 14 mm edge length, pyrophyllite gaskets and WC cubes with an 8 mm truncated-edge length (TEL). Details on calibration are provided in Righter et al. (2013) (multi-anvil) and Filiberto et al. (2008) (piston cylinder). Experiments were heated for varying durations depending on the temperature to allow for chemical equilibrium (Table 2). At the end of the experiments, recovered samples were mounted in epoxy and polished under methanol to prevent the dissolution and loss of sulfide phases.

## 2.3 Analytical methods:

Run products were analyzed using a Cameca SX100 and a FEG-JEOL 8530F electron probe microanalyzers (EPMA) at NASA JSC, to characterize their texture and chemical compositions. Sample images were collected using back-scattered electron microscopy, while chemical compositions were acquired using wavelength dispersive spectroscopy. Analyzes with the EPMA were performed with an accelerating voltage of 15 kV, and a beam current of 15 nA for silicates and sulfides, and 20 nA for metals.

We used the following standards: oligoclase for Al and Na, diopside for Ca, rutile for Ti, chromite for Cr, rhodonite for Mn, various glass standards (like GOR132 komatiite and VG568 rhyolite) for Mg, Si and Fe in silicates, troilite for S, and metals for Fe, Ni, Cr, S, Si, Co, in metallic and sulfide phases. Counting time on the peak and background was 20s and 10s, respectively. Since almost all liquid phases formed dendritic textures during the quench, chemical analyses of these phases were performed with a defocused electron beam of 5 to 30 μm diameter.

## 2.4 Thermodynamic modeling with MELTS and pMELTS

Thermodynamic modeling using the MELTS and pMELTS algorithms (Ghiorso et al. 2002) was conducted to compare our experimental results with those predicted for magmatic silicate liquids in sulfur-free conditions. We used the average composition of the silicate melts obtained at 100% melting as the input composition (see Table 1), assuming it would closely resemble a homogenized silicate mantle of a planetary body formed with EH4 Indarch enstatite chondrites. At high pressure and temperature, chemical equilibration in our samples caused changes in both the silicate and metal compositions from the original chondritic starting material. These changes are due to (1) the partial oxidation of Si into $SiO_2$, leading to an increase of $SiO_2$ concentration in the silicate from 51.8 to 57.5 wt% (2) sulfur partitioning between silicate, metal and sulfide, resulting in a lowered S-content of the metal (from ~11 to ~0.4 wt%) and its solubility in the silicate melt (3 wt% S at 100% melting), (3) carbon incorporation into the metal phase due to contamination from the graphite capsule (1.5 wt%) and (4) the partial loss of volatile elements (S, Na and K) during heating (Collinet et al. 2020). We note however that changes (2) to (4) are also expected to occur in Mercury during its differentiation. Therefore, it is necessary to consider the chemical composition of the silicate fraction of our experimental runs as a proxy for a hypothetical Mercury mantle. Equilibrium melting was modeled using MELTS and pMELTS to predict phase proportions and chemical compositions under sulfur-free conditions and near the iron-wüstite buffer, considering a starting composition equivalent to the average silicate fraction of our samples that were totally melted (Table 1). A total of 1350 models resulted from MELTS and pMELTS calculations at pressures of 1 bar, 0.5, 1, 2 and 3 GPa and temperatures ranging from 1024 and 1900 °C (between the silicate solidus and liquidus temperatures). It should be noted that MELTS and pMELTS models do not account for the presence of reduced Si-rich metal and sulfide phases. Hence, since these models simulate conditions close but different than our experimental conditions, results will be useful to investigate the role on Mercury's melting properties of sulfur enrichment under extremely reduced conditions.

# 3. Results

## 3.1 Textures, phase relations and chemical equilibrium:

Experimental conditions for all experiments are detailed in Table 2. The chemical compositions of silicate melts and modal abundances, calculated by mass balance are listed in Table 3. Chemical compositions of the metal, sulfide and solid silicate phases are listed in Supplementary Table 1. For samples heated close to the liquidus, metallic phases exhibit spherical shapes ranging from 5 to a few hundred microns. In samples that are sulfide-saturated and run at 5 GPa, Fe-rich sulfides form shells around the metal blobs, similar to findings in previous studies conducted under comparable conditions (e.g. Boujibar et al. 2020). Samples heated at lower temperatures have liquid metallic and sulfide phases forming irregular pools located at the grain boundary (Fig.1a, d). In each sulfide-saturated sample, only one sulfide was observed, which compositions are discussed in section 3.3. These sulfides are likely liquid, as they form thin films around the metals and an observed 120° triple junction between two enstatite grains, metal and silicate melts (shown with the red arrow in Fig. 1a), suggesting textural equilibrium and partial wetting. Mg-Ca-Fe-bearing sulfides are known to have high liquidus temperatures (1600 to 2200°C at 1 bar, Pitsch et al. 2025). However, sulfides, here also contain 6- 20 wt% Mn and 2.5-21 wt% Cr, which are expected to be lower their melting temperatures. While two sulfides were observed in Namur et al. (2016b) for Mg-Ca-Fe sulfides, Pirotte et al. (2023) reported an experiment with only one sulfide, rich in Mg, with 4 wt% Mn, 4 wt% Cr and 0.9 wt% Ca. Experiments at 1 bar with Mg-Ca-Fe-Mn-Cr sulfides in McCoy et al. (1999) formed more than one phase. Therefore, the combination of high pressure and the presence of Mn and Cr likely plays a critical role in reducing the immiscibility gap for sulfides and possibly lowering their melting temperature. Silicate minerals predominantly consist of orthopyroxene, with olivine present in only four samples performed at the lowest pressure (0.5 GPa), and grain sizes ranging from 8 to 400 micrometers. Samples show significant crystal growth and euhedral textures for opx (Fig.1b), which directly results from our starting composition being rich in opx. The lack of other competing solid phases enabled the formation of larger opx like in enstatite achondrites (e.g. Udry et al. 2019). The silicate melts are either migrated at one end of the sample, forming a large pool of melt (Fig 1b-c), or located at the grain boundaries between solid phases (Fig. 1a,d). The metals and, in some cases, the silicate melts developed dendritic textures during the quenching process (Fig. 1b).

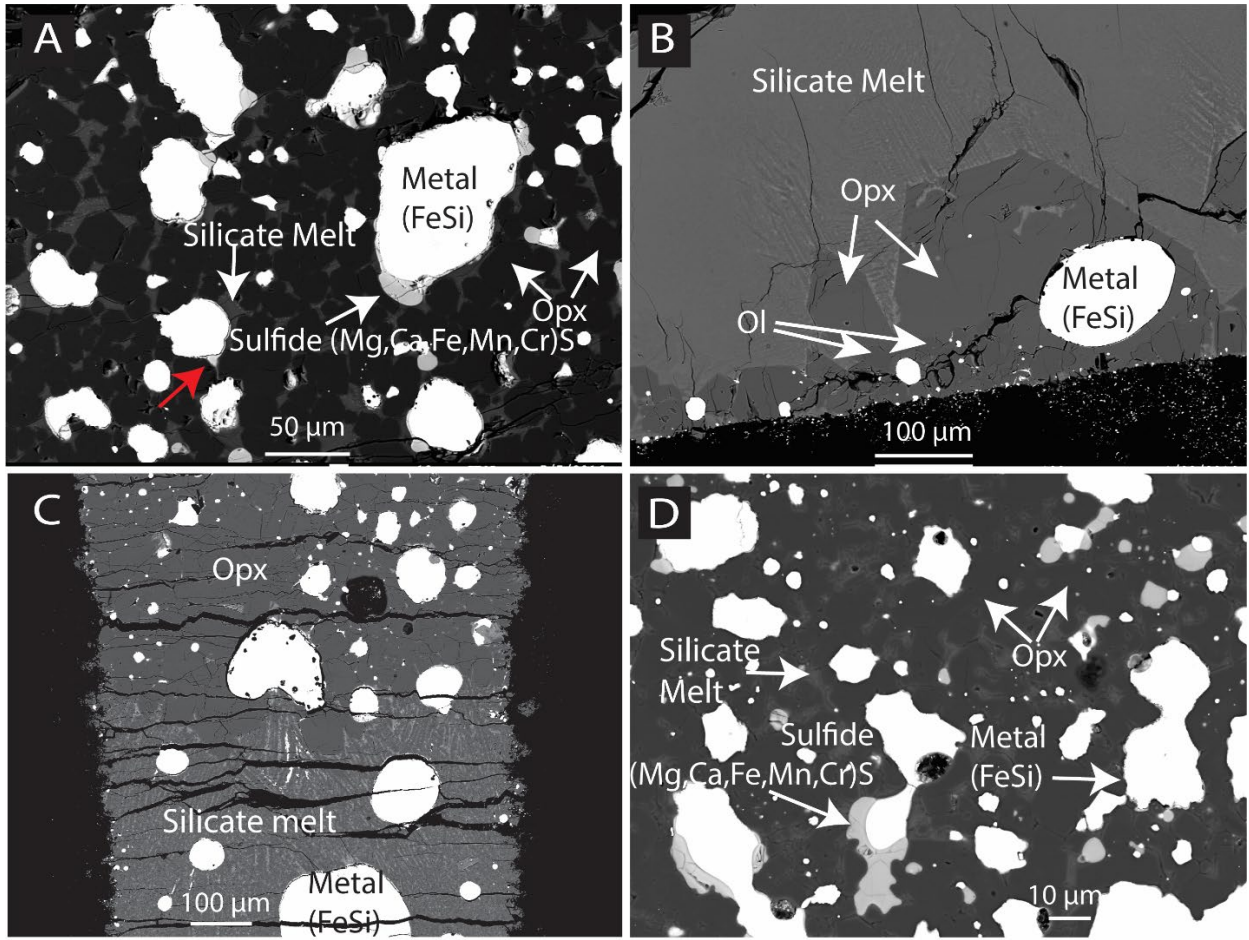

**Fig. 1.** Back-scattered electron images showing the textures of experimental charges. A: Sample #445 run at 2 GPa, 1450 °C) shows opx minerals and silicate melt at the grain boundary. The sample has also liquid FeSi-rich metal and Mg-Ca-Fe-Mn-Cr-rich sulfide phases located all throughout the sample. The red arrow shows the sulfide phase in a triple junction. B: Sample #988 run at 0.5 GPa and 1600 °C has large euhedral opx minerals and small olivine grains segregated from a large pool of silicate melt. Note the quench dendritic texture of the silicate melt. C: Sample #386 run at 3 GPa and 1780 °C presents opx grains on one side of the sample and the dendritic silicate melt on the other side. The FeSi-rich metal phase is embedded within both opx and silicate melt. D: Sample #1073 run at 2 GPa and 1500 °C shows similar textures as sample #445 (A).

All phases exhibit homogeneous compositions, and the minerals are not zoned. However, for three samples (#870, #878, at 1 GPa and #386 at 3 GPa), small fractions of interstitial silicate melt were found at the grain boundary, with compositions showing an enrichment in incompatible elements (Na, K, Al) relative to the larger pools of melts. Despite this, the orthopyroxenes in contact with these melts display homogeneous compositions without zoning. These interstitial melts are likely additional melting products of a slightly cooler portion of the sample. However, since the orthopyroxenes in these three samples have homogeneous compositions, these melts are not expected to have affected element distribution between the larger pools of melt and orthopyroxenes. In fact, the compositions of silicate melts in these three samples are similar to those without interstitial melts at comparable degrees of melting (#880 and #383, at 1 and 3 GPa, respectively). The absence of zoning, the euhedral textures of solid phases, and the homogeneous compositions all indicate that the chemical equilibrium was achieved in our charges.

Modal abundances derived from mass balance calculations show that the metal mass fraction remains relatively constant: 44.3 ± 2.5 wt%. Metals have relatively homogeneous compositions (Table 4), with only slightly higher sulfur-content at high pressures (on average 1.5 wt% compared to 0.3 wt% at lower pressures), which agrees with S increasingly siderophile behavior with increasing pressure (e.g. Boujibar et al. 2014). When sulfide is present, its mass fraction is relatively low at 5.5 ± 2.6 wt%. Additionally, orthopyroxene is stable in all of our partially melted samples, while olivine is only present in three samples conducted at the lowest pressure (0.5 GPa) and the highest temperatures (1450 to 1600 °C) (see Table 2 & Fig. S1). Both phases have chemical composition close to the forsterite and enstatite endmembers.

## 3.2 Oxygen fugacity

To investigate the silicate melting properties under the reducing conditions of Mercury, it is crucial to buffer samples at a low oxygen fugacity. In this study, we used Fe-Si-Ni metal as an internal redox buffer, which produces chondritic compositions (Berthet et al. 2009). Oxygen fugacity was calculated relative to iron-wüstite (Fe/FeO) buffer (ΔIW), considering the equilibrium:

$$\text{Fe} + \tfrac{1}{2}\,\text{O}_2 = \text{FeO} \tag{1}$$

with

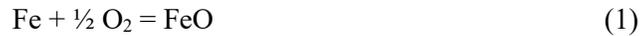

$$\Delta IW = \log f_{O_2}^{exp} - \log f_{O_2}^{IW\ buffer} = 2 * \log(\frac{x_{FeO}^{silicate}}{x_{Fe}^{metal}} * \frac{\gamma_{FeO}^{silicate}}{\gamma_{Fe}^{metal}}) \tag{2}$$

where $x_{FeO}^{silicate}$ and $x_{Fe}^{metal}$ are the FeO and Fe concentrations measured in the liquid silicate and metal phases, respectively. $\gamma_{FeO}^{silicate}$ is the activity coefficient of FeO in the silicate, considered equal to 1.7 following previous estimates (O'Neill & Eggins 2002), while $\gamma_{Fe}^{metal}$ is the Fe activity coefficient in the metal, calculated as a function of the metallic phase composition, using the online metal activity calculator from Norris Scientific (Wade et al. 2012). Oxygen fugacity was also calculated following the methodology described in Cartier et al. (2014), considering the equilibrium:

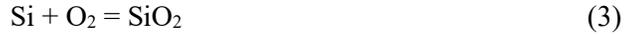

$$Si + O_2 = SiO_2 \qquad\qquad (3)$$

As in previous studies (Namur et al. 2016a & 2016b; Rose-Weston et al. 2009), the two $f$O$_2$ calculations do not systematically yield the same result, due to uncertainties in determining the activity coefficients of FeO and SiO$_2$ in the silicate melt. Differences between the two $f$O$_2$ calculations show broad correlations with these two variables (Fig. S2), which vary with pressure and temperature. Calculations using the equilibrium (3) is recommended when FeO concentration is less than 0.1 wt% (Cartier et al. 2014, Namur et al. 2016a). In the present study, FeO ranges from 0.18 to 1.33, with an average concentration of 0.6 ±0.3 wt%, which exceeds the range preventing the consideration of equilibrium (1). Among all our experiments, oxygen fugacity varies from IW-4.9 to IW-2.7 with an average of IW-3.7 ±0.6.

### 3.3 Chemical composition of sulfides

Our experimental results show a range of sulfide compositions, which can include relatively high concentrations of the predominantly lithophile elements, Mg (up to 22 wt%) and Ca (up to 13 wt%) (Table 4). The relative proportions of Mg and Ca vary, with a molar Mg/Ca ratio ranging from 2 to 20 (Fig. 2a). At 5 GPa, sulfides are rich in Fe, Cr and Ti, with compositions close to pure FeS and concentrations of Ca and Mg below their detection limits. As pressure decreases, the abundance of Mg and Ca appear to increase at the expanse of the other elements (Fig. 2a). However, this trend could also be the result of changes with temperature or chemical compositions, since the temperature range for experiments run at higher pressures are higher than those at lower pressures, and chemical compositions vary with pressure and temperature. Therefore, we performed linear regressions in order to better asses controlling factors of sulfide compositions (see below).

In all samples produced in our study, Fe is positively correlated with Cr and negatively correlated with Mg. As noted in previous findings by Nittler et al. (2023), Cr and Mg are negatively correlated (Fig. 2b-d). Results from McCoy et al. (1999)'s experiments performed on the natural EH4 chondrite at atmospheric pressure are not plotted because their samples have 2 sulfides present. While their data for the Mg-rich sulfides fit within the Fe-Mg linear correlation, they are outliers within the Fe-Cr and Mg-Cr linear fits, showing a consistent depletion in Cr compared to our experimental results. The reason for this discrepancy on Cr in McCoy et al. (1999) data is likely due to the presence of two sulfides and partitioning of Cr between them. Linear regressions between the concentrations (in wt%) of all three elements using our experimental data yield:

$$X_{Cr}^{sulfide} = 0.1262 * X_{Fe}^{sulfide} + 1.7671 \qquad (3)$$

$$X_{Mg}^{sulfide} = -0.5339 * X_{Fe}^{sulfide} + 25.044 \qquad (4)$$

$$X_{Cr}^{sulfide} = -0.2393 * X_{Mg}^{sulfide} + 7.7897 \qquad (5)$$

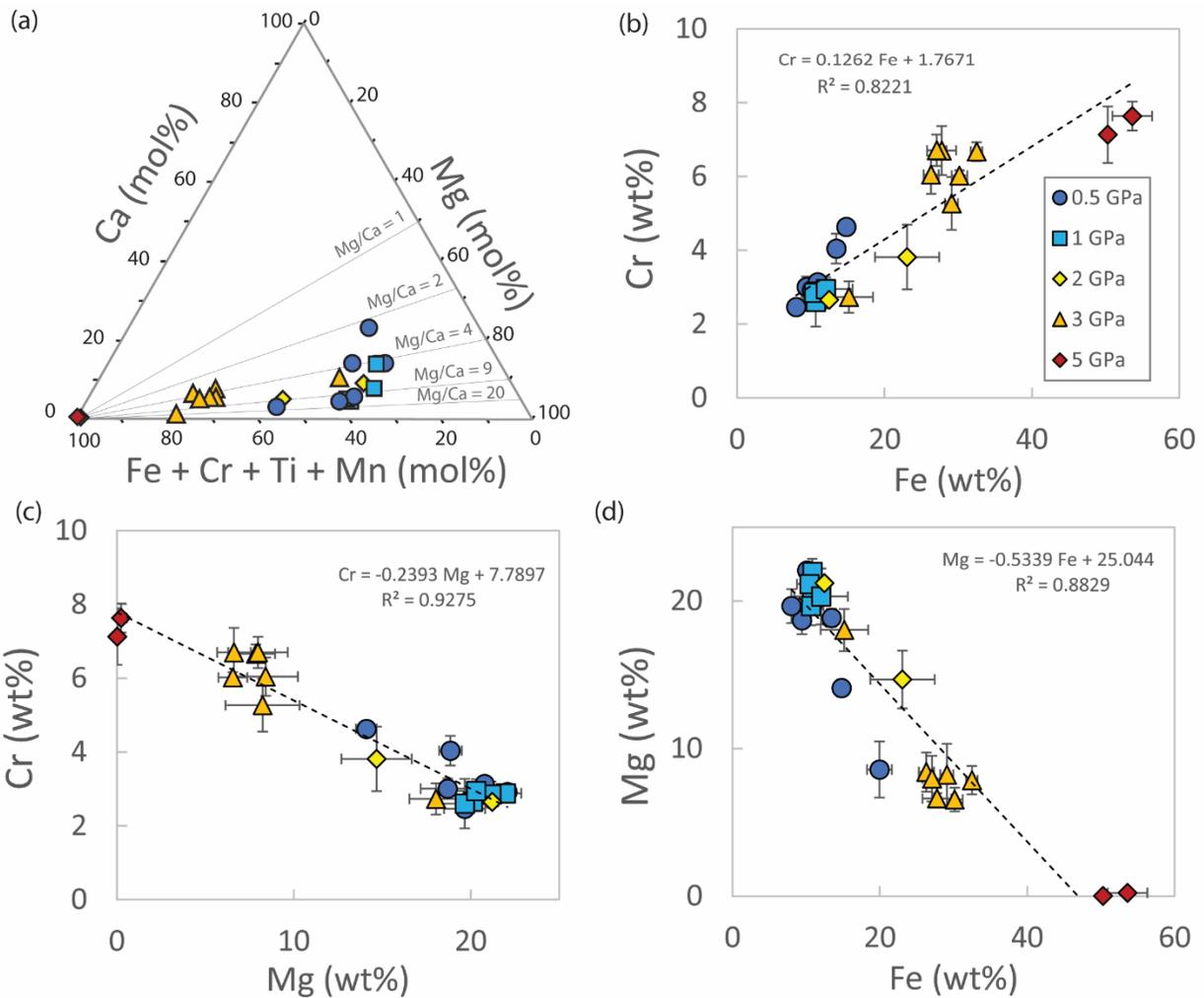

**Fig 2.** A: Mg-Ca-[Fe+Cr+Ti+Mn] ternary diagram showing the compositions of sulfides present in our experimental charges. Legend is shown in B. Correlation between concentrations of Fe and Cr (B), Cr and Mg (C), and Fe and Mg (D) in sulfides and linear regression fits shown in dotted lines with their corresponding equations and coefficients of determination ($R^2$).

Experimental data show that surprisingly, increasing oxygen fugacity enhance the partition of Mg and Ca into sulfides within the range of $f$O$_2$ investigated (IW-5 to IW-2.5) (Fig. 3). To examine the effects of pressure, temperature, $f$O$_2$ and chemical composition on Mg and Ca partition between sulfide and silicate ($D_{Mg}^{sulf/sil} = X_{Mg}^{sulf} / X_{Mg}^{sil}$ and $D_{Ca}^{sul/sil} = X_{Ca}^{sulf} / X_{Ca}^{sil}$, respectively, where X refers to concentrations in wt% in the silicate melt or sulfide), we performed a linear regression using data from this study and those of McCoy et al. (1999). While pressure is found insignificant for both Mg and Ca, temperature only affects Mg partitioning (Fig. 3a). Additionally, as SiO$_2$ concentration in the silicate melt increases, Ca becomes more chalcophile (Fig. 3b). The resulting equations predicting these partition coefficients are:

$$D_{Mg}^{sul/sil} = -0.48\ (\pm 0.35) + 0.53\ (\pm 0.06) * \Delta IW + 3964 (\pm 526) * 1/T \qquad (6)$$

$$D_{Ca}^{sul/sil} = -11.89\ (\pm 1.51) + 0.21 (\pm 0.06) * \Delta IW + 7.09 (\pm 0.85) * \log\left(X_{SiO_2}^{sil}\right) \quad (7)$$

Where $X_{SiO_2}^{sil}$ is the concentration of SiO$_2$ in the silicate melt in wt%. The quality of the fit and statistical parameters are shown in Fig. 3, Fig. S3 and Supplementary Table 2. The effect of oxygen fugacity is twice as strong for Mg compared to Ca, and the relationship between the silicate melt composition and Ca partitioning is discussed below in section 4.2. Altogether, our results show that relatively high $f$O$_2$ favors the formation of Mg- and Ca-rich sulfides, while low T and high SiO$_2$ content in the silicate melt increase Mg and Ca partitioning into sulfides, respectively.

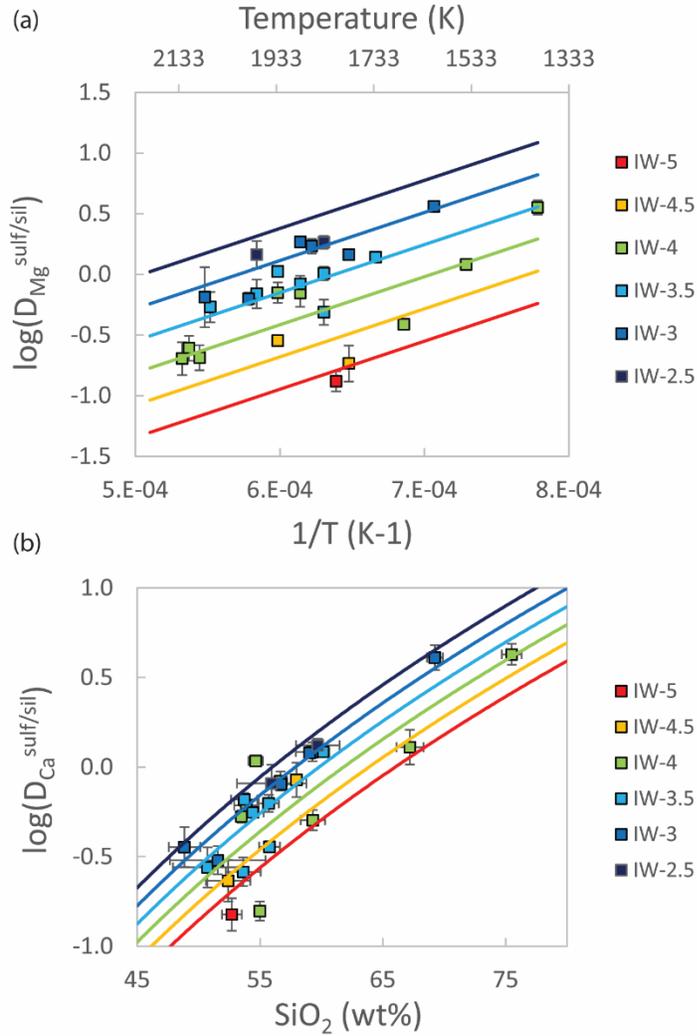

**Fig. 3** Logarithm of Mg (a) and Ca (b) partition coefficients between sulfide and silicate as a function of the inverse of temperature (a), and SiO$_2$ content of the silicate melt (b). The colors of the symbols correspond to the oxygen fugacity during the experiment. The error bars show the one σ standard deviation on chemical compositions and propagated uncertainties. The colored lines and curves show modeled partition coefficient using Eq. 6-7, respectively.

### 3.4 Silicate melt composition

The chemical compositions of the silicate melts exhibit trends consistent with previous work, showing an increase in MgO with the degree of silicate melting (F), which aligns with its compatibility with Ca-poor pyroxene and olivine (Fig. 4). Incompatible elements such as aluminum, sodium and potassium decrease in concentration with increasing F. Some of our samples appear to have experienced varying levels of volatile loss: samples #432 and #433 show decreased Na abundance at 2 GPa, and other samples exhibit reduced K, although K concentrations are already very low in starting compositions.

Despite significant alkali loss in samples produced by McCoy et al. (1999) at 1 bar, we included their data in our analysis. Silica concentration remains constant at 55 wt% $SiO_2$ when F exceeds 25 wt%, while its behavior at lower F varies with pressure. With decreasing F, $SiO_2$ increases at P < 2GPa and decreases at P > 2 GPa. This trend is consistent with previous work showing that the effect of alkalis on the silica content of melts changes with pressure (Hirschmann et al. 1998). This effect is attributed to the decrease in the degree of polymerization of silicate melts with pressure, which reduces the ability of alkalis to break up polymerized silica tetrahedra. However, this trend is also seen in alkali-free systems, where a thermal divide similarly affects the melt silica content (see below). CaO gradually decreases with F at pressures of 2 GPa or higher, while at lower pressure, it first increases with F up to 30 wt% melting and then decreases at higher F.

Resulting pMELTS compositions for partial melts have similarities and differences with those produced in experiments (see Fig. S4-S8). For all pressures investigated, models accurately predict the concentrations of $Al_2O_3$, $Na_2O$ and $K_2O$, while melt MgO concentrations coincide for pressures higher than 1 GPa only. At lower pressures, pMELTS models underestimate MgO content, which is likely results from the predicted larger stability of olivine relative to orthopyroxene in pMELTS (see below). In addition, at 0.5 and 1 GPa and between 15 and ~70 wt% melting, CaO and $SiO_2$ contents are underestimated and overestimated, respectively (Fig. S5-S6). At 3 GPa, CaO abundance is underestimated at F< 20%, while the trend for $SiO_2$ concentration with F is broadly concordant with experiments but slightly lower (Fig. S8). Models at 2 GPa show comparable concentrations for CaO and $SiO_2$ except for a partial melt from experiment #445 which has a higher CaO content than the broad trend (Fig. S7). The differences in composition are further discussed below.

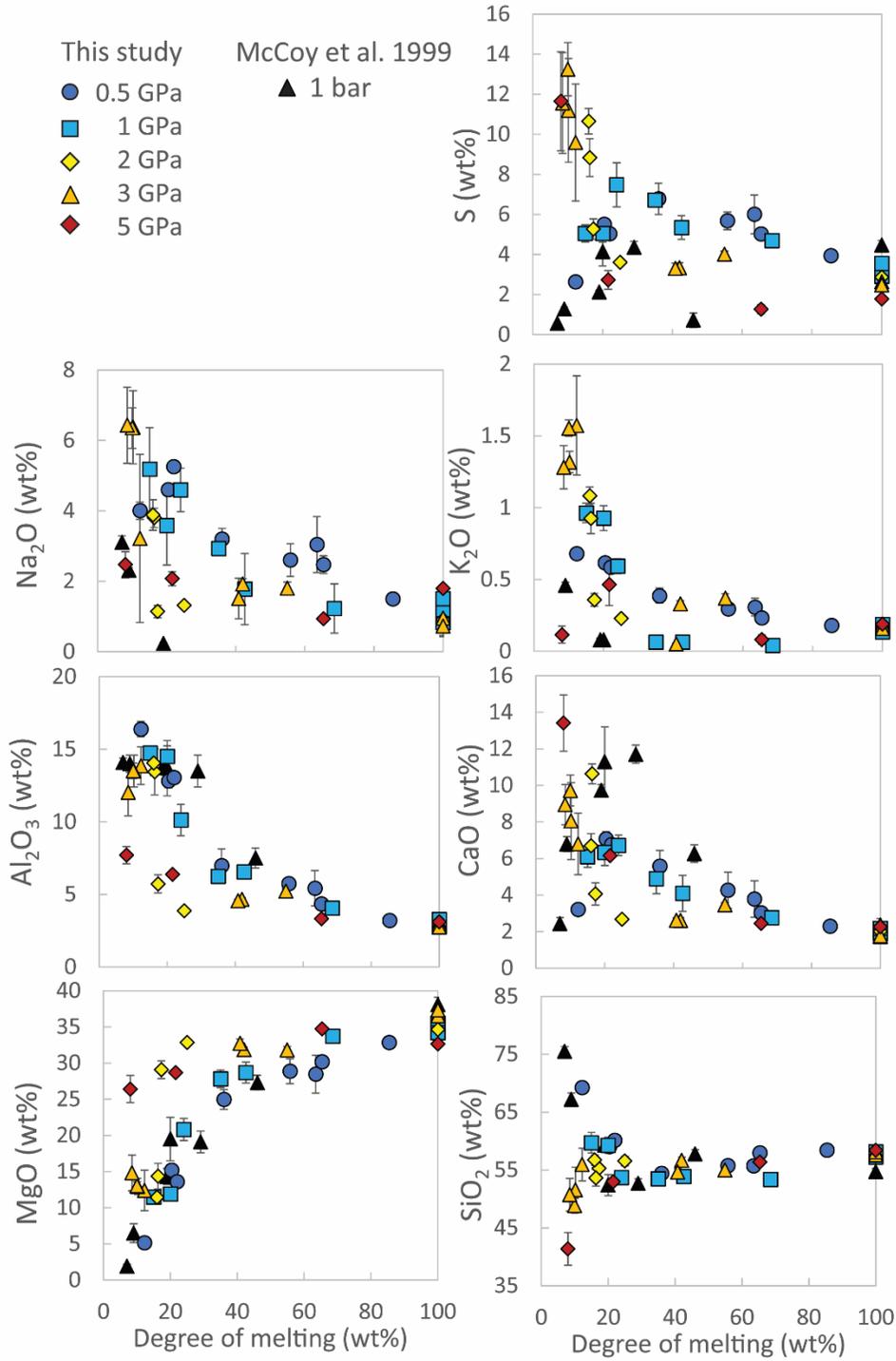

**Fig. 4.** Chemical compositions of our silicate melts as a function of the degree of silicate melting. Data from McCoy et al. (1999) are also shown for comparison.

# 4. Discussion

## 4.1 Silicate phase relations of a reduced EH chondritic mantle

To better understand the origin of these trends, melt compositions were plotted in a forsterite-quartz-Ca-Tschermak (Fo-Qz-CaTs) pseudoternary diagram, projected from diopside (Fig. 5). Due to the low $fO_2$, silicate phases are depleted in Fe, Cr and Ti, because of the siderophile behavior of Fe and chalcophile behavior of Cr and Ti (e.g. Cartier et al. 2020, Nittler et al. 2023, Righter et al. 2023). Consequently, our bulk silicate compositions are sulfur-rich silicates in a system close to CMASN. Silicate melts compositions at 1 bar (McCoy et al. 1999) generally follow the liquidus boundaries of the CMAS system at 1 bar (e.g. Longhi 1987), although they show an expansion of the enstatite stability field relative to quartz (Fig. 5a). Phase relations and melt compositions at 0.5 to 2 GPa also show an expansion of the enstatite stability field over forsterite when compared to the CMAS system (Liu & Presnall 2000). This effect can be seen in Figure 5, with the positions of the silicate melts in equilibrium with enstatite, which are within the stability field of forsterite in the CMAS system. It is important to note that all our samples contain only enstatite as the solid phase, except for three samples partially melted at 0.5 GPa with F > 55%. Experiments performed at 0.5 to 2 GPa indicate that low-temperature invariant points are shifted toward the Ca-Ts apex. This shift is likely due to the enrichment of our compositions in Na, as previously observed in the CMASN system (Walter & Presnall 1994) and in natural pyroxenites (Lambart et al. 2013). Thermodynamic modeling using MELTS (at 1 bar) and pMELTS (at higher pressures) software, which consider the complexity of multi-components systems (Ghiorso et al. 2002) yields similar results with a displacement of liquidus boundaries beyond invariant points of simpler systems (Fig. 5).

At 3 and 5 GPa, almost all melts are located to the left of the enstatite-Ca-Ts line (dotted line in Fig. 5b). This trend mimics the role of this join as a thermal divide which was found in previous work to influence melt compositions at pressures higher than ~3 GPa, where partial melts should fall on the same side of the thermal divide as the bulk composition (O'Hara & Yoder 1963, Milholland & Presnall 1998). However, this join was only found to act as a thermal divide when clinopyroxene and garnet are residual phases. Therefore, this barrier is likely only be apparent, since sample #374 melted at 3 GPa has a melt has a composition located to the right of this barrier. At high pressures, our results indicate an expansion of the enstatite field relative to spinel (or garnet at 5 GPa) when compared to predictions from liquidus boundaries in the CMAS and CMFAS systems (double lines in Fig.5b). This expansion is also predicted by pMELTS models at 3 GPa (Fig. 5b, Fig.S5). At low F (8-10%), silicate melts become critically silica-undersaturated, with compositions plotting left of the Fo-An-Di plane.

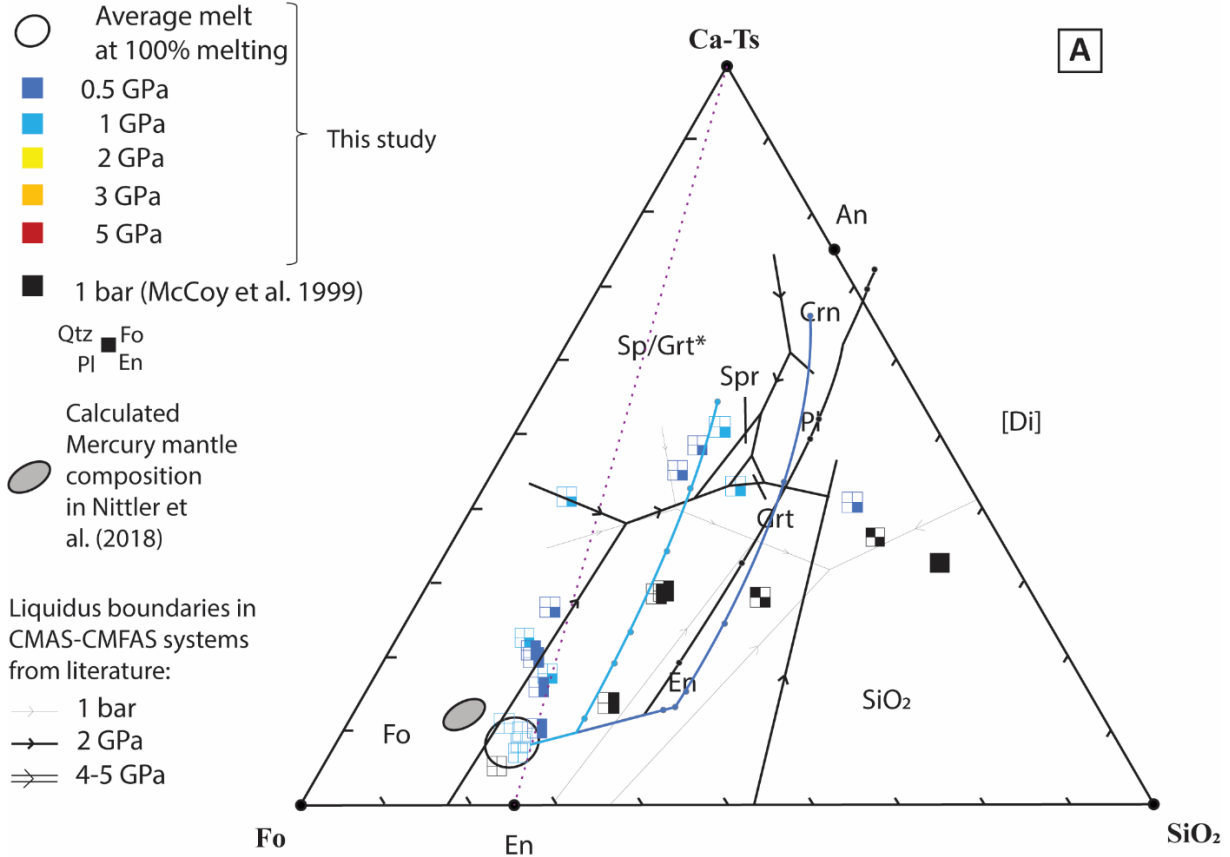

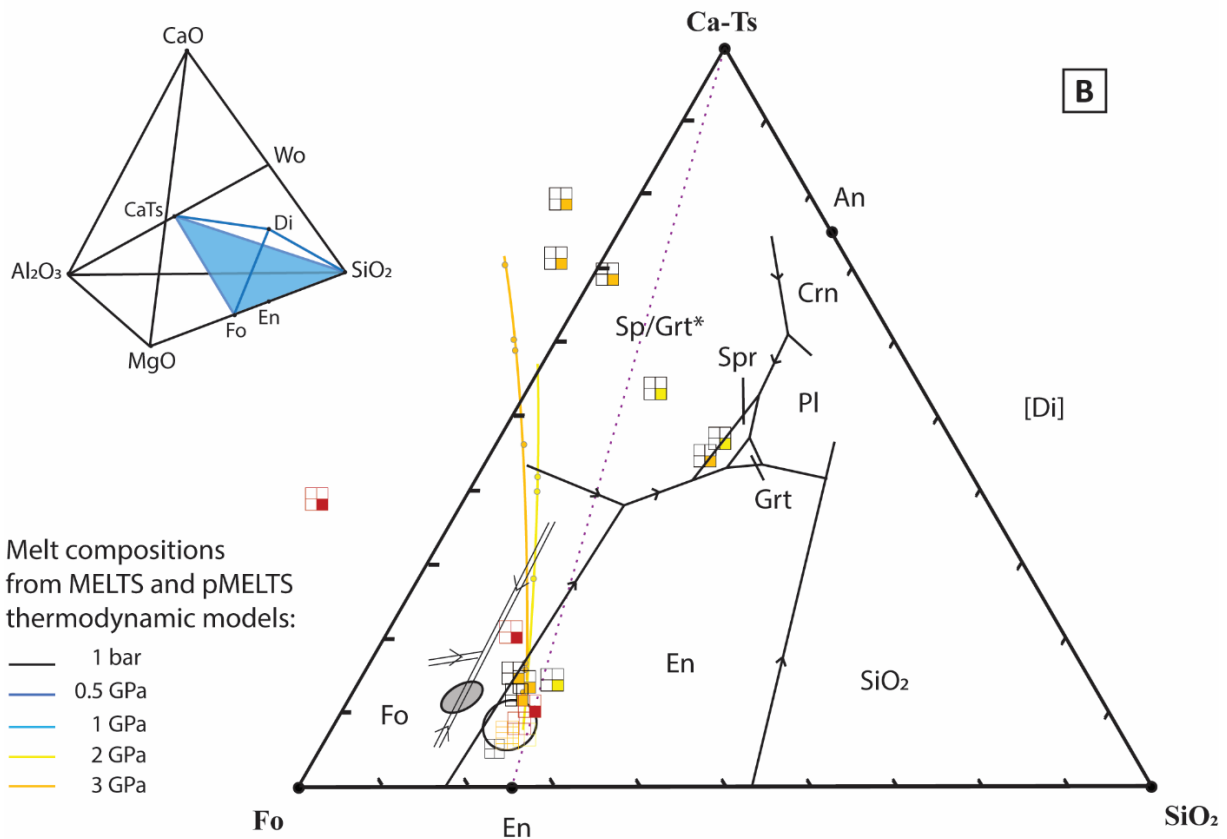

**Fig. 5.** Pseudoternary diagram projected from diopside onto the forsterite-Ca-Ts-quartz plane in the CMAS system (see inset), showing the composition of our silicate melts (squares) and phases present (colored square quarters, see legend) at pressures lower (A) and equal or higher (B) than 2 GPa. Results from McCoy et al. 1999 are also plotted to show phase relations at 1 bar. The open ellipse depicts the compositions of the silicate melt obtained at 100% melting, which are equivalent to the bulk composition. The grey ellipse shows the estimated composition of Mercury's mantle in Nittler et al. (2018). For comparison, phase boundaries at one bar (in CMAS system: Andersen 1915, Longhi 1987), 2 GPa (in CMAS system: Liu & Presnall 2000), and 4-5 GPa (in CMFAS system: Herzberg & Zhang 1997, with a pyrolitic terrestrial mantle: Walter 1998) are shown as thin grey lines, think black lines, and fine double lines, respectively. Silicate melt compositions obtained with pMELTS thermodynamic models over temperature ranges of our experiments are shown as colored lines. Small dots on these lines represent results at temperatures of our experiments. Projections are performed following the method described in O'Hara (1972). Our experimental data suggest the enstatite stability field expands relative to all other phases, when comparing results with those in the CMAS system.

## 4.2 Role of sulfur in silicate phase equilibria at low $f$O₂

Our results show that the liquidus field of enstatite expands relative to forsterite at 0.5 and 1 GPa, which agrees with previous findings on multiple saturation points in reduced S-rich systems (Namur et al. 2016b). This explains why pMELTS overpredicts the stability of olivine compared to enstatite at 0.5 and 1 GPa (see Fig. S5-6), since pMELTS does not account for the presence of sulfide phases and solubility in silicate melts and is not calibrated for extremely reduced conditions. This major change in S-rich reduced systems is likely due to the presence of CaS and MgS complexes in the silicate melts, which increase the activity of $SiO_2$. This effect has also been observed in a recent study investigating sulfide speciation in Mercurian magmas (Anzures et al. 2025). As a result, Mg becomes less available to form Mg-rich phases like forsterite (Eq. 8) and instead favor enstatite crystallization:

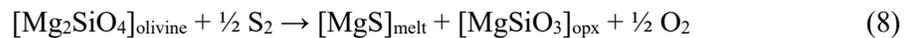

$$[Mg_2SiO_4]_{olivine} + \tfrac{1}{2}\,S_2 \rightarrow [MgS]_{melt} + [MgSiO_3]_{opx} + \tfrac{1}{2}\,O_2 \qquad (8)$$

In the Fo-Qz-CaTs ternary diagram, reaction (8) explains the observed shift of high temperature melt compositions to the left of the cotectic lines for CMAS system and to the left of the compositions predicted by MELTS and pMELTS (Figure 5a). Equations (9-11) can account for the compositional changes observed at 3 and 5 GPa, specifically the enrichment of melts in Ca and Al, relative to melts formed in S-poor and more oxidized conditions (pMELTS predictions) (Fig. 5b):

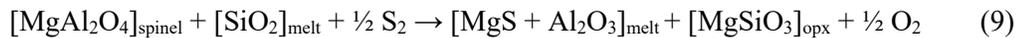

$$[MgAl_2O_4]_{spinel} + [SiO_2]_{melt} + \tfrac{1}{2}\,S_2 \rightarrow [MgS + Al_2O_3]_{melt} + [MgSiO_3]_{opx} + \tfrac{1}{2}\,O_2 \qquad (9)$$

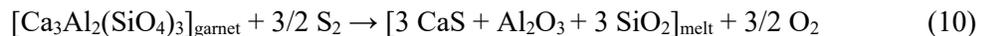

$$[Ca_3Al_2(SiO_4)_3]_{garnet} + 3/2\,S_2 \rightarrow [3\,CaS + Al_2O_3 + 3\,SiO_2]_{melt} + 3/2\,O_2 \qquad (10)$$

$$[Mg_3Al_2(SiO_4)_3]_{garnet} + \tfrac{1}{2} S_2 \rightarrow 2 [MgSiO_3]_{opx} + [SiO_2 + MgS + Al_2O_3]_{melt} + 3/2 O_2 \quad (11)$$

Spinel is predicted by pMELTS to be stable for the coolest sample produced at 3 GPa, yet none of our samples contain spinel. However, this may result from spinel overprediction by pMELTS in Cr-bearing compositions (Asimow et al. 1995). At pressures of 3 GPa and lower, the presence of Mg-rich sulfides (up to 22 wt%) likely reduces MgO activity in the silicate melt, which prevented olivine crystallization. Future studies should include experiments at lower F (<8-9% at 3-5 GPa and <12-16% at 0.5-2 GPa) to determine whether the enstatite stability field expands relative to these other phases, such as feldspar and diopside.

The role of sulfur on expanding enstatite stability field aligns with previous studies using Raman and XANES spectroscopy, which detected CaS and MgS complexes (Namur et al. 2016b, Anzures et al. 2020, Pommier et al. 2023). Pommier et al. (2023) also identified Mg-S-Si and Ca-S-Si complexes using [29]Si NMR spectroscopy. In a sulfur-free system, the expansion of enstatite stability field over forsterite or spinel typically suggests an increase in the silicate melt polymerization (e.g. Kushiro 1975). However, because sulfur forms Mg-S-Si and Ca-S-Si complexes rather than Si-S-Si (as suggested by Pommier et al. 2023), the increased stability of enstatite does not necessarily indicate increased polymerization of the silicate melt. In fact, a recent experimental study showed that adding sulfur to silicate melts decreases their viscosity (Mouser et al. 2021), implying a decrease in their polymerization.

Additional changes occur due to the saturation of Ca- and Mg-rich sulfides. The spread of melt chemical compositions at low temperature (Fig. 5) is a direct result of the sulfide saturation in the silicate melt and the presence of Ca- and Mg-rich sulfides. For example, sample #955, run at 0.5 GPa, has the highest Ca abundance in the sulfide (13 wt%), and its silicate melt being very $SiO_2$-rich (like dacites) is shifted away from Ca-Ts toward quartz relative to pMELTS predictions. This is also the case for two samples from McCoy et al. (1999) that are also rich in $SiO_2$ (like dacites and rhyolites) having 10 and 12 wt% Ca in coexisting sulfides. This trend agrees with the correlation seen between Ca partitioning between sulfide and silicate and $SiO_2$ abundance of the silicate melt (see section 3.3 and Fig. 3b). Additionally, at 3 GPa, sample #374 run at a lower temperature than the rest of the 3 GPa-series of samples has a sulfide with the highest Mg and Ca concentrations (18 and 5.5 wt%, respectively) and a silicate melt shifted toward quartz apex (away from both forsterite and Ca-Ts endmembers) relative to the other samples and pMELTS predictions. In fact, this shift is also observed for samples run at 0.5, 1 and 2 GPa and low temperature, which all have relatively high Mg and Ca (Table 3) and varying silicate melt compositions.

Figure 6 shows how sulfide saturation influences Ca concentration in the silicate melt compositions, differently depending on the pressure. At pressures lower than 2 GPa, while sulfur solubility in silicate melts increases the silicate melt Ca-content at moderate to high F (20-60%) (see comparison with pMELTS predictions in Fig. S4-6), this trend reverses at lower F when sulfides become rich in Ca (Fig. 6a-b, d). In contrast, at 3 GPa, the Ca content in silicate melts remains high even at the lower F (Fig. 6c). In fact, at F ≤ 20%, for all pressures, CaO in the silicate and Ca in the sulfide show complementary trends: when the Ca concentration in the sulfide is high, its concentration in the silicate melt is low, and vice versa. As shown in section 3.3, Ca partitioning between sulfide and silicate increases with increasing $SiO_2$ content of the silicate melt (Fig. 3b). Therefore, the change in trends at pressures lower or higher than 2 GPa is due to those of $SiO_2$ in the silicate melt. As shown earlier, as F decreases, $SiO_2$ increases or decreases at pressure lower and higher than 2 GPa, respectively (section 3.4 & Fig. 4). The influence $SiO_2$ has on Ca partition between sulfide and silicate is likely linked to Ca speciation in the silicate melt. It may indicate that there is a limit to the extent of Ca-S-Si bonds in silicate melts, since large enrichments of the silicate melts in $SiO_2$ yield lower Ca content in the silicate relative to the sulfide. This may indicate that Ca-S bonds are more favorable than Ca-S-Si bonds in silicate melts. These results further highlight how silicate melt composition, sulfur solubility in silicate melts, and the partitioning of Ca and Mg into sulfides (when sulfide saturation is reached) are interconnected, adding complexity to silicate phase equilibria in sulfur-rich, reduced conditions. Altogether, both presence of Mg- and Ca-rich sulfides and sulfide solubility in silicate melts generate variability of silicate melt compositions in reduced S-rich systems.

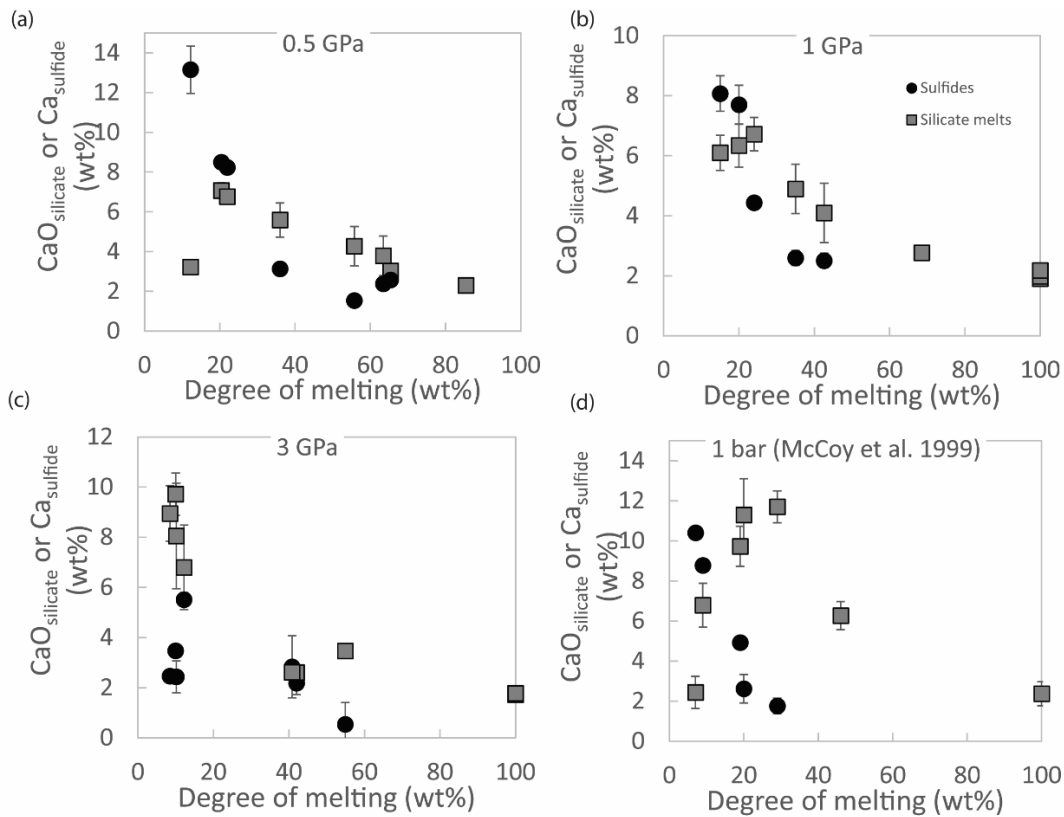

**Fig. 6** Comparison between CaO and Ca concentration in the silicate and sulfide melts, respectively, for samples at 0.5 (a), 1 (b) and 3 GPa (c) from this study, as well as those of McCoy et al. (1999) at 1 bar (d).

## 4.3 Sulfur and sulfides during Mercury's mantle-crust differentiation

Our experimental silicate melts and those of McCoy et al. (1999), exhibit a positive correlation between Ca/Si and S/Si (Fig. 7C), similar to the correlation observed on Mercury's surface, although not identical as both datasets do not overlap. Each set of experiments conducted using different devices (piston cylinder at 0.5-3 GPa, multi-anvil press at 5 GPa and gas-mixing furnace for McCoy et al. (1999)), follows specific correlation lines, likely reflecting the level of sulfur loss during heating. Previous studies have suggested that sulfur can be depleted in Mercury magmas during their ascent through the crust, during explosive eruptions and impact processes, which produced sulfur- and carbon-rich volcanic gases (Renggli et al. 2022, Deutsch et al., 2021). Sulfur loss during these processes may have led to a global sulfur depletion, explaining the differences between Mercury's surface S/Si and most of our experimental data (Fig. 7C). This loss of volatiles was recently proposed to have contributed to the formation of a transient atmosphere on Mercury (Deutsch et al. 2021). The correlation between Ca/Si and S/Si is observed across Mercury's surface, with the exception of the pyroclastic deposit Nathair Facula, a region northeast of Rachmaninoff basin (Weider et al. 2016, Nittler et al. 2020), where extreme loss of S-bearing volatiles during explosive volcanism shifted the composition away from the general trend. While Ca/Si-S/Si correlation has been previously attributed to the presence of oldhamite (CaS), it is important to note here that such a trend can occur without the presence of oldhamite, as primary silicate melts themselves can show a correlation between CaO and S. Therefore, if Mercury lavas remained uncrystallized after their deposit on the surface, their compositions could still exhibit a similar correlation.

Notably, if magmas ascend slowly, they can reequilibrate at shallow depth, which would affect magma composition and S solubility. S concentration in magmas at sulfide saturation decreases with decreasing temperature (Namur et al. 2016b). Any magma slowly ascending, cooling and decompressing will therefore have a lower S concentration and precipitate sulfides. Since precipitating sulfides are richer in Mg and Ca at lower temperature and with higher magma silica content, respectively (Fig. 3), they are expected to increase the Al and Si contents of the magma relative to Mg and Ca upon ascent. Comprehensive study of the effects of magma decompression or reequilibration upon ascent is beyond the scope of this contribution, so for our present purposes, we focus on the composition of primary magmas expected to derive from the partial melting of a primitive mantle close to enstatite chondrites in composition.

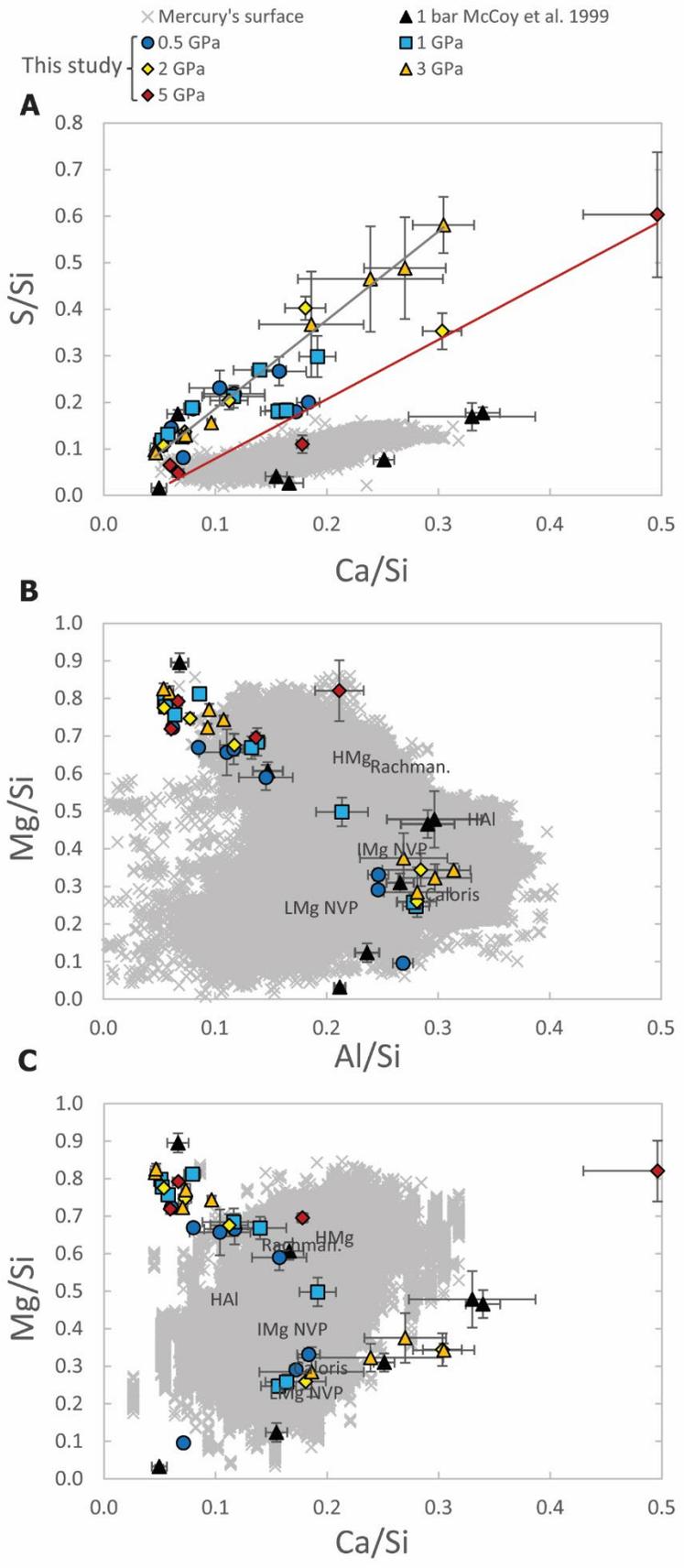

**Fig. 7** Elemental ratios (S/Si as a function of Ca/Si in (A), and Mg/Si as a function of Al/Si (B) and Ca/Si (C)) of silicate melts produced in this study from 0.5 to 5 GPa, as well as those from McCoy et al. 1999 (1 bar) (black triangles) and Mercury's surface compositions (grey crosses). In (A), the red and black lines are linear fits for samples at 5 and 3 GPa, respectively. In (B) and (C), the star represents the natural EH4 Indarch chondrite starting composition and the arrow shows the change of the bulk silicate composition with segregation of 12 wt% Si in the metal/core. Abbreviations are average compositions of terrane classes identified in Weider et al. (2015): HMg = High-Magnesium, HAl = High-Aluminum, LMg-NVP = Low-Magnesium Volcanic Plains, IMg-NVP = Intermediate-Magnesium Volcanic Plains, Caloris = Caloris basin, Rachman. = Rachmaninoff basin.

## 4.4 Comparison between EH chondrite partial melts and Mercury's surface composition

Our experimental results enable a direct comparison between the surface chemical compositions of Mercury and the partial melts of EH enstatite chondrites. The range of $Na_2O$ of produced melts fall within the estimates for Mercury's surface (Peplowski et al. 2014), with Na/Si ratios of 0.03-0.21 in the experimental melts, compared to 0.08-0.25 on Mercury (see Fig. S10). The melts produced span a variety of fields, including picritic basalts, basaltic andesites, andesites, basaltic trachy-andesites, phono-tephrites and dacites, while those of McCoy et al. (1999) extend into the rhyolitic field (Fig. 8). This demonstrates the variability of compositions that can be generated from a EH chondritic starting composition across a range of pressures and temperatures, covering nearly all compositions observed on Mercury's surface, in terms of $SiO_2$ and total alkalis. While IUGS classification diagrams, like the TAS diagram (and diagram for mafic and ultramafic rocks below) are designed for terrestrial igneous rocks, they may not be perfectly adapted to Mercury's surface rocks, most of which contain high MgO contents. Therefore, the TAS diagram in Fig. 8 also includes fields for picrites, komatiites and boninites which encompass the MgO range found in Mercury's rocks.

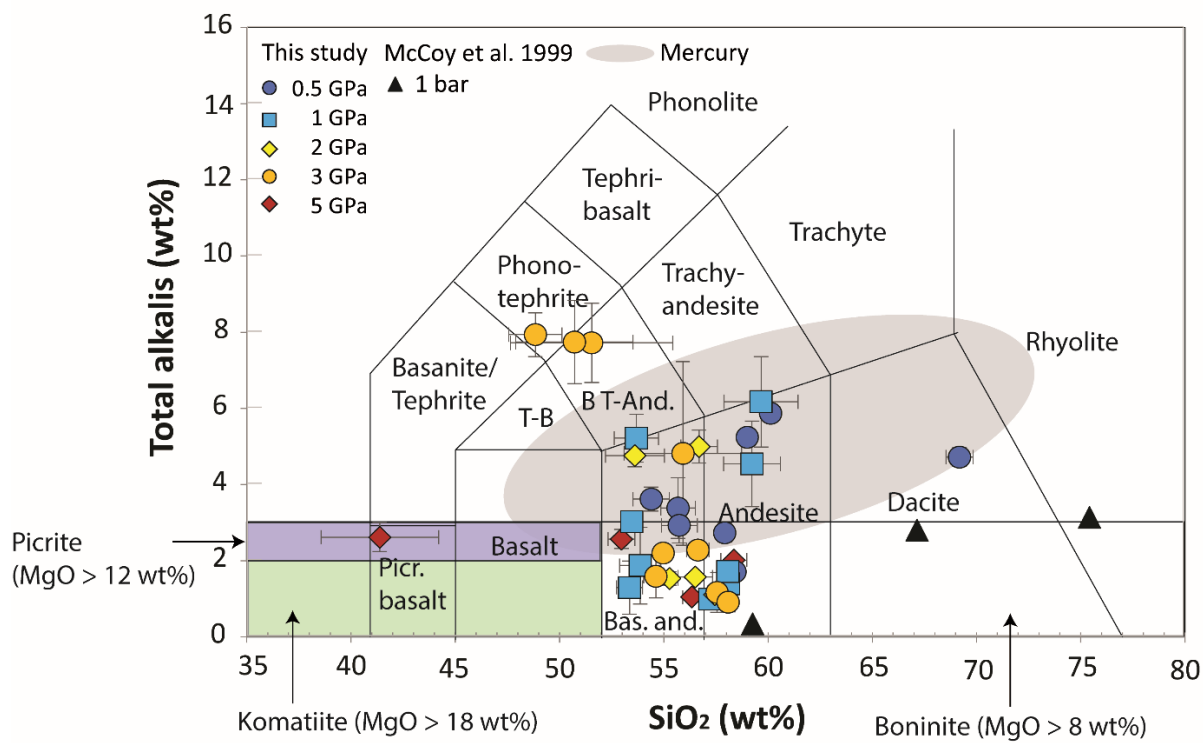

**Fig. 8** Comparison of experimental melt compositions with those of Mercury's surface (grey area) in a total alkali to silica diagram with fields for the classification of volcanic rocks (Le Maitre 2002, Le Bas 2000).

We further investigated which areas of Mercury's surface can be explained by the partial melting of a EH chondritic mantle, considering the abundances of major cations (Mg, Si, Al and Ca). Partial melting at varying pressures and temperatures results in a diverse range of melt compositions. Additionally, EH4 Indarch compositions are rich in alkalis, creating more variability in melt composition compared to a simple CMAS system. The solubility of sulfur in magmas also alters melt compositions due to the presence of CaS and MgS complexes, which yield compositions different from those predicted for sulfur-free or oxidized conditions. Therefore, our experimental data are valuable for testing a EH chondritic model for Mercury by examining whether melting at various pressures and temperatures can account for the observed variation in Mercury's surface geochemistry. We compare experimental results on major element ratios with previously defined geochemical terranes (Weider et al. 2015, Peplowski & Stockstill-Cahill 2019), and available surface data compositions (Nittler et al. 2020) (Fig. 7). Our results show that experimental melts have Mg/Si, Al/Si and Ca/Si ratios that overlap with many surface data points. Specifically, the Intermediate-Mg Northern Volcanic Plains (IMg NVP) have compositions that closely match those of melts produced at 0.5-1 GPa (40-80 km depth).

To quantitatively compare our results with Mercury's surface data, we first interpolated our experimental data to cover all ranges of F from the minimum experimental F (12, 15, 16, 9 and 8 % melting at 0.5, 1, 2, 3 and 5 GPa, respectively) to 100% melting. We achieved this by fitting linear and polynomial functions to the experimental compositions for MgO, $Al_2O_3$, $SiO_2$ and CaO (see Fig. S11-12). We then performed a regression analysis by comparing surface ratios (from unsmoothed maps of Nittler et al. (2020)) with those determined experimentally, selecting only pixels where all three ratios (Mg/Si, Al/Si and Ca/Si) were measured by MESSENGER. We chose results where the sum of all relative residuals for the three ratios was minimized. These residuals were also used to estimate the level of agreement or disagreement between experimental and surface data:

$$Relative\ residual = \frac{|experimental\ ratio - observed\ ratio|}{observed\ ratio} \qquad (12)$$

Results show that 96, 74 and 26 % of Mercury's mapped surface in all three ratios (% of quarter-degree pixels in the cylindrical projections) has a composition similar to experimental partial melts with a maximum relative error of 50, 30 and 20%, respectively. For comparison, the relative errors on surface ratios measured by MESSENGER mostly range from 1 to 30% (see Fig. S15), and our experimental melts cover a similar range. The resulting pressures and degrees of melting for each pixel that matches experimental data are shown in Fig. 9. Geochemical maps derived from MESSENGER mission are compared with those of the closest experimental melts in composition in Fig. 10. Maps of residuals for each ratio and their distribution are illustrated in Fig. S13. Our numerical modeling indicates that most surface compositions match melts formed at 0.5 and 3 GPa with degrees of melting of 15 and 30 %, respectively (Fig. 9). Regions that closely match experimental data are parts of the northern volcanic plains (NVP), where magmas could have been formed at 0.5-1 GPa (40-80 km depth) with ~15% melting. Some southern NVP areas close to the high-Al rich region match melts formed at 3 GPa (240 km depth). These regions have higher crater densities than the rest of the NVP (Weider et al. 2015) and could be the result of older volcanic deposits formed at greater depths. It should also be noted that some NVP areas, specifically at the very high latitudes (>80°N), poorly fit with experimental melts, showing relatively high silica-contents with low Mg/Si and Al/Si (Fig. S15). These discrepancies may indicate specific chemical reservoirs and possibly multiple differentiation stages of their mantle sources, as noted by Charlier et al. (2013) and Namur et al. (2016b). Interestingly, the most silica-rich NVP at the highest latitudes, which have compositions poorly matching experimental data correspond to areas where spacecraft measurements errors are largest (40-50%, Fig. S15A). In addition, several areas of Mercury's surface have a higher Al/Si than experimental melts (ranging between 0.3 and 0.4 compared to 0.2-0.3), including the Caloris basin and its surrounding smooth plains, the High-Al geochemical terrane (HAl) and areas south of it, as well as the South pole. These areas may have formed from differentiated mantle sources.

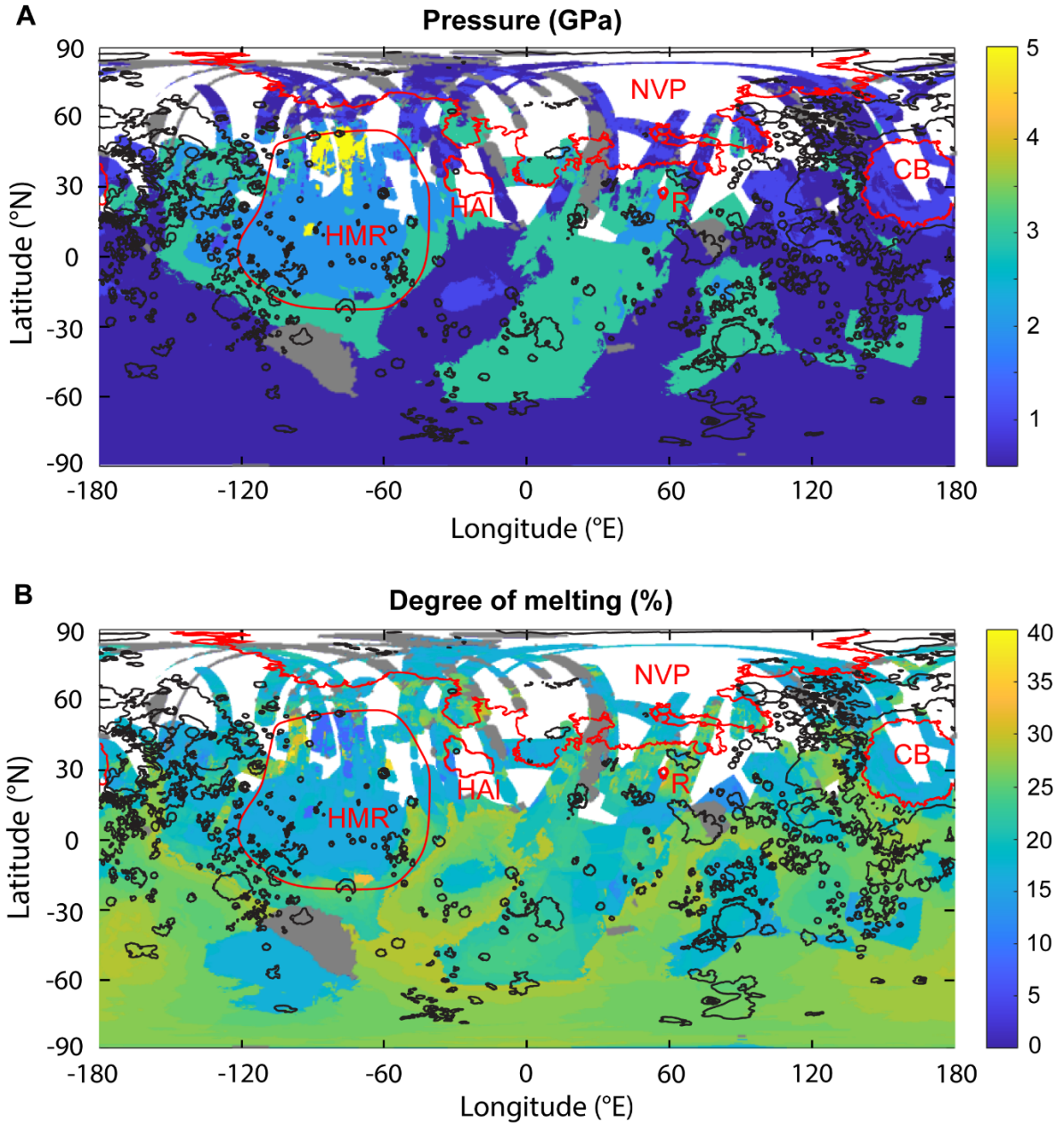

**Fig. 9.** Maps of Mercury showing pressures and degrees of melting (wt%) of EH chondrites, with chemical compositions matching the most closely with Mercury's surface chemical compositions with a relative residual up to 50% (shown in Fig. **S13**). White areas represent surface areas where Ca/Si was not measured by MESSENGER XRS spectrometer, and grey areas are those poorly matching partial melts of EH enstatite chondrites. Black outlines represent contours of smooth plains (Denevi et al. 2013, Nittler et al. 2020), while red outlines correspond to geochemical terranes identified in previous studies (Weider et al., 2015).

HMg = High-Magnesium, HAl = High-Aluminum, NVP = Volcanic Plains, CB = Caloris basin, R = Rachmaninoff basin.

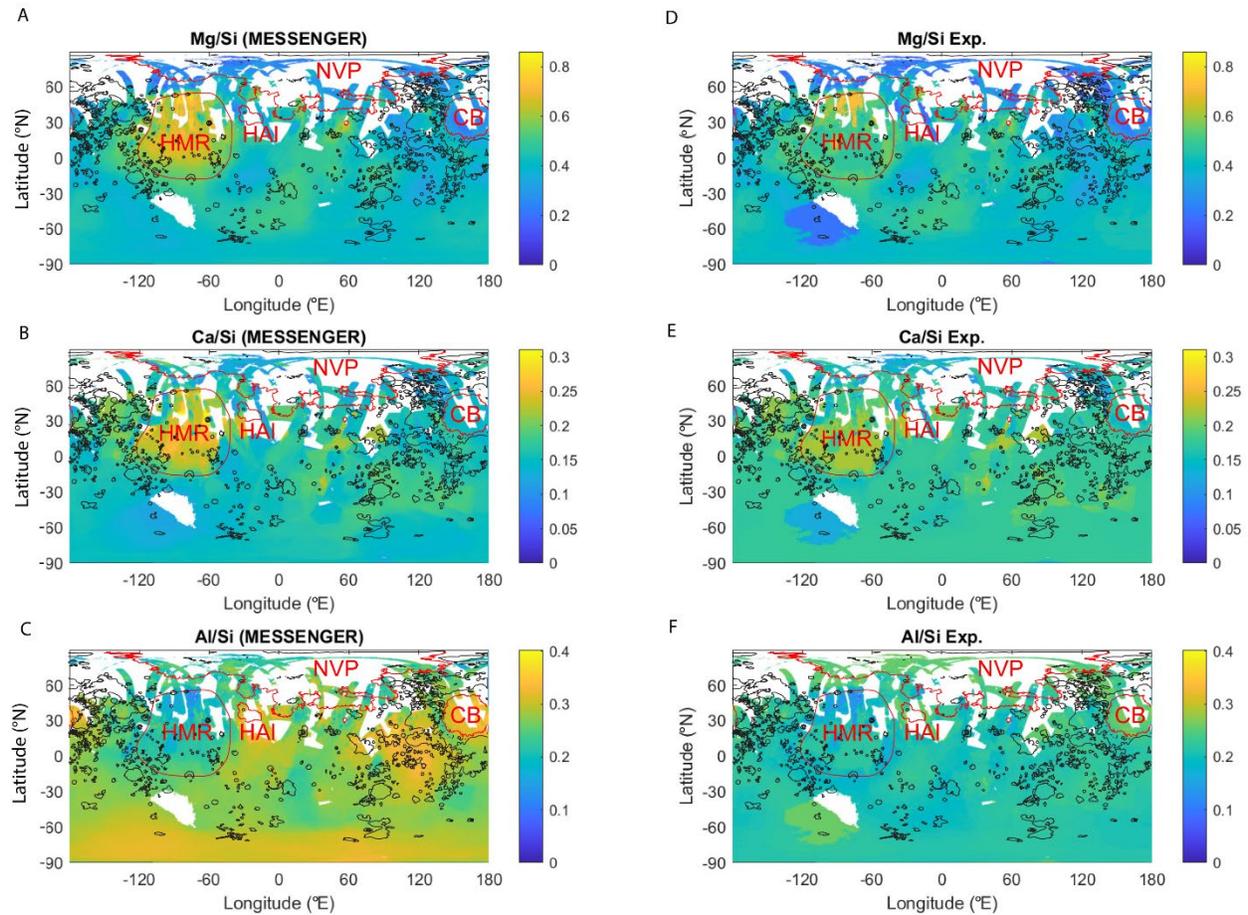

**Fig. 10.** Maps of Mg/Si, Ca/Si and Al/Si of Mercury as mapped by the MESSENGER mission (A-C) compared to those derived from experiments presented in this study that match the most closely to Mercury composition (D-F). See Fig.9 caption for outline and abbreviation descriptions.

The intermediate range of Ca/Si and relatively low Mg/Si of Caloris basin is consistent with the pressure range and degree of melting observed in NVP terranes, despite a higher Al/Si than NVPs and experimental melts. The circum-Caloris plains suggest similar pressures (0.5-1 GPa, 40-80 km depth) of melting. However, the northern and eastern plains suggest 15% melting, while the southern and western plains, which are slightly richer in Al, are consistent with a higher F (25%). Interestingly, melts matching the geochemical composition of Caloris basin and surrounding smooth plains are found in many areas of Mercury, especially in the Southern hemisphere, which has been mapped at very low spatial resolution (ranging from several hundreds to 2000 km). However, given the different Al/Si range of Caloris basin and surrounding plains with melts generated from a putative EH chondritic mantle, more complex differentiation processes may be required to explain their composition.

The Mg/Si ratio of the high-magnesium region displays a concentric pattern with an offset center, being very rich in Mg in its northernmost part at 50°N and 75°E (large yellow area in Fig.9A). Our results indicate that the northernmost part of the HMR, where Mg is the highest, matches compositions of melts formed at 5 GPa (the pressure at the core-mantle boundary) and low degrees of melting (5-10%). Recent studies suggested that diamond may exist as a thin layer at the core-mantle boundary (Xu et al. 2024). If diamond is present at the mantle's base, the northernmost area of the HMR is the most likely location for its discovery on Mercury's surface. Surrounding this area, most of the HMR has surface ratios that correspond to melts formed at 2 GPa with ~15% melting. The outer parts of the HMR and areas extending to Beethoven crater are associated with melts formed at 3 GPa and ~25% melting. These findings align with previous work by Namur et al. (2016b), suggesting melting at higher depth for the HMR. Interestingly, some NW parts of the HMR characterized by low Ca/Si indicate high degrees of melting (35 to 70%).

Previous work has proposed that the HMR is a remnant of a very large, degraded impact basin that excavated or melted the mantle possibly down to the core-mantle boundary, producing fractures that would allow large amounts of magma to penetrate the crust. This scenario was supported by the thin crust and low topography of the northern HMR, as well as the sharp topographic contrast with the adjacent smooth plains (Weider et al. 2015). However, the absence of impact-related structures and associated ejecta contradicts this hypothesis (Whitten et al. 2014, Frank et al. 2017). Modeling such a large impact predicts ejecta deposits extending to the planet's antipode (Frank et al. 2017). Alternatively, the HMR composition might result from high-pressure and high-temperature melting at an early stage of crust formation (Namur et al. 2016b), or from differentiated mantle sources with complex differentiation processes (Frank et al. 2017). Since melting products of a primitive enstatite chondrite composition at high pressures have compositions similar to those of the HMR, differentiated mantle sources are not required to explain its geochemical signature. However, our results cannot rule out either impact scenario or melting of an older warmer mantle. If the HMR resulted from a large impact, other areas of Mercury, such as the intercrater plains and heavily crater terranes (IcP-HCT), with compositions similar to melts formed at 3 GPa (at approximately the same latitudes as the HMR, and 0-30°E, 100°E and 160 °E) might represent impact ejecta whose geochemical signatures have been modified by mixing with younger terrains formed from shallower sources. Overall, our results agree with those of Namur et al. (2016b), suggesting shallow and deep sources for the younger NVPs and older high-Mg terranes, respectively.

## 4.5 Role of silicate melt mixing or impact gardening

Geochemical maps of Mercury's surface were constructed with spatial resolutions ranging from a few hundred to a couple of thousand kilometers (Nittler et al. 2020). As a result, each pixel in these maps likely represent mixtures of two or more compositions. Furthermore, the heavily cratered surface of Mercury may lead to mixing of local rocks with distant impact ejecta due to impact gardening. Additionally, if melting occurs at different depths (polybaric melting) and magmas mix prior to eruption, the surface composition could reflect combinations of various compositions. To address this, we performed linear regressions to assess whether surface compositions could result from the mixing of two silicate melts with different compositions. Since mixing modeling requires concentrations rather than ratios, we first converted ratios into concentrations in wt% by including Fe/Si and S/Si maps (Nittler et al. 2020) and calculating O by stoichiometry, with S as $S^{2-}$. Mass balance calculations were then performed to fit Mercury's surface ratios with mixtures of experimentally derived compositions. For example, for Mg, the calculation used the following formula:

$$X_{mix}^{Mg} = f_1 * X_{Exp1}^{Mg} + f_2 * X_{Exp2}^{Mg} \qquad (13)$$

where $X_{Exp1}^{Mg}$ and $X_{Exp1}^{Mg}$ are Mg concentrations of two silicate melts derived from the partial melting of EH4 Indarch chondrites or a mantle of comparable composition, at two sets of pressure and temperature. $f_1$ and $f_2$ (with $f_2 = 1 - f_1$) are the mass fractions of these two melts. To select the best results, we minimized the sum of the relative residuals for Mg/Si, Al/Si and Ca/Si (Eq. 7). To improve the efficiency of the linear regressions, we added a geometric filtering step. In a scatter plot, points representing perfect mixtures of two reference points lie on a line defined by these two points. Thus, the two vectors $\vec{v_1}$ and $\vec{v_2}$, defined by the mixture point and each of the two reference points are collinear. The scalar product $\vec{v_1}.\vec{v_2}$ should therefore be close to zero. The filter involved selecting only combinations of two experimental melts where the scalar product of the vectors, defined by the Mercury surface datapoint and each experimental melt, was close to zero. Scalar products were calculated using their coordinate definition, which we equate with their geometric definitions, to determine the angle between the two vectors:

$$\theta = \widehat{\vec{v_1} \, \vec{v_2}} = \cos^{-1} \frac{\vec{v_1}.\vec{v_2}}{\|\vec{v_1}\| * \|\vec{v_2}\|} \qquad (14)$$

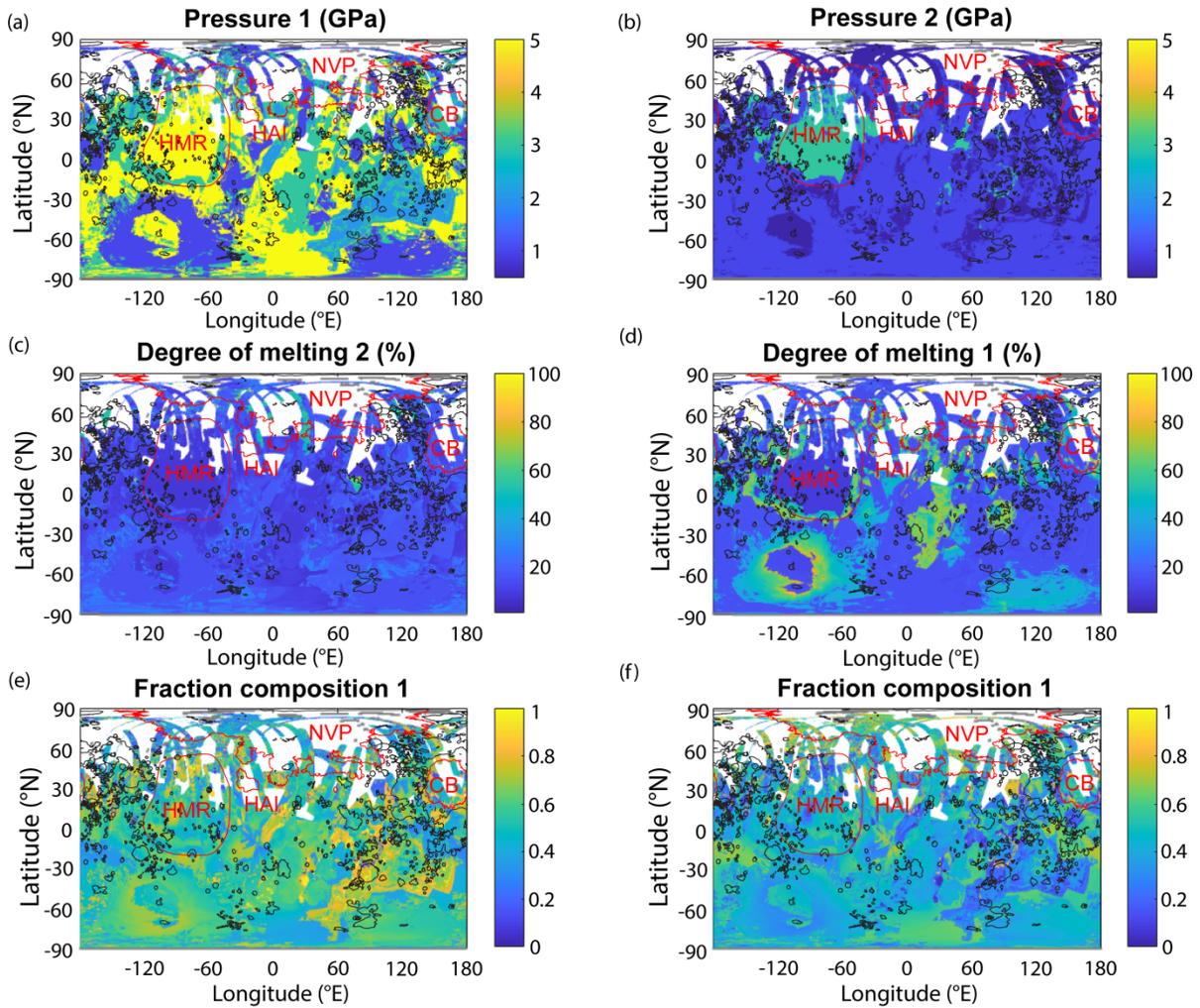

**Fig.11** Maps of Mercury showing pressures (a-b) and degrees of melting (wt%) (c-d) of two melts of EH chondrites, which mixture of chemical compositions match the most closely with Mercury's surface chemical compositions, with a relative residual up to 50% (shown in Fig. S14). (e-f) indicate the mass fraction of each melt. White areas represent surface areas where Ca/Si was not measured by MESSENGER XRS spectrometer, and grey areas are those poorly matching mixtures of partial melts of EH enstatite chondrites. Outlines abbreviations are described in Fig. 9.

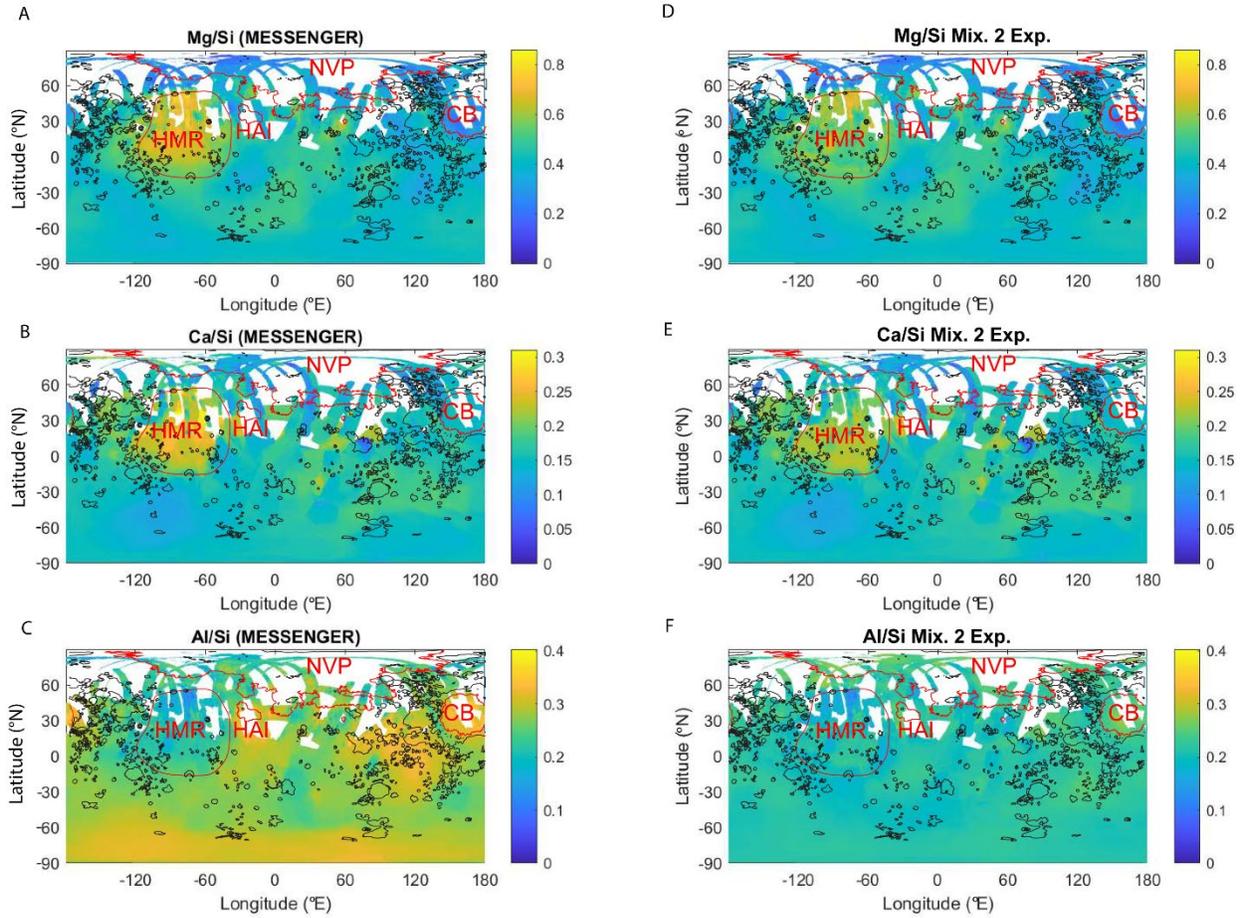

**Fig. 12.** Maps of Mg/Si, Ca/Si and Al/Si of Mercury as mapped by the MESSENGER mission (A-C) compared to those derived from mixtures of melts from experiments matching the most closely compositions of Mercury (D-F).

We filtered out combinations of points where the angle between vectors is close to 180±90 degrees. Figure 11 plots maps for the pressure and temperature of both experimental melts, as well as their mass fractions, while Figure 12 shows the comparison between observed (MESSENGER data) and calculated compositions from experimental data. Residual maps for the best fits can be found in Fig. S14. Table 5 summarizes results for all the fitting for the direct comparison (discussed in section 4.4) and comparison with mixtures of melts. While, Ca/Si is better resolved with the mixture model, Al/Si remains lower in the modeled compositions relative to the compositions of the South pole, Caloris basin, the High-Al region and their surrounding areas (Fig. 12). Results indicate a generally lower pressure range for the NVPs and other areas in the southern hemisphere. One striking finding from the mixture modeling is that most of Mercury's surface shows one modeled melt composition similar to melts formed at 3 or 5 GPa (Fig. 11a) and the other similar to melts formed at 0.5 or 1 GPa. Additionally, almost all the HMR now suggests compositions resulting from a mixture of melts from 3 and 5 GPa. If large impacts, such as the one forming the Caloris basin and possibly the HMR, excavated mantle materials, this component may represent impact ejecta from these deep-sourced materials.

Using mixture models to analyze Mercury's surface helps reduce discrepancies between calculated and observed Ca/Si over several areas, including the NVPs, a large area southwest of the HMR extending from 30° to 60°S, and circum-Caloris plains (Fig. 10,12). Additionally, many areas show better fits for Mg/Si, including regions east of the HAl, a large area southwest of the HMR, several parts of the NVPs and regions in the northwest circum-Caloris basin plains. Residuals for Al/Si ratios remain high for the southern hemisphere mapped at a lower spatial resolution, as well as some areas like the equatorial plains close the HAl and at 80°-100°E (Fig. S14b).

## 4.6 Mantle sources and heterogeneities

Previous experimental work investigating the origin of Mercury's surface examined the pressures and temperatures of melting of Mercury's surface lavas, by identifying multiple saturation points (Namur et al. 2016b, Vander Kaaden & McCubbin, 2016). This methodology is derived from studies on the generation of Earth's mid-ocean ridge basalts (MORBs), which are known to form by adiabatic decompression melting, producing multiple increments of melts over a range of pressures. In these conditions, the average pressure and temperature of melting along the entire column where magmas formed can be inferred by identifying the multiple saturation point of the basalt (pressure and temperature where the melt is in equilibrium with multiple solid phases) (Asimow & Longhi, 2004). However, this type of melting process for Mercury is questionable because (1) the presence of large flood volcanic plains suggests higher degrees of partial melting (15-30%) than those of MORB formation (Vander Kaaden & McCubbin, 2016), (2) the possible formation of Mercury from enstatite chondrites, which are extremely rich in enstatite, precludes the systematic equilibrium of melts with multiple solid silicate phases. The present study offers the opportunity to test whether Mercury's volcanic surface materials can be formed by the batch melting of an enstatite-rich mantle similar to EH4 enstatite chondrites. The normative mineralogy for the proposed primitive Mercury's mantle composition (average bulk silicate in our experiments, Table 1) comprises 67% enstatite, 14% olivine, 12% plagioclase and 6% diopside (Fig. 13). Previous mantle composition estimates (based on multiple saturation points) suggested a lherzolitic mantle (Nittler et al. 2018) with slight variations in mineralogy for the NVPs and IcP-HCT, showing 38-40% enstatite, 32-35% olivine, 17-20% plagioclase and 5-10% diopside.

While our results agree with the shallow and deep sources for the young NVPs and old High-Mg terranes (like in Namur et al., 2016b), they offer several advantages: (1) they are closer to chondritic compositions, and therefore, more easily reconciled with meteoritic data, (2) they do not require significant mantle heterogeneities while still accounting for a large fraction of Mercury's geochemical variability, (3) they consider the effects of Mg-Ca-bearing sulfides on mantle properties, and (4) they involve large ranges of degrees of melting. Additionally, magma production mechanisms could include both adiabatic decompression batch melting, due to the rise of hot mantle plumes, and non-adiabatic melting involving external sources such as large impacts. The latter effects can explain the high pressures of melting observed in the HMR and IcP-HCT, as well as the local high degrees of melting (80-90%) seen in the mixing model. Throughout the investigated pressure-temperature range, residual phases were dominated by enstatite and only experiments run at 0.5 GPa (corresponding to 40-80 km depth) and close to the liquidus (1450 to 1600 °C) included olivine. This implies that if Mercury accreted enstatite chondrite-like material, as the crust formed, the residual mantle was also likely enriched in enstatite. Hence, accounting for 12 wt% Si in the core (a value close to the median of geophysical estimates, Goossens et al. 2022), Mercury may be closer in composition to enstatite chondrites than Earth. Enstatite chondrites and Earth share very similar isotopic compositions, but Earth is more depleted in Si and enriched in Mg and refractory elements, including Al and Ca (e.g. Javoy et al. 2010). A few scenarios have been proposed to explain these discrepancies, such as the collisional erosion of a silica-rich protocrust (Boujibar et al. 2015), or the accretion of olivine-rich chondrules found in enstatite chondrites (Marrocchi et al. 2025).

Previous studies have suggested a relatively rapid cessation of mantle convection after 3-4 Gyr due to Mercury's small mantle thickness compared to other terrestrial planets (e.g., Tosi et al. 2013). Therefore, it remains important to consider Mercury's mantle stratification in the context of magma ocean crystallization. Besides graphite, which is suggested to have formed a primary flotation crust, Mercury's low FeO-content resulted in very small density contrasts (Vander Kaaden & McCubbin, 2015). Given the extensive stability of enstatite, as the magma ocean crystallized, enstatite would have fractionated for a prolonged period before the magma ocean shifted toward compositions other than pyroxenite (specifically olivine websterite). To transition from a pyroxenite to a peridotite composition, the magma ocean would need 80, 70 and 60% enstatite to fractionate at 5, 3 and 1 GPa, respectively (Fig. 13). Therefore, if cooling of the magma ocean led to stratification and significant differentiation, these are more likely to have occurred near the surface or close to the crust-mantle boundary than in the deeper layers. In addition, fractionation of sulfides with variable densities could allow for the development of density contrasts and overturn of sulfide-bearing mantle (Mouser & Dygert, 2023). Our work shows that sulfides become richer in Mg and Ca with decreasing temperature, and increasing silica, respectively. Both conditions are more likely to occur at shallow depths (melts produced at low pressure and temperature are richer in silica).

Hence, local heterogeneities could be present in the uppermost layers of the mantle, possibly generating melt compositions like the Al-rich surface areas that are poorly matching melts produced in this study. Notably, many areas having a higher Al/Si than those from partial melts of EH chondrites are from the Southern hemisphere mapped at the lowest spatial resolution. Future data collected by Bepi-Colombo mission in the Southern hemisphere will improve our understanding of Mercury's surface and inferred mantle composition.

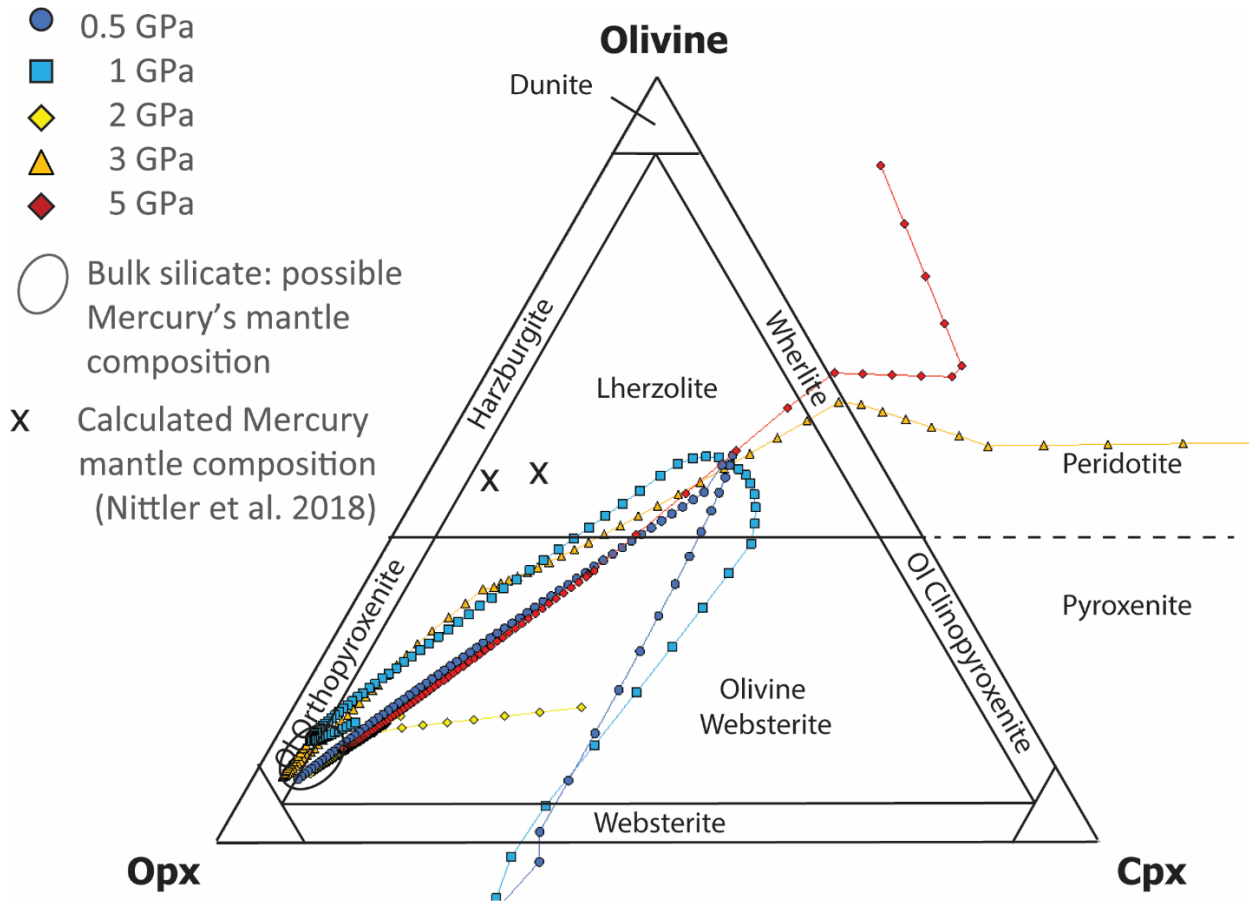

**Fig. 13** CIPW normative mineralogy of silicate melts interpolated from experimental results (colored symbols) (see Fig. S12-12), reported in a diagram for ultramafic rocks (Streckeisen 1974). The mineralogy of our starting EH chondritic composition (ellipse) which may represent Mercury's mantle composition is shown along that previously estimated in Nittler et al. (2018) (crosses). Melts close to the ellipse are formed at the highest degree of melting.

# 5. Conclusion

Findings from our study can be summarized with the following:

- Partial melting of a reduced EH4 Indarch enstatite chondrite (with $Si/SiO_2$ of 0.18) shows that the stability field of enstatite expands over olivine. This is interpreted as the result of Ca-S and Mg-S complexes in the silicate melt, increasing $SiO_2$ activity and favoring enstatite crystallization.

- The correlation between Ca and S concentrations in silicate melts forms a trend similar to the global trend observed on Mercury's surface, but with lower S abundances, likely arising from global S degassing during volcanic eruptions.

- Additionally, at low temperatures and high $SiO_2$ of the silicate melt, sulfides are enriched in Mg and Ca, respectively (up to 13 and 22 wt%, respectively). These sulfides generate a larger range of compositions than expected in sulfide-undersaturated conditions.

- Silicate melts produced by partial melting of this reduced EH4 chondrite show a variety of compositions similar to those observed on Mercury. These include Mg-rich compositions at high pressures (2 to 5 GPa) like in the high-magnesium region, and Si-rich compositions at lower pressures (0.5 to 1 GPa) like the northern volcanic plains. The comparison of these melts with Mercury's surface indicates that a large fraction of Mercury's surface has a composition similar to those predicted from experiments, suggesting that Mercury's mantle may be predominantly pyroxenitic. Terranes with mismatched composition have higher Al-content and include the Al-rich region, Caloris basin, and areas of the Southern hemisphere. The high-magnesium region shows a chemical composition indicative of melting at high pressures up to the base of the mantle. Several areas of intercrater plains and heavily cratered terranes might be mixtures of melts generated at the core-mantle boundary and melts from lower pressures. Our results align with both shallow and deep sources for Mercury's young NVPs and older High-Mg terranes, respectively.

- A mantle with an enstatite chondrite composition can explain some of the geochemical variability of Mercury's surface, especially for the High-Mg region and the NVPs. However, heterogeneities in the mantle would still be required to explain Al enrichments in the Caloris basin and around the High-Al region.

- If Mercury formed with materials like enstatite chondrites, most of its mantle preserved its pyroxenite composition due to the extensive stability of enstatite. If mantle stratification occurred during magma ocean crystallization, it is more likely to be more significant in the layers close to the surface, or driven by sulfide fractionation, since sulfides would yield more density contrasts.

## Acknowledgements

This work was funded by the NASA Research Initiative Award 80NSSC24K0785 and NASA Postdoctoral Program fellowship awarded to A.B, an RTOP from the NASA Cosmochemistry program awarded to K.R, and support from NASA Planetary Science Division. We acknowledge Lisa Danielson, Anne Peslier, Loan Le, and Kent Ross for support with electron microprobe and sample preparation. We thank the associate Paolo Sossi, as well as Bernard Charlier and an anonymous reviewer for their constructive feedback.

## Author contribution

**Asmaa Boujibar:** Conceptualization, Methodology, Data curation and analysis, Funding acquisition, Writing- Original draft preparation **Kevin Righter**: Funding acquisition, Writing- review & editing. **Emmanuel Fontaine**: Software. **Max Collinet**: Writing - review & editing. **Sarah Lambart:** Writing - review & editing. **Larry Nittler:** Writing- reviewing and editing, **Kellye Pando:** Data curation and editing.

## Declaration of competing interests

The authors declare that they have no known competing financial interests or personal relationships that could have appeared to influence the work reported in this paper.

## Declaration of generative AI in scientific writing

During the preparation of this work the authors used ChatGPT in order to improve the flow and readability of the text. After using this tool/service, the authors reviewed and edited the content as needed and take full responsibility for the content of the publication.

## Data availability

Experimental data are available in tables in the manuscript and supplementary material.

# Tables

**Table 1** Chemical compositions of the EH4 Indarch chondrite (Wiik 1956, Berthet et al. 2009), starting powder, average and standard deviation for bulk compositions from fully melted samples (in wt%).

|  | Natural EH4 Indarch chondrite (Wiik 1956) | Starting powder | Average bulk silicate composition | Standard deviation bulk composition |
|---|---|---|---|---|
| $SiO_2$ | 61.8 | 51.8 | 57.47 | 1.19 |
| MgO | 30.6 | 38.6 | 35.53 | 1.76 |
| $Al_2O_3$ | 2.5 | 3.2 | 2.98 | 0.23 |
| CaO | 1.7 | 2.1 | 2.03 | 0.23 |
| $Na_2O$ | 1.8 | 2.2 | 1.13 | 0.39 |
| $K_2O$ | 0.19 | 0.24 | 0.17 | 0.02 |
| $Cr_2O_3$ | 0.82 | 1.04 | 0.04 | 0.01 |
| $MnO_2$ | 0.44 | 0.68 |  |  |
| $TiO_2$ | 0.11 | 0.13 | 0.13 | 0.01 |
| MnO |  |  | 0.33 | 0.06 |
| S |  |  | 2.99 | 0.80 |
| FeO |  |  | 0.26 | 0.09 |
| Fe | 82.42 | 72.53 | 80.10 | 0.73 |
| Ni | 4.55 | 4.00 | 4.67 | 0.84 |
| Co | 0.2 | 0.17 | 0.k,59 | 0.04 |
| S | 12.83 | 11.29 | 0.39 | 0.62 |
| Si |  | 12.00 | 11.94 | 0.33 |
| Mn |  |  | 0.10 | 0.07 |
| Cr |  |  | 0.69 | 0.10 |
| C* |  |  | 1.51 | 0.59 |
| Silicate (wt%) | 59 | 50 | 54 |  |
| Metal + Sulfide (wt%) | 41 | 50 | 46 |  |
| $Si/SiO_2$ | 0 | 0.23 | 0.18 |  |

**Table 2** Experimental conditions and run products.

| Run | Pressure (GPa) | Temp.[1] (°C) | Heating duration | ΔIW[2] | ΔIW[3] | F[4] (wt%) | Phase proportions (wt%)[5] |
|---|---|---|---|---|---|---|---|
| #309 | 5 | 1600 | 1h30 | -4.1 | -6.2 | 100 | 45 met, 55 sil melt |
| #311 | 5 | 1700 | 1h30 | -4.0 | -5.5 | 65 | 48 met, 34 sil melt, 18 opx |
| #315 | 5 | 1550 | 2h | -3.9 | -6.5 | 22 | 10 sulf, 39 met, 11 sil melt, 40 opx |
| #314 | 5 | 1600 | 1h30 | -2.9 | -6.2 | 8 | 13 sulf, 37 met, 4 sil melt, 46 opx |
| #384 | 3 | 1850 | 30 min | -4.0 | -4.3 | 100 | 46 met, 54 sil melt |
| #385 | 3 | 1880 | 15 min | -4.0 | -4.4 | 100 | 47 met, 53 sil melt |
| #382 | 3 | 1750 | 2h | -4.2 | -4.9 | 55 | 6 sulf, 43 met, 28 sil melt, 23 opx |
| #386 | 3 | 1780 | 1h30 | -3.8 | -4.4 | 42 | 7 sulf, 43 met, 21 sil melt, 29 opx |
| #383 | 3 | 1800 | 1h | -4.1 | -4.3 | 41 | 9 sulf, 42 met, 20 sil melt, 29 opx |
| #374 | 3 | 1600 | 1h30 | -2.7 | -4.7 | 12 | 7 sulf, 44 met, 6 sil melt, 43 opx |

| | | | | | | |
|---|---|---|---|---|---|---|
| #388 | 3 | 1735 | 40 min | -3.2 | -4.0 | 10 | 8 sulf, 43 met, 5 sil melt, 44 opx |
| #387 | 3 | 1620 | 1h30 | -3.1 | -4.8 | 10 | 8 sulf, 43 met, 5 sil melt, 44 opx |
| #389 | 3 | 1720 | 1h | -3.4 | -4.3 | 9 | 9 sulf, 44 met, 4 sil melt, 43 opx |
| #431 | 2 | 1640 | 1h30 | -4.2 | -4.7 | 100 | 47 met, 53 sil melt |
| #432 | 2 | 1600 | 2h | -4.3 | -5.4 | 25 | 48 met, 13 sil melt, 39 opx |
| #433 | 2 | 1550 | 3h | -4.0 | -5.4 | 17 | 48 met, 9 sil melt, 43 opx |
| #445 | 2 | 1450 | 4h | -3.4 | -6.1 | 16 | 9 sulf, 42 met, 8 sil melt, 41 opx |
| #1073 | 2 | 1500 | 3h | -3.2 | -5.5 | 16 | 6 sulf, 44 met, 8 sil melt, 42 opx |
| #869 | 1 | 1500 | 3h | -4.4 | -7.2 | 100 | 47 met, 53 sil melt |
| #870 | 1 | 1600 | 1h30 | -3.7 | -6.3 | 68.5 | 46 met, 37 sil melt, 17 opx |
| #878 | 1 | 1500 | 4h | -3.8 | -6.9 | 42.6 | 1 sulf, 45 met, 23 sil melt, 31 opx |
| #880 | 1 | 1550 | 3h | -4.0 | -6.1 | 35 | 1 sulf, 42 met, 20 sil melt, 37 opx |
| #872 | 1 | 1550 | 2h30 | -3.7 | -5.8 | 24 | 2 sulf, 41 met, 14 sil melt, 43 opx |
| #871 | 1 | 1450 | 4h30 | -2.7 | -7.4 | 15 | 3 sulf, 44 met, 8 sil melt, 45 opx |
| #874 | 1 | 1475 | 6h30 | -3.5 | -6.5 | 20 | 3 sulf, 45 met, 10.5 sil melt, 41.5 opx |
| #963 | 1 | 1650 | 1h30 | -4.6 | -5.3 | 100 | 46 met, 54 sil melt |
| #964 | 1 | 1600 | 1h30 | -4.6 | -5.5 | 100 | 47 met, 53 sil melt |
| #955 | 0.5 | 1250 | 20h15 | -3.0 | -8.1 | 12 | 7 sulf, 44 met, 6 sil melt, 43 opx |
| #952 | 0.5 | 1400 | 7h | -3.1 | -7.2 | 20 | 6 sulf, 45 met, 10 sil melt, 39 opx |
| #953 | 0.5 | 1350 | 9h | -3.3 | -7.7 | 22 | 7 sulf, 43 met, 11 sil melt, 39 opx |
| #949 | 0.5 | 1500 | 3h | -3.3 | -6.7 | 36 | 5 sulf, 45 met, 18 sil melt, 32 opx |
| #948 | 0.5 | 1600 | 1h30 | -3.4 | -5.9 | 64 | 2 sulf, 46 met, 33 sil melt, 9 opx, 10 ol |
| #951 | 0.5 | 1450 | 5h | -3.6 | -6.3 | 56 | 4 sulf, 44 met, 29 sil melt, 19 opx, 4 ol |
| #966 | 0.5 | 1550 | 1h30 | -4.4 | -5.8 | 65 | 6 sulf, 42 met, 34 sil melt, 11 opx, 7 ol |
| #988 | 0.5 | 1600 | 1h45 | -4.9 | -5.4 | 86 | 45 met, 47 sil melt, 1 opx, 7 ol |

[1] Temp. = peak temperatures at which samples were left to equilibrate for the given heating duration. [2] Oxygen fugacity ($f$O$_2$) in log units relative to iron-wustite buffer calculated using Fe-Fe equilibria (see text for more details). [3] $f$O$_2$ in log units relative to iron-wustite buffer calculated using Si-SiO$_2$ equilibria (following Cartier et al .2014). [4] F = degree of silicate melting. We note that temperature-degree of melting relationship is not reproducible among different samples. [5] Phase proportions are calculated by mass balance: met = metal, sulf = sulfide, sil melt = silicate melt, opx = orthopyroxene, ol = olivine.

**Table 3** Average chemical compositions of silicate melts and their 1 σ standard deviations.

| Silicate melts | N[a] | SiO$_2$ | TiO$_2$ | Al$_2$O$_3$ | Cr$_2$O$_3$ | FeO | MnO | MgO | CaO | Na$_2$O | K$_2$O | S | NiO | Total |
|---|---|---|---|---|---|---|---|---|---|---|---|---|---|---|
| #309 | 8 | 58.39 | 0.12 | 3.11 | 0.03 | 0.28 | 0.25 | 32.66 | 2.27 | 1.80 | 0.19 | 1.78 | n.m. | 100.02 |
| | | 0.62 | 0.02 | 0.06 | 0.01 | 0.06 | 0.02 | 0.47 | 0.03 | 0.08 | 0.01 | 0.15 | | |
| #311 | 7 | 56.37 | 0.12 | 3.32 | 0.03 | 0.35 | 0.23 | 34.75 | 2.45 | 0.93 | 0.08 | 1.27 | n.m. | 99.29 |
| | | 0.44 | 0.02 | 0.18 | 0.02 | 0.10 | 0.03 | 0.36 | 0.19 | 0.09 | 0.03 | 0.22 | | |
| #315 | 9 | 52.99 | 0.19 | 6.38 | 0.03 | 0.37 | 0.30 | 28.69 | 6.16 | 2.07 | 0.46 | 2.73 | n.m. | 99.04 |
| | | 0.64 | 0.02 | 0.11 | 0.01 | 0.11 | 0.04 | 0.40 | 0.15 | 0.20 | 0.15 | 0.47 | | |
| #314 | 11 | 41.36 | 0.25 | 7.71 | 0.07 | 1.33 | 0.90 | 26.41 | 13.41 | 2.47 | 0.12 | 11.65 | n.m. | 99.88 |
| | | 2.83 | 0.07 | 0.59 | 0.03 | 0.32 | 0.28 | 1.88 | 1.54 | 0.37 | 0.06 | 2.47 | | |
| #384 | 7 | 57.60 | 0.12 | 3.00 | 0.04 | 0.36 | 0.30 | 36.54 | 1.73 | 0.95 | 0.19 | 2.64 | 0.01 | 102.15 |
| | | 0.45 | 0.01 | 0.18 | 0.01 | 0.09 | 0.11 | 0.72 | 0.16 | 0.52 | 0.02 | 0.37 | 0.02 | |
| #385 | 6 | 58.13 | 0.11 | 2.76 | 0.04 | 0.38 | 0.27 | 37.34 | 1.77 | 0.72 | 0.16 | 2.48 | 0.02 | 102.95 |
| | | 0.40 | 0.02 | 0.07 | 0.02 | 0.08 | 0.04 | 0.60 | 0.09 | 0.26 | 0.02 | 0.07 | 0.03 | |
| #382 | 8 | 54.99 | 0.17 | 5.22 | 0.04 | 0.30 | 0.41 | 31.82 | 3.46 | 1.80 | 0.37 | 4.01 | n.m. | 100.59 |
| | | 0.17 | 0.02 | 0.16 | 0.02 | 0.03 | 0.02 | 0.45 | 0.12 | 0.18 | 0.03 | 0.16 | | |
| #386 | 6 | 56.67 | 0.17 | 4.66 | 0.03 | 0.38 | 0.38 | 31.86 | 2.61 | 1.92 | 0.33 | 3.33 | n.m. | 100.67 |
| | | 0.54 | 0.01 | 0.14 | 0.01 | 0.08 | 0.08 | 0.33 | 0.15 | 0.16 | 0.02 | 0.25 | | |
| #383 | 7 | 54.65 | 0.22 | 4.57 | 0.01 | 0.33 | 0.36 | 32.75 | 2.62 | 1.51 | 0.05 | 3.31 | n.m. | 100.37 |
| | | 0.60 | 0.02 | 0.09 | 0.01 | 0.05 | 0.05 | 0.57 | 0.13 | 0.57 | 0.00 | 0.25 | | |
| #374 | 5 | 55.95 | n.m. | 13.87 | 0.05 | 1.21 | 0.13 | 12.39 | 6.80 | 3.22 | 1.57 | 9.59 | n.m. | 101.16 |
| | | 2.82 | | 1.29 | 0.03 | 0.34 | 0.08 | 2.81 | 1.68 | 2.39 | 0.35 | 2.92 | | |
| #388 | 10 | 51.56 | 0.28 | 13.50 | 0.06 | 0.82 | 0.55 | 12.94 | 8.05 | 6.37 | 1.32 | 11.19 | n.m. | 101.07 |
| | | 3.90 | 0.10 | 1.09 | 0.03 | 0.39 | 0.26 | 1.12 | 2.10 | 1.04 | 0.07 | 2.59 | | |
| #387 | 7 | 48.85 | 0.35 | 13.52 | 0.09 | 0.92 | 0.85 | 13.04 | 9.72 | 6.35 | 1.55 | 13.25 | n.m. | 101.88 |
| | | 1.28 | 0.06 | 0.53 | 0.03 | 0.14 | 0.23 | 0.57 | 0.84 | 0.58 | 0.06 | 1.33 | | |
| #389 | 12 | 50.73 | 0.27 | 12.03 | 0.05 | 0.63 | 0.64 | 14.82 | 8.95 | 6.43 | 1.28 | 11.56 | n.m. | 101.62 |
| | | 2.82 | 0.11 | 1.62 | 0.04 | 0.28 | 0.49 | 2.48 | 1.10 | 1.08 | 0.15 | 2.51 | | |
| #431 | 7 | 57.48 | 0.13 | 2.75 | 0.05 | 0.28 | 0.31 | 34.65 | 2.02 | 0.91 | 0.16 | 2.87 | n.m. | 100.17 |

| | | | | | | | | | | | | | |
|---|---|---|---|---|---|---|---|---|---|---|---|---|---|
| | | 0.11 | 0.02 | 0.04 | 0.01 | 0.04 | 0.02 | 0.27 | 0.06 | 0.05 | 0.01 | 0.06 | | |
| #432 | 5 | 56.54 | 0.14 | 3.87 | 0.05 | 0.26 | 0.39 | 32.86 | 2.68 | 1.31 | 0.23 | 3.61 | n.m. | 100.15 |
| | | 0.82 | 0.02 | 0.03 | 0.01 | 0.04 | 0.04 | 0.39 | 0.20 | 0.08 | 0.01 | 0.11 | | |
| #433 | 10 | 55.30 | 0.19 | 5.72 | 0.06 | 0.35 | 0.56 | 29.08 | 4.06 | 1.14 | 0.36 | 5.27 | n.m. | 99.45 |
| | | 0.40 | 0.01 | 0.63 | 0.01 | 0.06 | 0.09 | 1.24 | 0.61 | 0.17 | 0.04 | 0.51 | | |
| #445 | 7 | 53.63 | 0.18 | 13.45 | 0.04 | 0.73 | 0.11 | 14.36 | 10.63 | 3.80 | 0.93 | 8.84 | n.m. | 102.3 |
| | | 1.42 | 0.04 | 1.59 | 0.02 | 0.18 | 0.02 | 1.78 | 0.54 | 0.27 | 0.11 | 0.94 | | |
| #1073 | 7 | 56.72 | 0.15 | 14.05 | 0.04 | 1.07 | 0.08 | 11.45 | 6.70 | 3.88 | 1.08 | 10.65 | n.m. | 100.56 |
| | | 0.88 | 0.01 | 0.85 | 0.01 | 0.17 | 0.03 | 0.86 | 0.66 | 0.43 | 0.06 | 0.64 | | |
| #869 | 9 | 57.23 | 0.12 | 2.76 | 0.05 | 0.24 | 0.38 | 35.52 | 1.90 | 0.82 | 0.14 | 2.92 | n.m. | 102.09 |
| | | 0.44 | 0.02 | 0.08 | 0.02 | 0.03 | 0.05 | 0.28 | 0.20 | 0.08 | 0.02 | 0.19 | | |
| #870 | 9 | 53.37 | 0.22 | 4.05 | 0.02 | 0.25 | 0.48 | 33.73 | 2.76 | 1.23 | 0.04 | 4.69 | n.m. | 100.85 |
| | | 0.63 | 0.01 | 0.16 | 0.01 | 0.03 | 0.04 | 0.56 | 0.25 | 0.70 | 0.00 | 0.13 | | |
| #878 | 7 | 53.90 | 0.26 | 6.55 | 0.02 | 0.42 | 0.29 | 28.64 | 4.10 | 1.78 | 0.06 | 5.34 | n.m. | 100.27 |
| | | 0.99 | 0.03 | 0.31 | 0.01 | 0.07 | 0.11 | 1.45 | 0.99 | 1.01 | 0.00 | 0.59 | | |
| #880 | 8 | 53.49 | 0.27 | 6.25 | 0.02 | 0.36 | 0.39 | 27.83 | 4.89 | 2.93 | 0.06 | 6.73 | n.m. | 103.23 |
| | | 0.23 | 0.02 | 0.14 | 0.01 | 0.10 | 0.13 | 1.24 | 0.82 | 0.20 | 0.00 | 0.17 | | |
| #872 | 9 | 53.72 | 0.17 | 10.12 | 0.03 | 0.56 | 0.12 | 20.81 | 6.72 | 4.60 | 0.59 | 7.48 | n.m. | 101.20 |
| | | 1.07 | 0.02 | 1.08 | 0.02 | 0.12 | 0.02 | 1.54 | 0.56 | 0.62 | 0.05 | 1.10 | | |
| #871 | 7 | 59.71 | 0.06 | 14.74 | 0.02 | 0.70 | 0.05 | 11.48 | 6.10 | 5.18 | 0.96 | 5.05 | n.m. | 101.53 |
| | | 1.78 | 0.02 | 0.44 | 0.00 | 0.10 | 0.02 | 0.80 | 0.58 | 1.18 | 0.07 | 0.42 | | |
| #874 | 8 | 59.27 | n.m. | 14.52 | 0.03 | 0.72 | 0.06 | 11.89 | 6.34 | 3.59 | 0.93 | 5.05 | n.m. | 99.86 |
| | | 1.37 | | 0.71 | 0.02 | 0.16 | 0.02 | 0.92 | 0.72 | 1.12 | 0.09 | 0.44 | | |
| #963 | 6 | 58.15 | 0.13 | 2.86 | 0.03 | 0.21 | 0.31 | 35.19 | 1.98 | 1.20 | 0.17 | 3.22 | n.m. | 103.45 |
| | | 0.21 | 0.02 | 0.05 | 0.01 | 0.03 | 0.03 | 0.21 | 0.07 | 0.06 | 0.01 | 0.10 | | |
| #964 | 7 | 58.11 | 0.12 | 3.27 | 0.04 | 0.18 | 0.36 | 34.22 | 2.18 | 1.50 | 0.19 | 3.56 | n.m. | 101.96 |
| | | 0.13 | 0.01 | 0.04 | 0.01 | 0.03 | 0.02 | 0.21 | 0.13 | 0.06 | 0.02 | 0.16 | | |
| #955 | 8 | 69.25 | 0.03 | 16.39 | 0.01 | 0.69 | 0.01 | 5.16 | 3.22 | 4.01 | 0.68 | 2.64 | n.m. | 102.08 |
| | | 0.64 | 0.01 | 0.53 | 0.01 | 0.09 | 0.01 | 0.77 | 0.13 | 0.25 | 0.03 | 0.09 | | |
| #952 | 7 | 59.04 | 0.12 | 12.82 | 0.04 | 0.71 | 0.06 | 15.21 | 7.07 | 4.60 | 0.62 | 5.51 | n.m. | 103.05 |
| | | 0.30 | 0.01 | 0.48 | 0.02 | 0.22 | 0.02 | 0.75 | 0.39 | 0.18 | 0.03 | 0.20 | | |

| Sulfides | Nᵃ | Mg | Si | Na | Al | K | Ca | Ti | S | Cr | Mn | Fe | O | Tot |
|---|---|---|---|---|---|---|---|---|---|---|---|---|---|---|
| #953 | 8 | 60.16 | 0.12 | 13.05 | 0.04 | 0.66 | 0.06 | 13.62 | 6.76 | 5.25 | 0.59 | 5.04 | n.m. | 102.82 |
|  |  | 0.44 | 0.01 | 0.24 | 0.01 | 0.15 | 0.02 | 0.61 | 0.21 | 0.08 | 0.02 | 0.16 |  |  |
| #949 | 7 | 54.42 | 0.20 | 6.99 | 0.06 | 0.57 | 0.22 | 24.97 | 5.59 | 3.20 | 0.39 | 6.78 | n.m. | 100.00 |
|  |  | 0.87 | 0.02 | 1.15 | 0.03 | 0.21 | 0.05 | 1.37 | 0.86 | 0.31 | 0.05 | 0.78 |  |  |
| #948 | 7 | 55.71 | 0.20 | 5.43 | 0.05 | 0.50 | 0.40 | 28.47 | 3.78 | 3.04 | 0.31 | 6.01 | n.m. | 100.91 |
|  |  | 0.83 | 0.02 | 1.21 | 0.02 | 0.23 | 0.08 | 2.61 | 0.99 | 0.80 | 0.06 | 0.97 |  |  |
| #951 | 7 | 55.77 | 0.25 | 5.74 | 0.10 | 0.48 | 0.45 | 28.88 | 4.27 | 2.60 | 0.30 | 5.69 | n.m. | 101.69 |
|  |  | 0.85 | 0.02 | 0.40 | 0.03 | 0.13 | 0.11 | 1.72 | 0.99 | 0.46 | 0.04 | 0.44 |  |  |
| #966 | 7 | 57.96 | 0.20 | 4.35 | 0.10 | 0.29 | 0.63 | 30.20 | 3.03 | 2.47 | 0.23 | 5.03 | n.m. | 104.50 |
|  |  | 0.25 | 0.02 | 0.06 | 0.01 | 0.05 | 0.04 | 0.37 | 0.15 | 0.25 | 0.01 | 0.13 |  |  |
| #988 | 15 | 58.45 | 0.14 | 3.20 | 0.04 | 0.18 | 0.41 | 32.86 | 2.30 | 1.50 | 0.18 | 3.94 | n.m. | 101.23 |
|  |  | 0.21 | 0.02 | 0.19 | 0.01 | 0.02 | 0.03 | 0.34 | 0.12 | 0.14 | 0.01 | 0.17 |  |  |

ᵃN = Number of analysis. n.m. = not measured.

**Table 4** Averages and 1σ standard deviations for the chemical compositions of sulfides. The average and standard deviation composition of all metals from all samples are also shown to assess that the presence of a metal phase in our charges have limited effects on silicate-sulfide phase equilibria (see section 2.1 for more details). Chemical compositions of metals for individual samples are given in Supplementary Table 1.

| Sulfides | Nᵃ | Mg | Si | Na | Al | K | Ca | Ti | S | Cr | Mn | Fe | O | Tot |
|---|---|---|---|---|---|---|---|---|---|---|---|---|---|---|
| #315 | 6 | 0.01 | 0.07 | 0.14 | 0.02 | 0.04 | b.d.l. | 2.01 | 36.96 | 7.13 | 4.73 | 50.30 | n.m. | 101.40 |
|  |  | 0.01 | 0.06 | 0.12 | 0.01 | 0.04 |  | 0.70 | 0.66 | 0.76 | 0.35 | 0.61 |  |  |
| #314 | 6 | 0.23 | 0.96 | 0.09 | 0.02 | 0.02 | 0.02 | 1.05 | 34.46 | 7.63 | 1.70 | 53.60 | n.m. | 99.80 |
|  |  | 0.31 | 0.44 | 0.08 | 0.03 | 0.03 | 0.04 | 0.69 | 1.39 | 0.39 | 0.66 | 2.69 |  |  |
| #382 | 5 | 6.55 | 0.09 | 0.50 | 0.01 | n.m. | 0.54 | 0.96 | 39.77 | 6.02 | 15.12 | 30.16 | n.m. | 99.73 |
|  |  | 0.80 | 0.05 | 0.08 | 0.00 |  | 0.13 | 0.18 | 1.01 | 0.15 | 1.09 | 1.07 |  |  |
| #386 | 4 | 7.86 | 0.25 | 0.68 | 0.01 | b.d.l. | 2.18 | 0.68 | 40.73 | 6.67 | 10.64 | 32.48 | n.m. | 102.19 |
|  |  | 0.95 | 0.14 | 0.03 | 0.02 |  | 0.50 | 0.11 | 0.17 | 0.25 | 1.20 | 0.75 |  |  |
| #383 | 5 | 6.62 | 1.44 | 1.51 | 0.31 | 0.04 | 2.84 | 1.01 | 37.60 | 6.70 | 11.17 | 27.75 | n.m. | 96.98 |
|  |  | 0.37 | 1.19 | 0.29 | 0.41 | 0.08 | 0.87 | 0.17 | 1.60 | 0.67 | 0.53 | 1.97 |  |  |
| #374 | 7 | 18.03 | 0.44 | 0.85 | 0.02 | 0.00 | 5.52 | 1.31 | 44.81 | 2.73 | 9.36 | 15.17 | n.m. | 98.26 |
|  |  | 1.43 | 0.52 | 0.09 | 0.02 | 0.01 | 0.45 | 0.07 | 1.80 | 0.43 | 0.75 | 3.23 |  |  |

| # | | | | | | | | | | | | | | |
|---|---|---|---|---|---|---|---|---|---|---|---|---|---|---|
| #388 | 3 | 8.41 | 1.45 | 1.31 | 0.19 | b.d.l. | 2.43 | 0.87 | 38.71 | 6.04 | 10.91 | 26.31 | n.m. | 96.64 |
| | | 1.33 | 0.74 | 0.26 | 0.07 | | 1.23 | 0.19 | 1.46 | 0.52 | 3.71 | 1.04 | | |
| #387 | 3 | 8.23 | 1.65 | 1.10 | 0.33 | 0.03 | 3.48 | 0.85 | 37.90 | 5.27 | 9.70 | 29.08 | n.m. | 97.62 |
| | | 2.08 | 0.78 | 0.13 | 0.24 | 0.04 | 0.27 | 0.11 | 0.58 | 0.72 | 4.50 | 0.89 | | |
| #389 | 4 | 7.97 | 0.59 | 1.69 | 0.16 | 0.02 | 2.47 | 0.96 | 39.43 | 6.71 | 11.41 | 27.08 | n.m. | 98.49 |
| | | 1.55 | 0.56 | 0.32 | 0.16 | 0.01 | 0.64 | 0.16 | 1.21 | 0.43 | 1.92 | 1.29 | | |
| #445 | 7 | 14.68 | 0.05 | 0.42 | 0.00 | 0.01 | 2.77 | 1.07 | 43.28 | 3.81 | 12.04 | 23.07 | n.m. | 101.19 |
| | | 1.96 | 0.02 | 0.05 | 0.01 | 0.01 | 0.32 | 0.05 | 0.83 | 0.87 | 1.98 | 4.37 | | |
| #1073 | 8 | 21.21 | 0.17 | 1.24 | 0.01 | 0.01 | 5.38 | 1.31 | 45.79 | 2.65 | 10.40 | 12.46 | n.m. | 100.64 |
| | | 0.33 | 0.08 | 0.35 | 0.00 | 0.01 | 0.05 | 0.02 | 0.69 | 0.03 | 0.11 | 0.24 | | |
| #878 | 5 | 20.11 | 0.08 | 0.55 | 0.01 | 0.01 | 2.50 | 0.90 | 45.20 | 2.66 | 15.34 | 10.80 | 0.87 | 99.32 |
| | | 0.46 | 0.02 | 0.41 | 0.01 | 0.01 | 0.25 | 0.05 | 1.26 | 0.20 | 0.96 | 0.63 | 1.88 | |
| #880 | 6 | 19.66 | 0.45 | 0.44 | 0.09 | 0.01 | 2.59 | 0.90 | 44.87 | 2.60 | 16.09 | 10.66 | 0.79 | 99.16 |
| | | 1.30 | 0.57 | 0.08 | 0.15 | 0.01 | 0.25 | 0.05 | 0.87 | 0.67 | 0.82 | 1.68 | 1.43 | |
| #872 | 6 | 21.98 | 0.09 | 0.49 | 0.00 | b.d.l. | 4.43 | 1.09 | 47.73 | 2.88 | 10.40 | 10.81 | n.m. | 99.91 |
| | | 0.87 | 0.03 | 0.02 | 0.01 | | 0.14 | 0.01 | 0.63 | 0.21 | 0.43 | 0.52 | | |
| #871 | 5 | 21.15 | 0.22 | 0.57 | 0.03 | 0.02 | 8.07 | 1.05 | 46.76 | 2.85 | 6.46 | 10.41 | n.m. | 97.60 |
| | | 0.93 | 0.25 | 0.10 | 0.06 | 0.03 | 0.59 | 0.06 | 1.58 | 0.35 | 0.34 | 1.68 | | |
| #874 | 6 | 20.30 | 0.17 | 0.66 | 0.04 | 0.02 | 7.70 | n.m. | 46.17 | 2.95 | 6.78 | 12.02 | 1.94 | 98.74 |
| | | 1.89 | 0.17 | 0.13 | 0.05 | 0.01 | 0.65 | | 2.23 | 0.32 | 0.33 | 3.63 | 2.02 | |
| #955 | 7 | 18.70 | 0.73 | 0.64 | 0.05 | 0.00 | 13.15 | 1.11 | 45.56 | 3.01 | 6.06 | 9.42 | n.m. | 98.41 |
| | | 0.95 | 0.75 | 0.06 | 0.09 | 0.00 | 1.19 | 0.05 | 1.74 | 0.28 | 0.43 | 1.04 | | |
| #952 | 7 | 22.06 | 0.11 | 0.52 | 0.01 | 0.01 | 8.49 | 1.09 | 48.43 | 2.92 | 7.11 | 10.14 | n.m. | 100.87 |
| | | 0.29 | 0.03 | 0.02 | 0.00 | 0.01 | 0.21 | 0.01 | 0.20 | 0.09 | 0.29 | 0.45 | | |
| #953 | 6 | 18.85 | 0.09 | 0.57 | 0.01 | 0.02 | 8.23 | 1.18 | 47.05 | 4.04 | 7.17 | 13.44 | n.m. | 100.64 |
| | | 0.62 | 0.04 | 0.05 | 0.00 | 0.01 | 0.19 | 0.01 | 0.27 | 0.40 | 0.26 | 1.02 | | |
| #949 | 6 | 20.77 | 0.18 | 0.44 | 0.00 | 0.00 | 3.12 | 1.00 | 47.20 | 3.13 | 13.43 | 10.92 | n.m. | 100.20 |
| | | 0.39 | 0.12 | 0.03 | 0.00 | 0.01 | 0.15 | 0.03 | 0.37 | 0.20 | 0.42 | 0.77 | | |
| #948 | 5 | 19.65 | 0.27 | 0.42 | 0.01 | 0.01 | 2.38 | 0.92 | 45.63 | 2.46 | 19.66 | 8.01 | n.m. | 99.43 |
| | | 1.15 | 0.33 | 0.04 | 0.01 | 0.01 | 0.14 | 0.02 | 0.87 | 0.10 | 0.63 | 0.33 | | |
| #951 | 7 | 14.10 | 0.04 | 0.29 | 0.00 | 0.01 | 1.53 | 0.87 | 44.08 | 4.63 | 20.19 | 14.78 | n.m. | 100.52 |
| | | 0.58 | 0.01 | 0.04 | 0.00 | 0.01 | 0.07 | 0.01 | 0.13 | 0.20 | 0.79 | 0.88 | | |
| #966 | 3 | 8.58 | 2.72 | 1.33 | 0.42 | 0.04 | 2.57 | 0.89 | 38.00 | 21.18 | 8.85 | 19.95 | n.m. | 104.52 |
| | | 1.90 | 0.55 | 0.17 | 0.14 | 0.02 | 0.16 | 0.06 | 0.50 | 1.82 | 0.56 | 1.67 | | |

| | | Si | S | Cr | Mn | Fe | Co | Ni | C[b] | | |
|---|---|---|---|---|---|---|---|---|---|---|---|
| Average all metals | | 11.60 | 0.47 | 0.63 | 0.04 | 80.28 | 0.38 | 4.69 | 1.93 | | 98.07 |
| | | 0.72 | 0.42 | 0.39 | 0.06 | 1.13 | 0.07 | 0.74 | 0.99 | | 0.99 |

[a]N = Number of analyses. [b]Carbon concentration (due to contamination from graphite capsule) is calculated as the subtraction of totals from 100 wt%. n.m. = not measured. b.d.l. = below detection limit.

**Table 5** Summary of results for pressure and degree of melting (inferred depth and temperature) for matching experimental melts with geochemical terranes, by comparing directly with experimental results and considering mixtures of two compositions. * Melting conditions for the second melt in the mixture model. NVP: Northern Volcanic Plains. HMR: High-Mg region. HAL: High Al terrane.

|  |  | Pressure (GPa) | Depth (km) | F (wt%) | Temperature (°C) | Pressure* (GPa) | Depth* (km) | F* (wt%) | Temperature* (°C) |
|---|---|---|---|---|---|---|---|---|---|
| Direct comparison | NVP | 0.5-1 | 40-80 | 15 | ~1300-1450 |  |  |  |  |
|  | HMR | 2-5 | 160-400 | 5-70 | ~1500-1750 |  |  |  |  |
|  | Caloris | 0.5-1 | 40-80 | 15 | ~1300-1450 |  |  |  |  |
|  | HAl | 1-3 | 80-240 | 10-25 | ~1550-1650 |  |  |  |  |
| Mixing model | NVP | 1-3 | 80-240 | 20-50 | ~1500-1700 | 0.5 | 40 | 10 | ~1250 |
|  | HMR | 3-5 | 240-400 | 10 | ~1700-1750 | 1-3 | 80-240 | 10-90 | ~1500-1800 |
|  | Caloris | 2-3 | 160-240 | 20 | ~1600-1650 | 0.5-1 | 40-80 | 20-40 | ~1400-1550 |
|  | HAl | 1-5 | 40-400 | 10-20 | ~1500-1750 | 0.5-3 | 40-240 | 20-60 | ~1350-1700 |

# Supplementary material

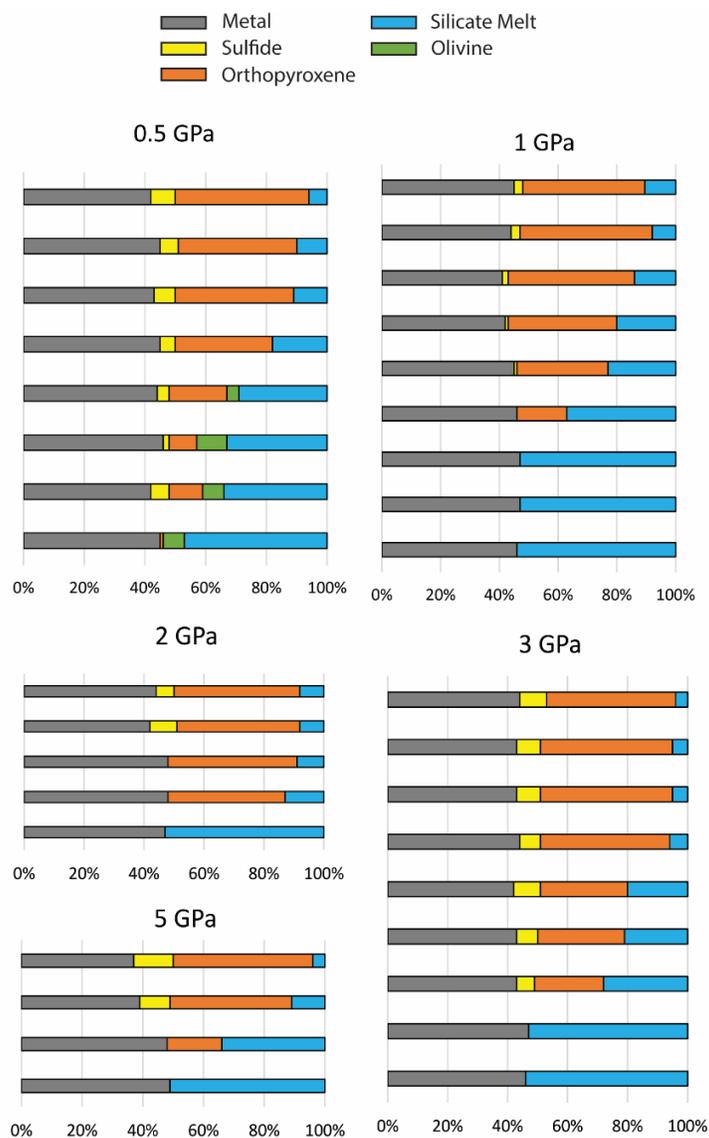

**Fig. S1** Phase proportions in our experimental charges (wt%).

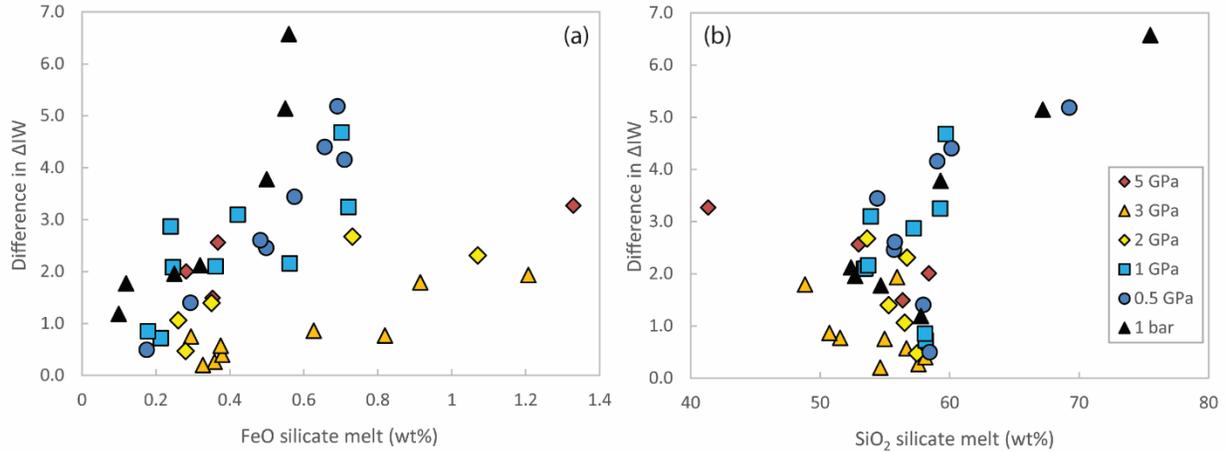

**Fig. S2** Difference between ΔIW (log oxygen fugacity relative to IW buffer, see text and Table 1) between the two calculations based on FeO-Fe and SiO₂-Si equilibria. This difference is plotted as a function of FeO (a) and SiO₂ (b) concentrations in the silicate melt.

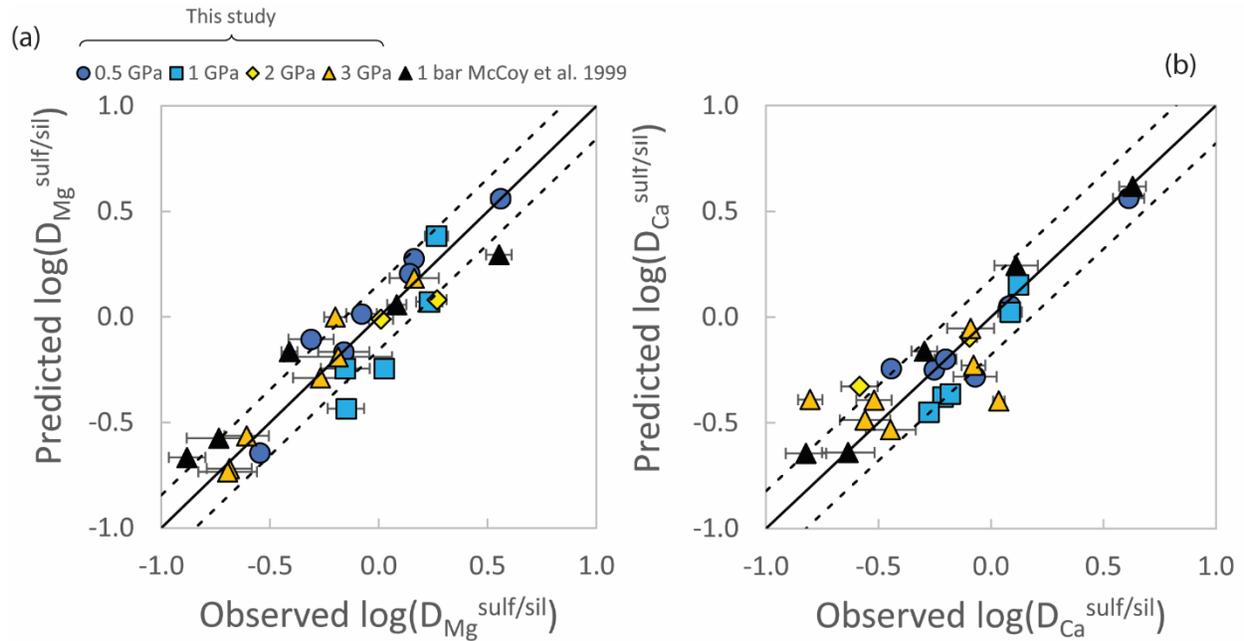

**Fig. S3** Comparison between predicted and observed partition coefficients of Ca (a) and Mg (b) between sulfide and silicate. Thick lines are 1-to-1 correspondence, and the dotted lines represent the 1 sigma errors on the regressions.

**Table S2** Results for the thermodynamic models predicting Mg and Ca partition coefficients between sulfide and silicate. a to c are coefficients described in Eq. 7 and main text, N is the number of experiments used in the models. σ is the 1 sigma error on the regression or coefficient estimate. F is the significance of the regressions and $R^2$ the proportion of the variance predictable by the models.

**Mg (sulfide/silicate)**

|  | Coefficient | σ | p-value |
|---|---|---|---|
| a | -0.475 | 0.345 | 0.18 |
| b (ΔIW) | 0.53 | 0.056 | 1.00E-07 |
| c (1/T) | 3964 | 526 | 2.00E-09 |
| N | 26 |  |  |
| σ | 0.1557 |  |  |
| $R^2$ | 0.8413 |  |  |
| F | 67 |  |  |
| p-value | 2E-10 |  |  |

**Ca (sulfide/silicate)**

|  | Coefficient | σ | p-value |
|---|---|---|---|
| a | -11.890 | 1.511 | 5.68E-08 |
| b (ΔIW) | 0.203 | 0.063 | 4.02E-03 |
| c (log(SiO$_2^{wt\%}$)) | 7.091 | 0.846 | 1.89E-08 |
| N | 26 |  |  |
| σ | 0.18 |  |  |
| $R^2$ | 0.78 |  |  |
| F | 42 |  |  |
| p-value | 2E-08 |  |  |

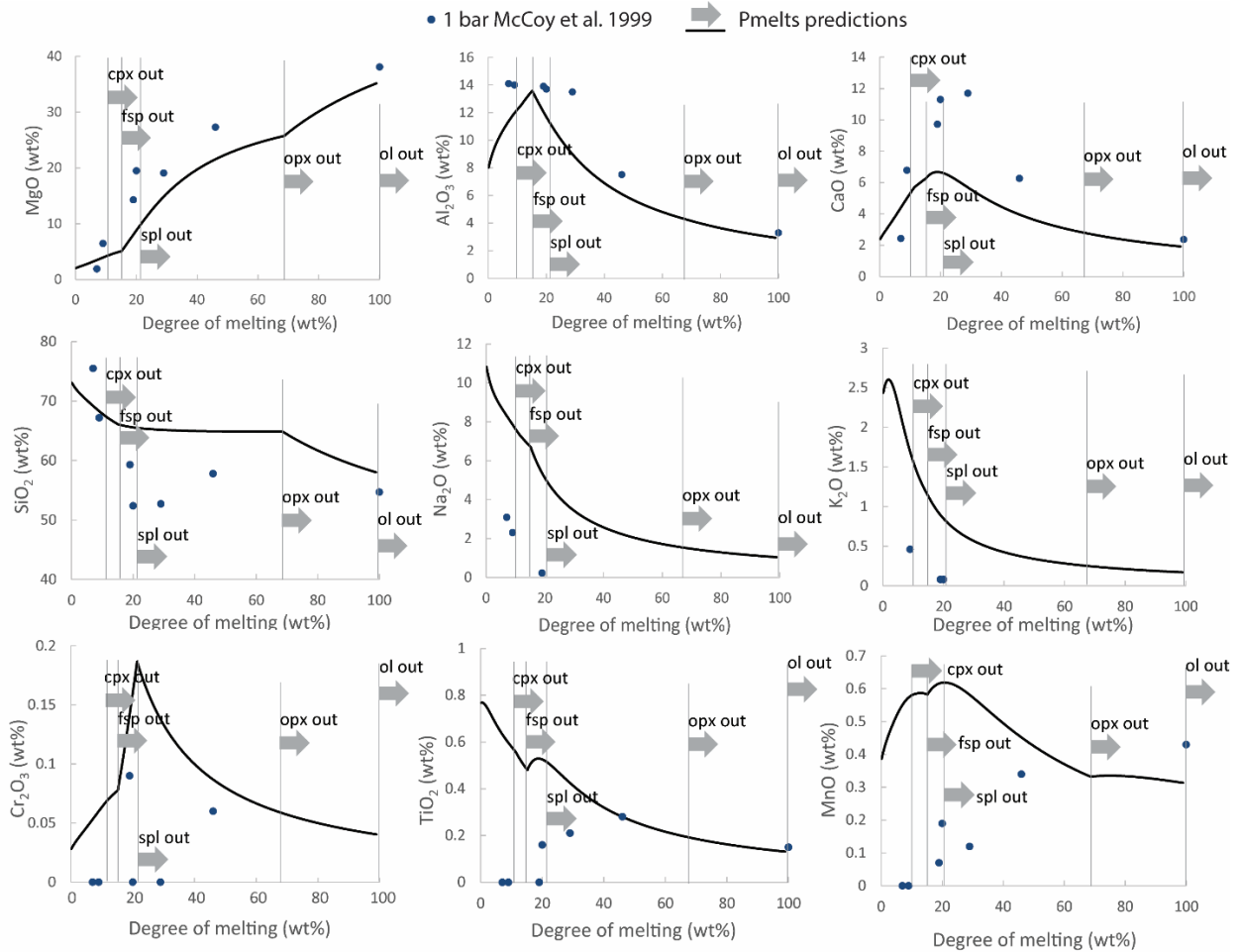

**Fig. S4** Change of the chemical compositions of the silicate melts as a function of the degree of melting (F). Circles are experimental data (from McCoy et al. 1999), and thick lines are *melts* predictions. Vertical lines and arrows show when a phase is predicted by *melts* to disappear at the F where the vertical lines are located. Spl = spinel. Fsp = feldspar. Opx = orthopyroxene. Cpx = Clinopyroxene. Ol = Olivine.

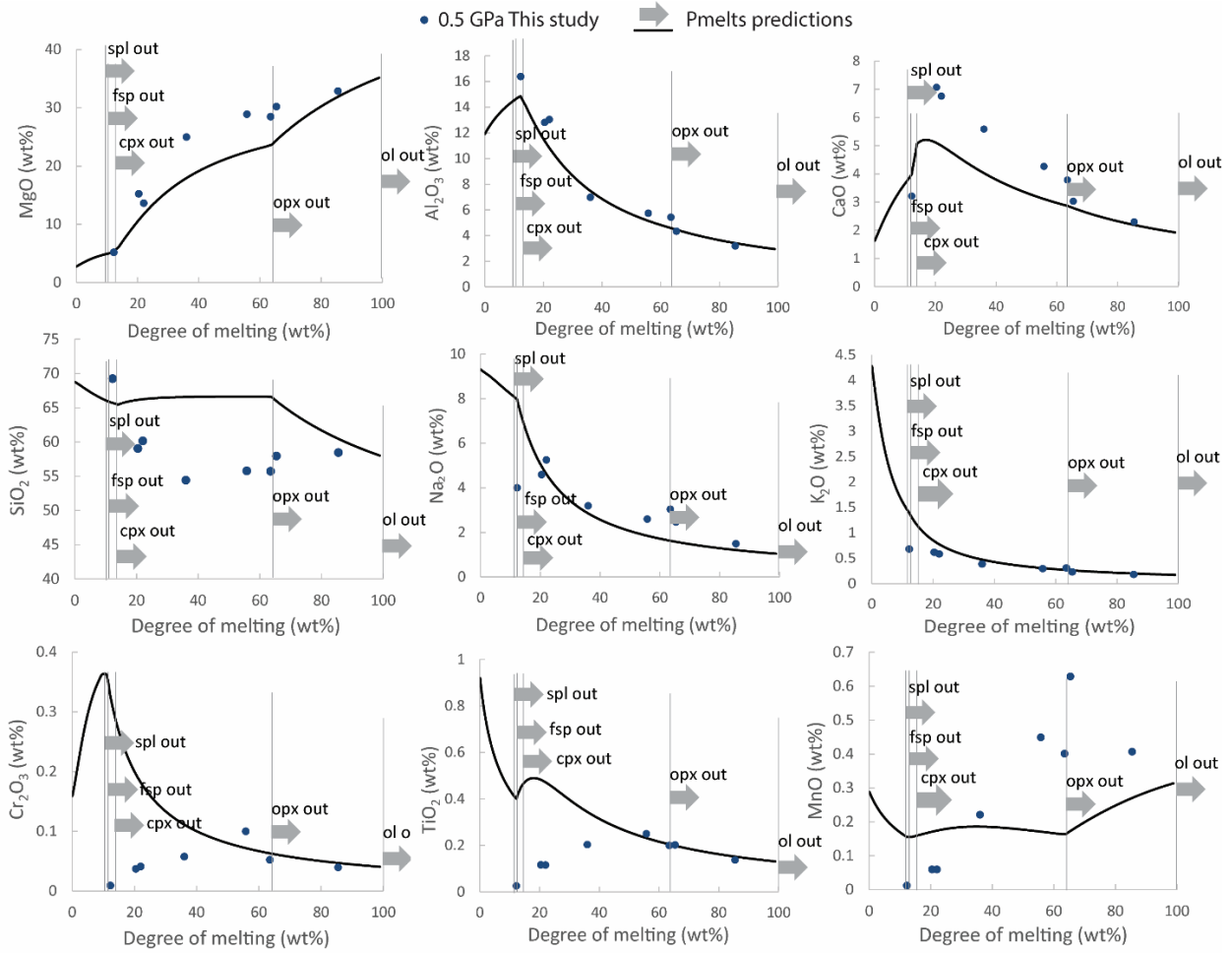

**Fig. S5** Same as Fig. S4 for 0.5 GPa. Circular symbols are our experimental data and thick lines are predictions from *pmelts*. Spl = spinel. Fsp = feldspar. Opx = orthopyroxene. Cpx = Clinopyroxene. Ol = Olivine.

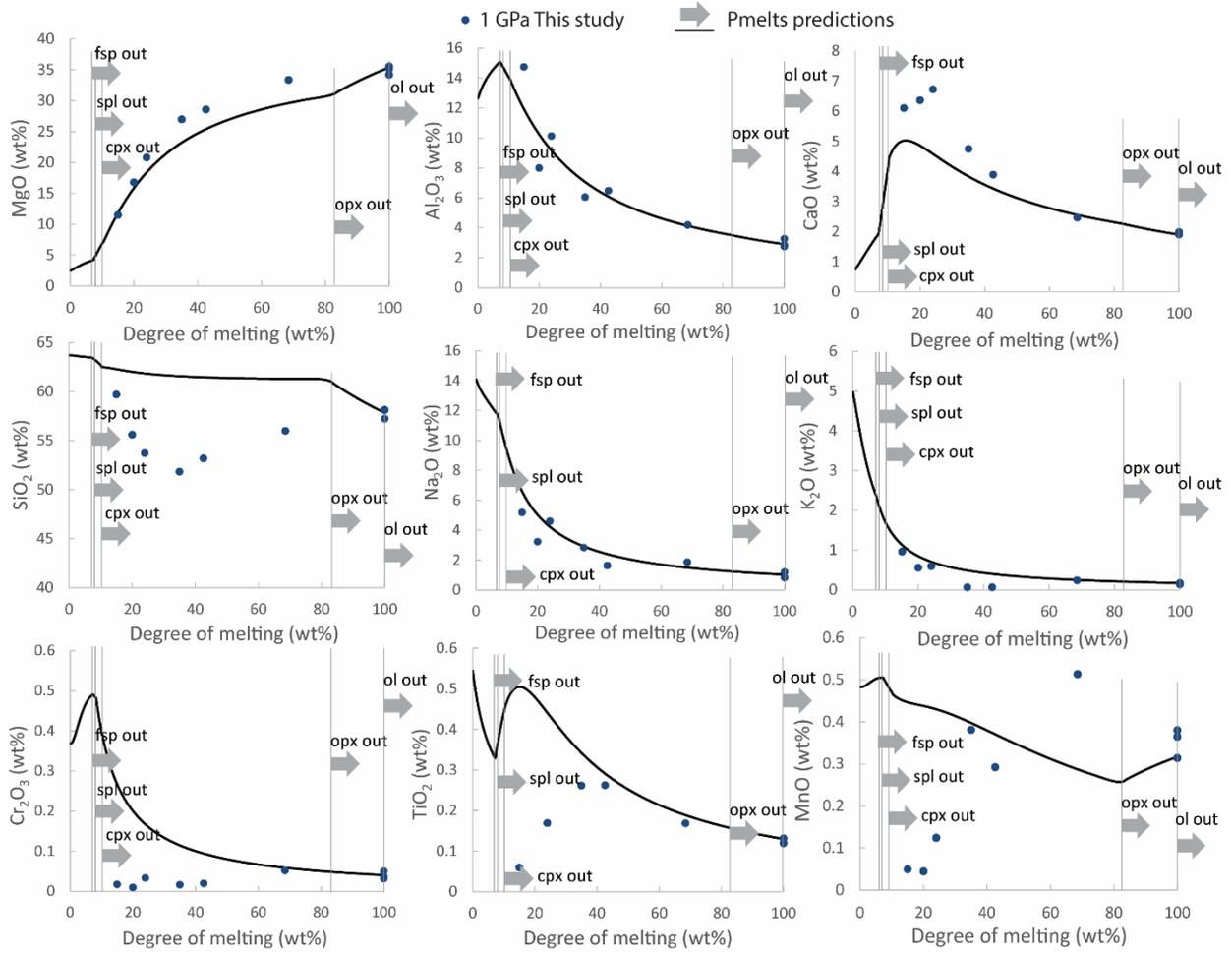

**Fig. S6** Same as Fig. S5 at 1 GPa.

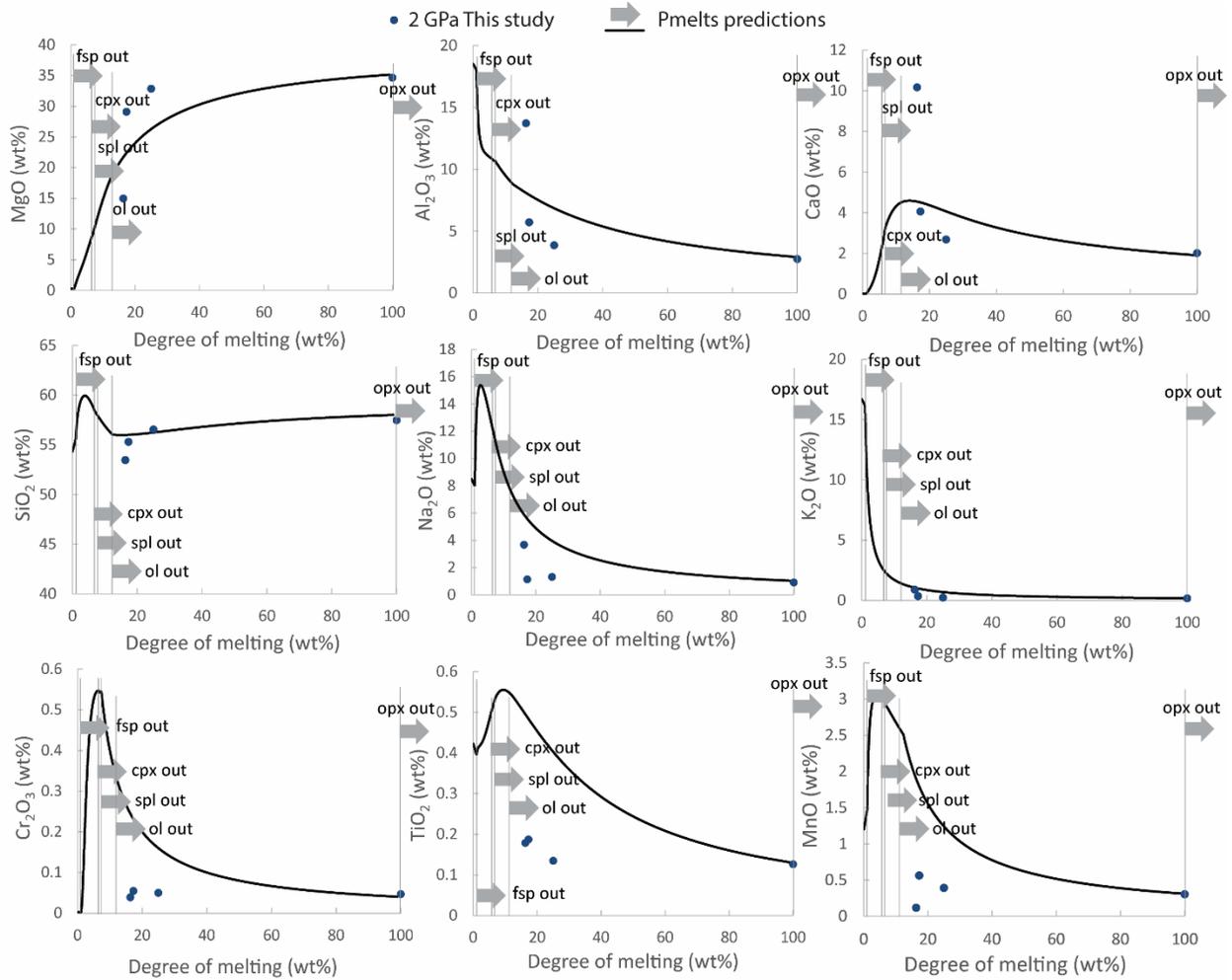

**Fig S7** Same as Fig. S5 at 2 GPa.

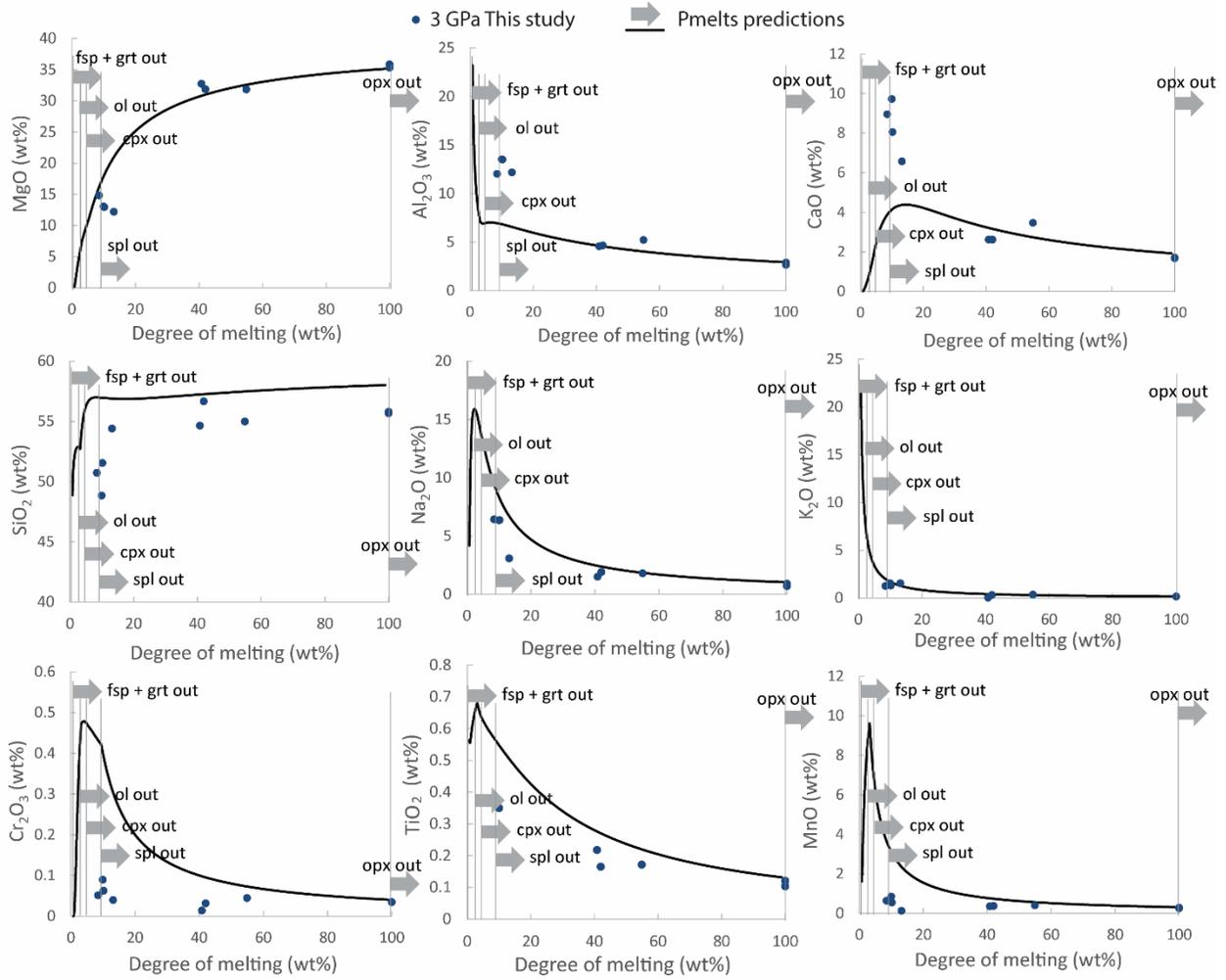

**Fig. S8** Same as Fig. S5 at 3 GPa.

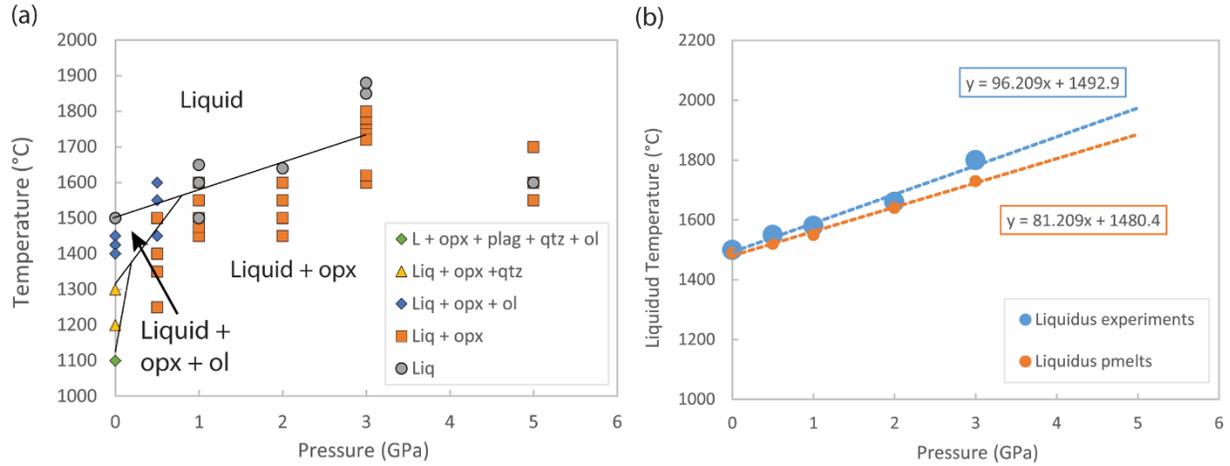

**Fig. S9** (a) Phase diagram for a reduced Indarch EH4 enstatite chondrite using experiments form this study from 0.5 to 5 GPa and those of McCoy et al. (1999). (b) Estimation of the liquidus from results shown in (a) (blue symbols and line) and from pmelts predictions (orange symbols and line). The equations are linear regression fits where x represents the pressure in GPa.

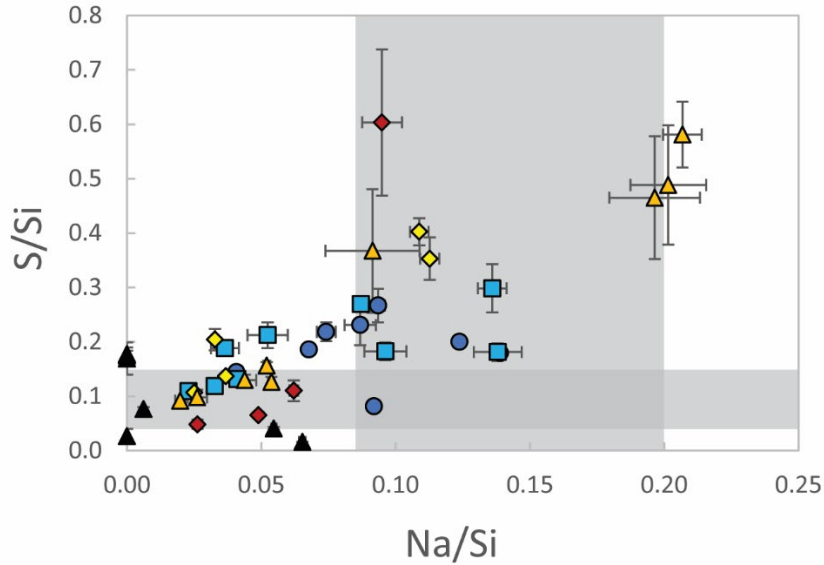

**Fig. S10** S/Si and Na/Si for our silicate melts and those of McCoy et al. 1999. Symbols are for the same pressures as those shown in Fig. 2, 4 and 7 in the main text. The grey areas show the ranges for both ratios measured on Mercury's surface by MESSENGER (Nittler et al 2020, Peplowski et al. 2014).

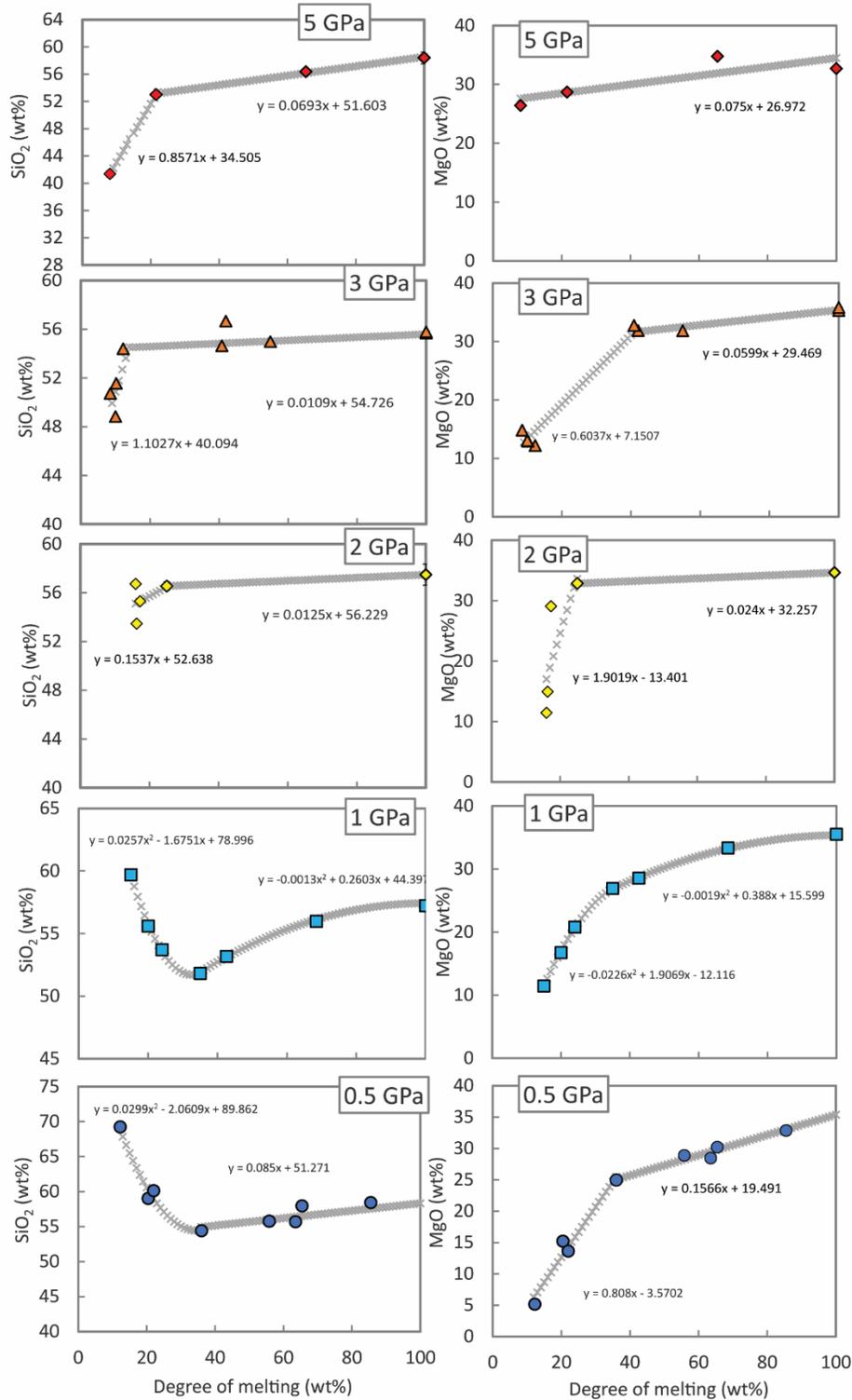

**Fig. S11** SiO$_2$ and MgO concentrations of silicate melts as a function of the silicate degree of melting, for our experimental data (filled symbols) and modeled concentrations interpolated from our experimental results (grey crosses). Equations are fits to the experimental data.

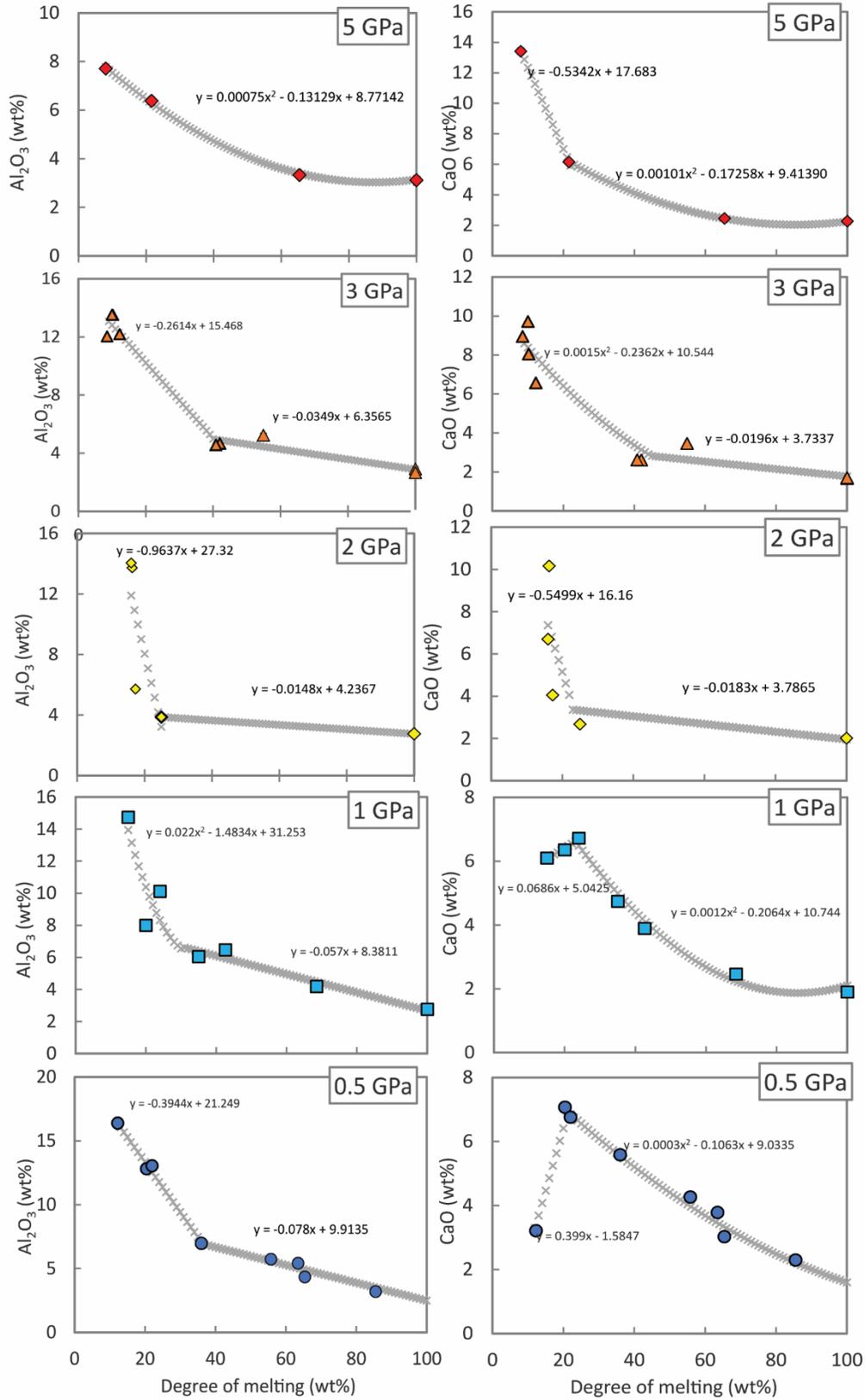

**Fig. S12** Same as Fig. S11 for $Al_2O_3$ and CaO.

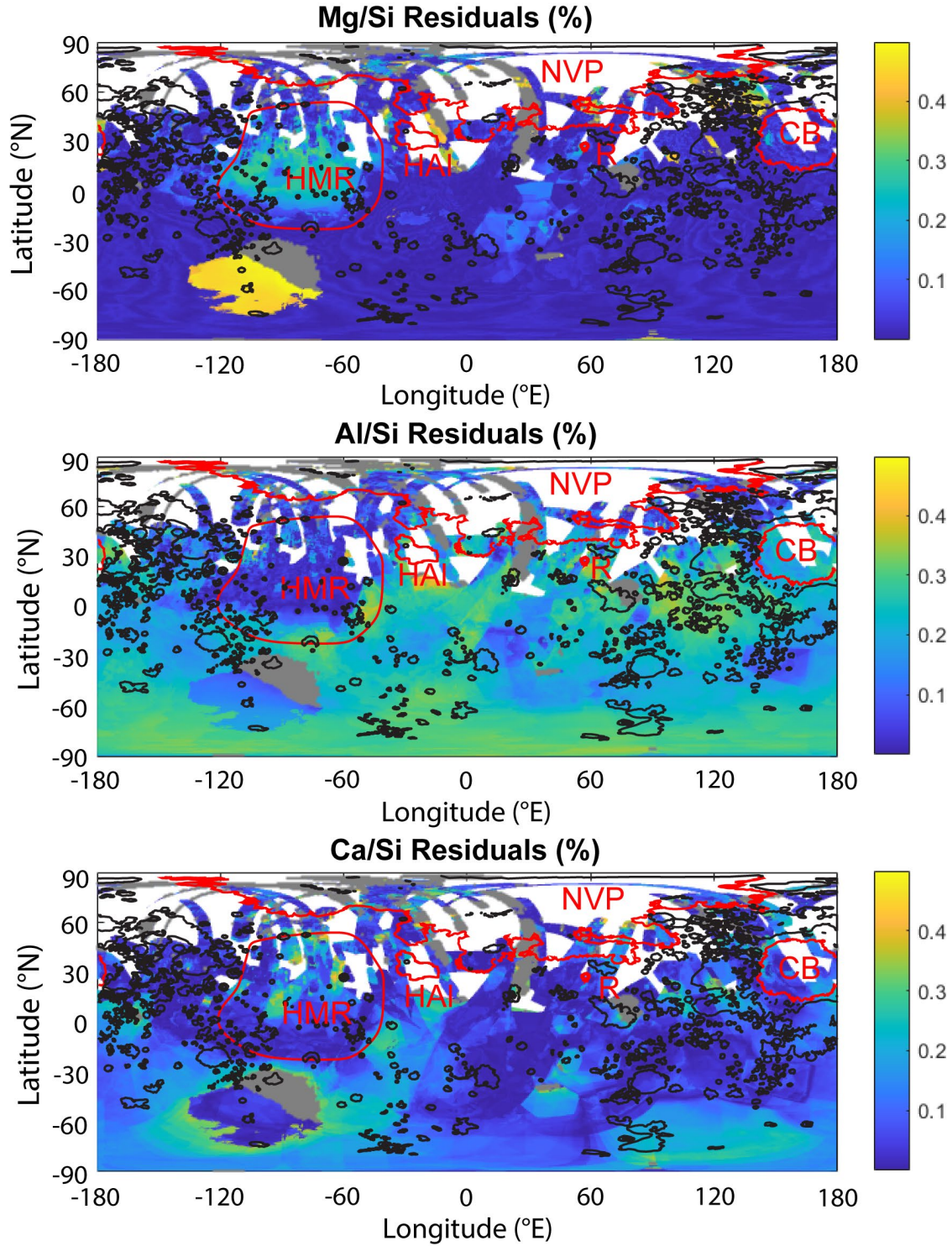

**Fig. S13** Maps of Mercury, showing residuals for fits between Mercury's surface and partial melts of EH enstatite chondrites, for Mg/Si (A), Al/Si (B), and Ca/Si (C). White areas represent surface areas where Ca/Si was not measured by MESSENGER XRS spectrometer, and grey areas are those poorly matching partial melts of EH enstatite chondrites. See Fig.7 for outline and abbreviation descriptions.

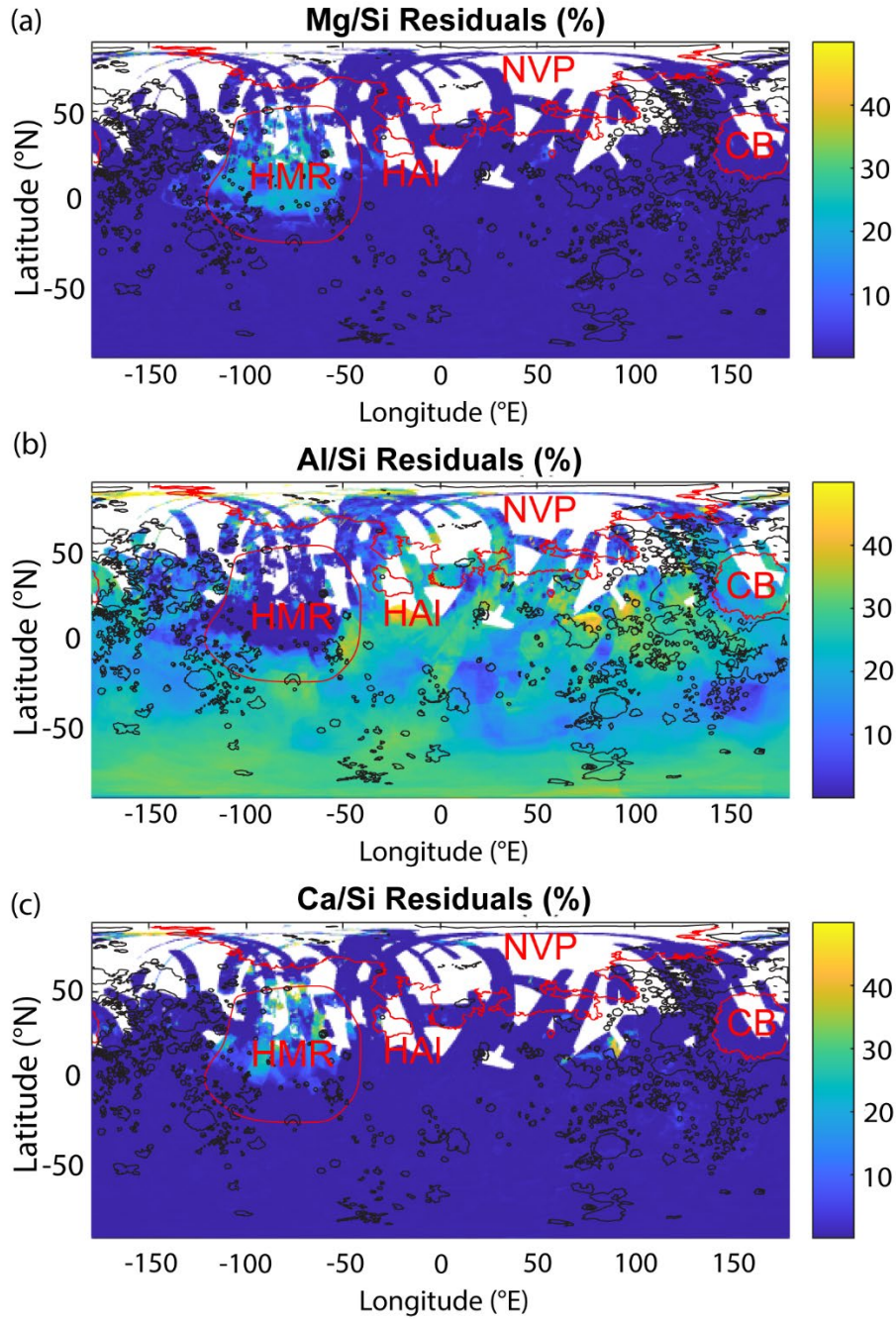

**Fig. S14** Maps of Mercury, showing residuals for fits between Mercury's surface and mixtures of partial melts of EH enstatite chondrites, for Mg/Si (a), Al/Si (b), and Ca/Si (c). See Fig.9 for outline and abbreviation descriptions.

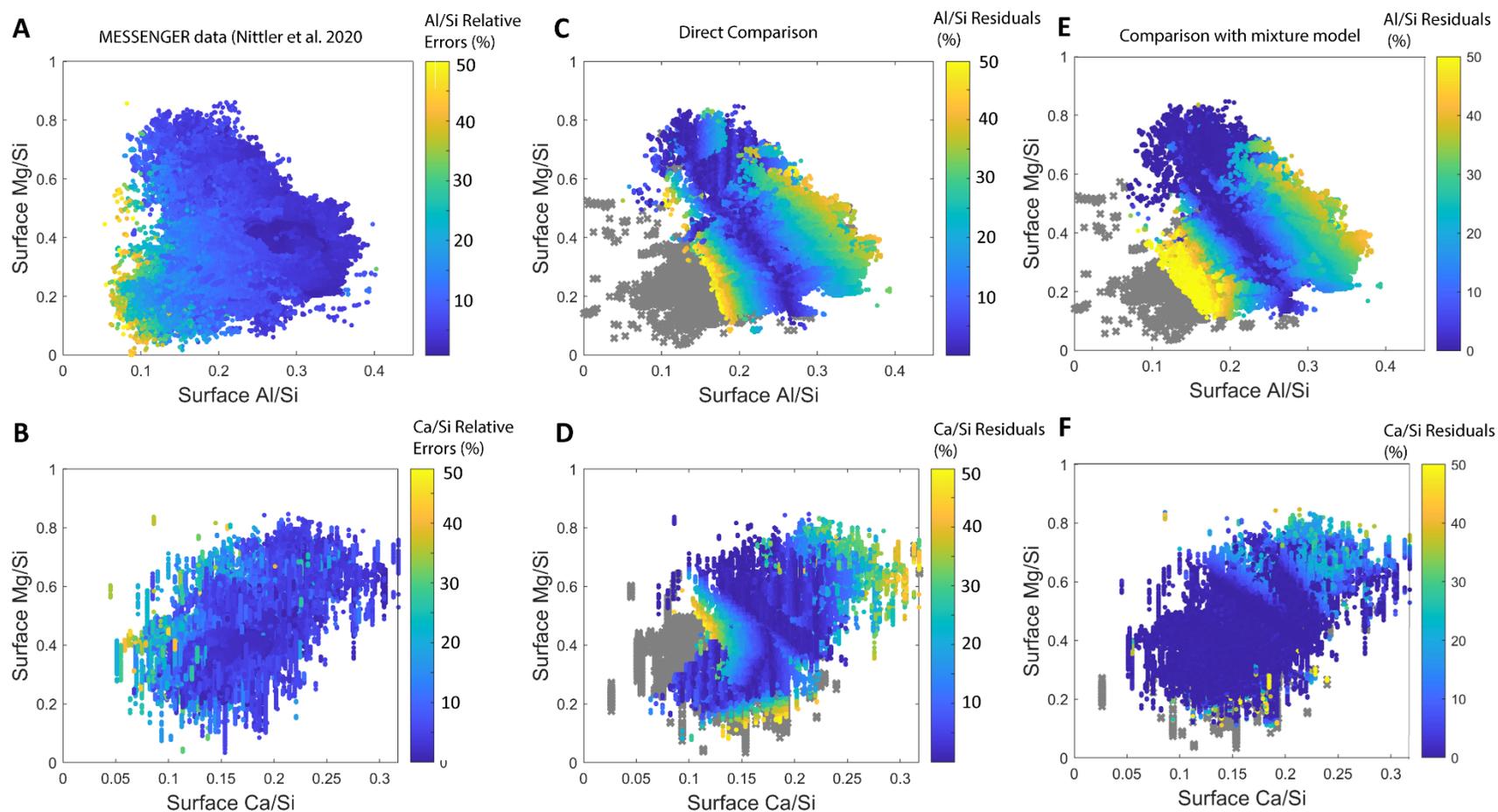

**Fig. S15** Mg/Si as a function of Al/Si (a, c, e) and Ca/Si (b, d, f). Symbol colors show the relative errors on MESSENGER X-ray spectrometer measurements (Nittler et al. 2020) (a, b), relative residuals for the direct comparison model between surface composition and surface composition (c, d) and relative residuals for the comparison with the mixture model (e, f). In (a-b) data at extremely low Al/Si are not shown because of their relative errors being higher than 50%. In (c-f), grey crosses are data where relative residuals are higher than 50 %.

Table S1 Chemical compositions of metals, orthopyroxenes and olivines.

Average chemical compositions of metals (wt%).

| | N# | Si | S | Cr | Mn | Fe | Co | Ni | Total | C* |
|---|---|---|---|---|---|---|---|---|---|---|
| **#309** | 7 | 12.45 | 1.90 | 0.66 | 0.17 | 79.17 | 0.46 | 3.98 | 98.78 | 1.22 |
| **#311** | 7 | 12.09 | 1.94 | 0.70 | 0.19 | 79.65 | 0.38 | 4.10 | 99.04 | 0.96 |
| **#315** | 7 | 11.96 | 1.15 | 0.48 | 0.05 | 78.78 | 0.43 | 5.42 | 98.27 | 1.73 |
| **#314** | 7 | 12.41 | 0.82 | 0.27 | 0.01 | 79.12 | 0.43 | 5.42 | 98.49 | 1.51 |
| **#384** | 6 | 11.61 | 0.32 | 0.58 | 0.02 | 81.20 | 0.37 | 3.43 | 97.52 | 2.48 |
| **#385** | 6 | 11.99 | 0.38 | 0.54 | 0.03 | 79.92 | 0.41 | 4.99 | 98.26 | 1.74 |
| **#382** | 7 | 12.05 | 0.57 | 0.48 | 0.01 | 80.14 | 0.43 | 4.60 | 98.27 | 1.73 |
| **#386** | 6 | 11.47 | 0.31 | 0.37 | b.d.l. | 79.49 | 0.46 | 4.88 | 96.99 | 3.01 |
| **#383** | 7 | 11.35 | 0.28 | 0.42 | b.d.l. | 80.90 | 0.35 | 4.86 | 98.16 | 1.84 |
| **#374** | 6 | 11.76 | 0.53 | 0.37 | 0.00 | 79.64 | 0.34 | 4.79 | 97.43 | 2.57 |
| **#388** | 6 | 11.84 | 0.38 | 0.41 | 0.02 | 79.54 | 0.48 | 5.08 | 97.75 | 2.25 |
| **#387** | 6 | 11.58 | 0.53 | 0.37 | 0.00 | 78.89 | 0.38 | 5.83 | 97.57 | 2.43 |
| **#389** | 6 | 11.11 | 0.40 | 0.42 | 0.00 | 80.01 | 0.36 | 4.82 | 97.12 | 2.88 |
| **#431** | 7 | 11.58 | 0.75 | 0.77 | 0.19 | 80.26 | 0.38 | 4.18 | 98.12 | 1.88 |
| **#432** | 7 | 11.60 | 0.28 | 0.56 | 0.01 | 80.80 | 0.40 | 5.07 | 98.74 | 1.26 |
| **#433** | 7 | 12.01 | 0.58 | 0.56 | 0.03 | 80.01 | 0.40 | 4.73 | 98.31 | 1.69 |
| **#445** | 7 | 12.65 | 0.37 | 0.28 | 0.01 | 81.82 | 0.28 | 4.36 | 99.76 | 0.24 |
| **#1073** | 7 | 11.75 | 0.45 | 0.45 | 0.01 | 80.59 | 0.28 | 4.36 | 97.88 | 2.12 |
| **#869** | 10 | 12.29 | 0.53 | 0.76 | 0.17 | 79.25 | 0.38 | 5.84 | 99.22 | 0.78 |
| **#870** | 17 | 12.91 | 0.48 | 0.73 | 0.09 | 78.48 | 0.43 | 5.20 | 98.31 | 1.69 |
| **#878** | 7 | 11.04 | 0.34 | 0.68 | b.d.l. | 81.24 | b.d.l. | 4.61 | 97.92 | 2.08 |
| **#880** | 8 | 11.19 | 0.25 | 0.66 | 0.01 | 80.42 | 0.34 | 5.75 | 98.61 | 1.39 |
| **#872** | 10 | 12.24 | 0.48 | 0.70 | 0.04 | 79.99 | 0.36 | 5.73 | 99.53 | 0.47 |
| **#871** | 9 | 12.67 | 0.21 | 0.58 | 0.01 | 80.46 | 0.38 | 5.46 | 99.76 | 0.24 |
| **#874** | 7 | 11.56 | 0.20 | 0.56 | 0.01 | 80.68 | 0.44 | 4.72 | 98.17 | 1.83 |
| **#963** | 7 | 11.85 | 0.11 | 0.75 | 0.06 | 80.63 | 0.33 | 5.32 | 99.06 | 0.94 |
| **#964** | 7 | 11.81 | 0.16 | 0.78 | 0.06 | 80.30 | 0.41 | 4.96 | 98.47 | 1.53 |
| **#955** | 8 | 10.25 | 0.11 | 0.58 | 0.01 | 81.50 | 0.40 | 3.34 | 96.19 | 3.81 |
| **#952** | 7 | 10.43 | 0.19 | 0.68 | b.d.l. | 79.75 | 0.42 | 4.74 | 96.22 | 3.78 |
| **#953** | 7 | 10.21 | 0.22 | 0.59 | b.d.l. | 82.52 | 0.45 | 3.32 | 97.31 | 2.69 |
| **#949** | 7 | 11.53 | 0.25 | 0.55 | 0.00 | 80.17 | 0.34 | 3.56 | 96.39 | 3.61 |
| **#948** | 7 | 11.32 | 0.31 | 0.75 | 0.04 | 79.09 | 0.38 | 4.16 | 96.05 | 3.95 |
| **#951** | 7 | 10.37 | 0.30 | 0.53 | b.d.l. | 82.10 | 0.39 | 3.39 | 97.08 | 2.92 |
| **#966** | 7 | 10.29 | 0.26 | 0.96 | 0.01 | 83.76 | 0.31 | 3.80 | 99.40 | 0.60 |
| **#988** | 10 | 10.62 | 0.09 | 2.66 | n.m. | 79.40 | 0.10 | 5.25 | 98.11 | 1.89 |

Standard deviations (1 sigma) for metal chemical compositions

|        | Si    | S     | Cr    | Mn   | Fe    | Co    | Ni    |
|--------|-------|-------|-------|------|-------|-------|-------|
| #309   | 0.51  | 1.40  | 0.27  | 0.17 | 1.57  | 0.03  | 0.19  |
| #311   | 0.61  | 1.61  | 0.35  | 0.28 | 1.46  | 0.03  | 0.27  |
| #315   | 0.32  | 0.31  | 0.07  | 0.02 | 0.79  | 0.05  | 0.63  |
| #314   | 0.11  | 0.18  | 0.06  | 0.01 | 0.35  | 0.02  | 0.32  |
| #384   | 0.30  | 0.04  | 0.06  | 0.01 | 0.64  | 0.03  | 0.47  |
| #385   | 0.21  | 0.09  | 0.08  | 0.02 | 0.22  | 0.05  | 0.21  |
| #382   | 0.10  | 0.28  | 0.10  | 0.02 | 0.43  | 0.01  | 0.05  |
| #386   | 0.41  | 0.09  | 0.06  |      | 0.58  | 0.03  | 0.60  |
| #383   | 0.15  | 0.16  | 0.05  |      | 0.53  | 0.04  | 0.40  |
| #374   | 0.06  | 0.12  | 0.03  | 0.01 | 0.36  | 0.01  | 0.06  |
| #388   | 0.22  | 0.15  | 0.08  | 0.01 | 0.50  | 0.08  | 0.43  |
| #387   | 0.06  | 0.11  | 0.10  | 0.01 | 0.51  | 0.03  | 0.20  |
| #389   | 0.16  | 0.25  | 0.07  | 0.02 | 0.36  | 0.01  | 0.30  |
| #431   | 0.18  | 0.53  | 0.06  | 0.13 | 2.59  | 0.11  | 1.18  |
| #432   | 0.19  | 0.14  | 0.08  | 0.01 | 1.63  | 0.11  | 1.21  |
| #433   | 0.16  | 0.32  | 0.13  | 0.02 | 0.64  | 0.02  | 0.26  |
| #445   | 0.98  | 0.22  | 0.13  | 0.01 | 1.43  | 0.09  | 0.91  |
| #1073  | 0.13  | 0.06  | 0.04  | 0.01 | 0.43  | 0.04  | 0.15  |
| #869   | 0.18  | 0.33  | 0.09  | 0.14 | 0.93  | 0.04  | 0.47  |
| #870   | 0.79  | 0.46  | 0.15  | 0.16 | 2.23  | 0.07  | 0.50  |
| #878   | 0.41  | 0.07  | 0.04  |      | 1.37  |       | 0.83  |
| #880   | 0.40  | 0.10  | 0.09  | 0.01 | 1.06  | 0.08  | 1.11  |
| #872   | 0.62  | 0.38  | 0.18  | 0.05 | 0.56  | 0.07  | 0.21  |
| #871   | 0.10  | 0.06  | 0.04  | 0.01 | 0.22  | 0.02  | 0.17  |
| #874   | 0.14  | 0.06  | 0.03  | 0.01 | 0.25  | 0.01  | 0.03  |
| #963   | 0.10  | 0.05  | 0.02  | 0.03 | 0.91  | 0.07  | 0.81  |
| #964   | 0.18  | 0.05  | 0.02  | 0.04 | 0.86  | 0.07  | 0.73  |
| #955   | 0.04  | 0.58  | 1.19  | 1.02 | 0.10  | 0.01  | 0.10  |
| #952   | 0.04  | 0.32  | 0.84  |      | 0.05  | 0.01  | 0.04  |
| #953   | 0.16  | 1.10  | 1.44  |      | 0.23  | 0.01  | 0.23  |
| #949   | 0.09  | 0.94  | 1.72  | 1.15 | 0.18  | 0.01  | 0.10  |
| #948   | 0.14  | 0.89  | 2.20  | 2.09 | 0.13  | 0.02  | 0.08  |
| #951   | 0.10  | 0.92  | 3.30  |      | 0.29  | 0.01  | 0.16  |
| #966   | 1.80  | 0.20  | 0.44  | 0.02 | 1.88  | 0.17  | 1.85  |
| #988   | 0.380 | 0.077 | 0.144 |      | 2.317 | 0.036 | 2.124 |

Average chemical compositions of orthopyroxenes (wt%)

| | N# | MgO | SiO2 | Na2O | Al2O3 | K2O | CaO | TiO2 | S | Cr2O3 | MnO | FeO | NiO | Total |
|---|---|---|---|---|---|---|---|---|---|---|---|---|---|---|
| **#311** | 7 | 38.42 | 59.57 | 0.21 | 0.77 | 0.00 | 0.37 | 0.02 | 0.01 | 0.01 | 0.08 | 0.26 | | 99.72 |
| **#315** | 25 | 39.08 | 58.04 | 0.51 | 1.60 | 0.10 | 0.79 | 0.03 | 0.15 | n.m. | 0.08 | 0.30 | | 100.75 |
| **#314** | 7 | 36.60 | 59.11 | 0.85 | 2.44 | 0.01 | 1.40 | 0.03 | 0.03 | n.m. | 0.06 | 0.56 | | 101.11 |
| **#382** | 5 | 40.18 | 58.91 | 0.11 | 0.78 | 0.00 | 0.38 | b.d.l. | 0.01 | 0.01 | 0.07 | 0.11 | | 100.56 |
| **#386** | 6 | 39.60 | 60.40 | 0.09 | 0.61 | 0.00 | 0.36 | 0.03 | b.d.l. | 0.01 | 0.08 | 0.33 | | 101.51 |
| **#383** | 8 | 39.82 | 59.31 | 0.12 | 0.58 | 0.00 | 0.34 | 0.03 | 0.00 | 0.01 | 0.08 | 0.21 | | 100.49 |
| **#374** | 6 | 38.23 | 59.74 | 0.35 | 1.67 | 0.00 | 1.02 | 0.04 | 0.01 | 0.02 | 0.04 | 0.93 | | 102.03 |
| **#388** | 7 | 39.38 | 60.45 | 0.16 | 1.05 | b.d.l. | 0.53 | 0.03 | 0.01 | 0.01 | 0.08 | 0.36 | | 102.05 |
| **#387** | 6 | 39.19 | 60.34 | 0.17 | 1.31 | 0.00 | 0.63 | 0.03 | b.d.l. | 0.00 | 0.07 | 0.30 | | 102.06 |
| **#389** | 6 | 39.32 | 60.37 | 0.12 | 0.85 | 0.01 | 0.48 | 0.02 | 0.01 | 0.01 | 0.07 | 0.12 | | 101.38 |
| **#432** | 6 | 38.80 | 59.96 | 0.05 | 0.32 | 0.00 | 0.24 | 0.02 | 0.00 | 0.01 | 0.08 | 0.10 | 0.00 | 99.59 |
| **#433** | 7 | 38.74 | 59.95 | 0.03 | 0.46 | b.d.l. | 0.30 | 0.03 | 0.01 | 0.01 | 0.08 | 0.04 | 0.01 | 99.66 |
| **#445** | 7 | 39.12 | 60.08 | 0.12 | 1.08 | 0.00 | 0.62 | 0.03 | 0.01 | 0.02 | 0.05 | 0.47 | | 101.61 |
| **#1073** | 7 | 38.80 | 57.93 | 0.18 | 1.08 | 0.01 | 0.70 | 0.04 | 0.09 | n.m. | 0.05 | 0.78 | | 99.69 |
| **#870** | 10 | 40.02 | 60.30 | 0.04 | 0.29 | 0.01 | 0.22 | 0.03 | 0.03 | 0.01 | 0.07 | 0.25 | | 101.26 |
| **#878** | 8 | 40.42 | 58.14 | 0.04 | 0.36 | 0.00 | 0.36 | 0.05 | 0.01 | 0.00 | 0.04 | 0.42 | | 99.86 |
| **#880** | 8 | 40.60 | 59.20 | 0.05 | 0.27 | 0.00 | 0.27 | 0.04 | 0.01 | 0.01 | 0.06 | 0.27 | | 100.78 |
| **#872** | 10 | 39.75 | 60.13 | 0.11 | 0.42 | 0.02 | 0.41 | 0.04 | 0.11 | 0.03 | 0.06 | 0.45 | | 101.53 |
| **#871** | 5 | 38.96 | 60.34 | 0.27 | 0.87 | 0.05 | 0.56 | 0.03 | 0.18 | 0.01 | 0.05 | 0.71 | | 102.01 |
| **#874** | 7 | 39.73 | 60.09 | 0.07 | 0.58 | 0.00 | 0.52 | 0.03 | n.m. | 0.02 | 0.04 | 0.66 | | 101.74 |
| **#955** | 5 | 38.80 | 59.94 | 0.10 | 0.42 | 0.01 | 0.29 | 0.03 | 0.06 | 0.02 | 0.03 | 0.85 | | 100.51 |
| **#952** | 6 | 40.81 | 59.67 | 0.02 | 0.20 | 0.00 | 0.19 | 0.04 | 0.02 | 0.01 | 0.02 | 0.66 | | 101.63 |
| **#953** | 6 | 40.81 | 59.61 | 0.02 | 0.19 | 0.00 | 0.19 | 0.04 | 0.01 | 0.01 | 0.02 | 0.65 | | 101.55 |
| **#949** | 6 | 40.54 | 59.81 | 0.02 | 0.17 | 0.01 | 0.16 | 0.02 | 0.01 | 0.02 | 0.04 | 0.51 | | 101.30 |
| **#948** | 10 | 40.15 | 58.14 | 0.03 | 0.11 | 0.00 | 0.20 | 0.02 | 0.02 | n.m. | 0.07 | 0.35 | | 99.11 |
| **#951** | 7 | 40.59 | 59.59 | 0.01 | 0.11 | 0.00 | 0.11 | 0.03 | 0.01 | 0.02 | 0.08 | 0.43 | | 100.95 |
| **#966** | 12 | 39.97 | 59.23 | 0.03 | 0.10 | 0.01 | 0.09 | 0.02 | 0.02 | n.m. | 0.10 | 0.33 | | 99.90 |
| **#988** | 7 | 40.15 | 60.99 | 0.01 | 0.11 | 0.00 | 0.07 | 0.01 | 0.00 | 0.01 | 0.07 | 0.12 | | 101.55 |

Standard deviations (1 sigma) for orthopyroxene compositions.

| | MgO | SiO2 | Na2O | Al2O3 | K2O | CaO | TiO2 | S | Cr2O3 | MnO | FeO |
|---|---|---|---|---|---|---|---|---|---|---|---|
| **#311** | 0.182 | 0.287 | 0.026 | 0.076 | 0.007 | 0.019 | 0.010 | 0.006 | 0.009 | 0.009 | 0.084 |
| **#315** | 0.902 | 0.905 | 0.105 | 0.296 | 0.146 | 0.208 | 0.017 | 0.177 | | 0.028 | 0.106 |
| **#314** | 0.314 | 0.647 | 0.153 | 0.215 | 0.006 | 0.162 | 0.028 | 0.020 | | 0.031 | 0.146 |
| **#382** | 0.190 | 0.477 | 0.037 | 0.097 | 0.007 | 0.020 | | 0.010 | 0.010 | 0.032 | 0.076 |
| **#386** | 0.111 | 0.187 | 0.016 | 0.096 | 0.006 | 0.025 | 0.012 | | 0.020 | 0.022 | 0.098 |
| **#383** | 0.218 | 0.240 | 0.022 | 0.097 | 0.008 | 0.035 | 0.015 | 0.008 | 0.008 | 0.022 | 0.099 |
| **#374** | 0.451 | 0.183 | 0.123 | 0.340 | 0.008 | 0.297 | 0.015 | 0.005 | 0.008 | 0.010 | 0.219 |
| **#388** | 0.267 | 0.139 | 0.016 | 0.150 | | 0.138 | 0.019 | 0.012 | 0.021 | 0.019 | 0.127 |
| **#387** | 0.100 | 0.218 | 0.018 | 0.187 | 0.004 | 0.061 | 0.016 | | 0.006 | 0.018 | 0.106 |
| **#389** | 0.151 | 0.246 | 0.011 | 0.156 | 0.006 | 0.031 | 0.015 | 0.006 | 0.013 | 0.010 | 0.028 |
| **#432** | 0.266 | 0.336 | 0.013 | 0.052 | 0.005 | 0.010 | 0.016 | 0.004 | 0.008 | 0.007 | 0.052 |
| **#433** | 0.186 | 0.235 | 0.008 | 0.074 | | 0.015 | 0.012 | 0.008 | 0.015 | 0.009 | 0.017 |
| **#445** | 0.315 | 0.549 | 0.013 | 0.188 | 0.000 | 0.135 | 0.011 | 0.007 | 0.009 | 0.017 | 0.071 |
| **#1073** | 0.399 | 1.029 | 0.064 | 0.221 | 0.010 | 0.208 | 0.014 | 0.138 | | 0.025 | 0.045 |
| **#870** | 0.313 | 0.368 | 0.027 | 0.060 | 0.013 | 0.042 | 0.014 | 0.048 | 0.015 | 0.035 | 0.145 |
| **#878** | 0.492 | 1.237 | 0.015 | 0.072 | 0.001 | 0.035 | 0.017 | 0.018 | 0.003 | 0.019 | 0.128 |
| **#880** | 0.376 | 0.337 | 0.022 | 0.066 | 0.001 | 0.040 | 0.019 | 0.005 | 0.004 | 0.028 | 0.091 |
| **#872** | 0.668 | 0.366 | 0.153 | 0.321 | 0.030 | 0.188 | 0.018 | 0.203 | 0.022 | 0.023 | 0.203 |
| **#871** | 1.590 | 0.501 | 0.330 | 0.448 | 0.071 | 0.200 | 0.013 | 0.245 | 0.012 | 0.031 | 0.135 |
| **#874** | 0.380 | 0.346 | 0.027 | 0.162 | 0.004 | 0.087 | 0.010 | | 0.015 | 0.021 | 0.148 |
| **#955** | 1.144 | 1.350 | 0.071 | 0.211 | 0.011 | 0.077 | 0.016 | 0.075 | 0.016 | 0.014 | 0.150 |
| **#952** | 0.220 | 0.253 | 0.006 | 0.043 | 0.008 | 0.025 | 0.023 | 0.013 | 0.016 | 0.011 | 0.089 |
| **#953** | 0.133 | 0.376 | 0.004 | 0.046 | 0.008 | 0.012 | 0.011 | 0.005 | 0.013 | 0.018 | 0.156 |
| **#949** | 0.225 | 0.319 | 0.011 | 0.038 | 0.006 | 0.025 | 0.013 | 0.009 | 0.006 | 0.013 | 0.259 |
| **#948** | 0.362 | 0.893 | 0.014 | 0.034 | 0.004 | 0.169 | 0.017 | 0.019 | | 0.032 | 0.176 |
| **#951** | 0.254 | 0.323 | 0.016 | 0.013 | 0.004 | 0.007 | 0.011 | 0.005 | 0.018 | 0.021 | 0.048 |
| **#966** | 0.201 | 0.277 | 0.019 | 0.027 | 0.007 | 0.019 | 0.016 | 0.019 | | 0.021 | 0.154 |
| **#988** | 0.057 | 0.178 | 0.009 | 0.039 | 0.003 | 0.006 | 0.015 | 0.005 | 0.009 | 0.008 | 0.112 |

Average compositions of olivines (wt%).

|      | N# | MgO | SiO2 | Na2O | Al2O3 | K2O | CaO | TiO2 | S | Cr2O3 | MnO | FeO | Total |
|------|----|-----|------|------|-------|-----|-----|------|---|-------|-----|-----|-------|
| **#951** | 6 | 57.81 | 42.08 | 0.01 | 0.02 | 0.01 | 0.07 | 0.02 | 0.04 | n.m. | 0.08 | 0.64 | 100.77 |
| **#966** | 14 | 57.31 | 42.24 | 0.03 | 0.02 | 0.00 | 0.06 | 0.01 | 0.02 | n.m. | 0.09 | 0.55 | 100.34 |
| **#988** | 7 | 57.58 | 43.39 | 0.00 | 0.01 | b.d.l. | 0.03 | 0.01 | 0.01 | 0.01 | 0.08 | 0.29 | 101.41 |

Standard deviations (1 sigma) for olivine compositions.

|      | MgO | SiO2 | Na2O | Al2O3 | K2O | CaO | TiO2 | S | Cr2O3 | MnO | FeO |
|------|-----|------|------|-------|-----|-----|------|---|-------|-----|-----|
| **#951** | 0.111 | 0.254 | 0.009 | 0.014 | 0.007 | 0.017 | 0.015 | 0.020 | | 0.015 | 0.079 |
| **#966** | 0.211 | 0.476 | 0.015 | 0.010 | 0.003 | 0.020 | 0.013 | 0.016 | | 0.023 | 0.319 |
| **#988** | 0.144 | 0.108 | 0.014 | 0.007 | | 0.010 | 0.010 | 0.003 | 0.014 | 0.011 | 0.103 |